\documentclass[longauth]{aa}  
\usepackage{amsmath,amssymb,amsfonts}
\usepackage{graphicx}
\usepackage{txfonts}
\usepackage{enumerate} 
\usepackage{multirow}
\usepackage{enumitem}
\usepackage{colordvi} 
\usepackage{bm}
\usepackage{color, colortbl}
\usepackage[T1]{fontenc} 
\usepackage{braket}
\usepackage{euclid}
\usepackage{placeins}
\usepackage{tabularx}
\usepackage{subcaption}
\usepackage{afterpage}
\usepackage[normalem]{ulem} 

\usepackage{booktabs}

\DeclareSIUnit \hGpc{\per\h\giga\parsec}
\DeclareSIUnit \hsolarmass{\per\h\solarmass}

\usepackage{hyperref}

\makeatletter
\renewcommand*\aa@pageof{, page \thepage{} of \pageref*{LastPage}}
\makeatother

\newcommand*{\hGpc}{\si{\per\h\giga\parsec}}
\newcommand*{\hsolarmass}{\si{\per\h\solarmass}}

\renewcommand{\bea}{\begin{eqnarray}}
\renewcommand{\eea}{\end{eqnarray}}
\newcommand{\bn}{{{\bm n}}}

\newcommand{\De}{\Delta}
\newcommand{\HH}{\mathcal{H}}
\newcommand{\al}{\alpha}
\newcommand{\bal}{\boldsymbol{\al}}

\newcommand{\de}{\delta}
\renewcommand{\ga}{\gamma}
\newcommand{\ka}{\kappa}
\newcommand{\Om}{\Omega}
\newcommand{\ds}{{\slash\hspace{-5pt}\partial}}

\newcommand{\wzero}{w_{0}}
\newcommand{\wa}{w_a}
\newcommand{\gradientsphere}{\boldsymbol{\nabla}_{\Omega}}
\newcommand{\laplaciansphere}{\Delta_{\Omega}}
\newcommand{\omegamatter}{\Om_{\text{m,0}}}
\newcommand{\ns}{n_{\text{s}}}
\newcommand{\fisher}{\tens{F}}
\newcommand{\covariance}{\tens{C}}

\definecolor{amaranth}{rgb}{0.9, 0.17, 0.31}
\definecolor{forestgreen(web)}{rgb}{0.13, 0.55, 0.13}
\definecolor{lavender(web)}{rgb}{0.9, 0.9, 0.98}
\definecolor{cosmiclatte}{rgb}{1.0, 0.97, 0.91}
\definecolor{jonquil}{rgb}{0.98, 0.85, 0.37}
\definecolor{khaki(x11)(lightkhaki)}{rgb}{0.94, 0.9, 0.55}
\definecolor{thistle}{rgb}{0.85, 0.75, 0.85}

\defcitealias{Blanchard:2019oqi}{EC20}

\usepackage[nameinlink,noabbrev]{cleveref}
\Crefname{equation}{Eq.}{Eqs}
\Crefname{section}{Sect.}{Sects}
\Crefname{figure}{Fig.}{Figs}
\crefname{equation}{Equation}{Equations}
\crefname{section}{Section}{Sections}
\crefname{figure}{Figure}{Figures}

\begin{document}

\title{\Euclid preparation: XIX. Impact of magnification on photometric galaxy clustering}

\author{Euclid Collaboration: F.~Lepori$^{1,7}$\thanks{\email{francesca.lepori2@uzh.ch}}, I.~Tutusaus$^{2,3}$, C.~Viglione$^{2,3}$, C.~Bonvin$^{1}$, S.~Camera$^{4,5,6}$, F.J.~Castander$^{2,3}$, R.~Durrer$^{1}$, P.~Fosalba$^{2,3}$, G.~Jelic-Cizmek$^{1}$, M.~Kunz$^{1}$
% ...
}

\institute{$^{1}$ Universit\'e de Gen\`eve, D\'epartement de Physique Th\'eorique and Centre for Astroparticle Physics, 24 quai Ernest-Ansermet, CH-1211 Gen\`eve 4, Switzerland\\
$^{2}$ Institute of Space Sciences (ICE, CSIC), Campus UAB, Carrer de Can Magrans, s/n, 08193 Barcelona, Spain\\
$^{3}$ Institut d’Estudis Espacials de Catalunya (IEEC), Carrer Gran Capit\'a 2-4, 08034 Barcelona, Spain\\
$^{4}$ INFN-Sezione di Torino, Via P. Giuria 1, I-10125 Torino, Italy\\
$^{5}$ Dipartimento di Fisica, Universit\'a degli Studi di Torino, Via P. Giuria 1, I-10125 Torino, Italy\\
$^{6}$ INAF-Osservatorio Astrofisico di Torino, Via Osservatorio 20, I-10025 Pino Torinese (TO), Italy\\
$^{7}$ Institute for Computational Science, University of Zurich, Winterthurerstrasse 190, 8057 Zurich, Switzerland\\
$^{8}$ AIM, CEA, CNRS, Universit\'{e} Paris-Saclay, Universit\'{e} de Paris, F-91191 Gif-sur-Yvette, France\\
$^{9}$ Instituto de F\'isica Te\'orica UAM-CSIC, Campus de Cantoblanco, E-28049 Madrid, Spain\\
$^{10}$ Universit\'e St Joseph; UR EGFEM, Faculty of Sciences, Beirut, Lebanon\\
$^{11}$ Institut de Recherche en Astrophysique et Plan\'etologie (IRAP), Universit\'e de Toulouse, CNRS, UPS, CNES, 14 Av. Edouard Belin, F-31400 Toulouse, France\\
$^{12}$ Departamento de F\'isica, FCFM, Universidad de Chile, Blanco Encalada 2008, Santiago, Chile\\
$^{13}$ Institute of Cosmology and Gravitation, University of Portsmouth, Portsmouth PO1 3FX, UK\\
$^{14}$ INAF-Osservatorio di Astrofisica e Scienza dello Spazio di Bologna, Via Piero Gobetti 93/3, I-40129 Bologna, Italy\\
$^{15}$ Max Planck Institute for Extraterrestrial Physics, Giessenbachstr. 1, D-85748 Garching, Germany\\
$^{16}$ INFN-Sezione di Roma Tre, Via della Vasca Navale 84, I-00146, Roma, Italy\\
$^{17}$ Department of Mathematics and Physics, Roma Tre University, Via della Vasca Navale 84, I-00146 Rome, Italy\\
$^{18}$ INAF-Osservatorio Astronomico di Capodimonte, Via Moiariello 16, I-80131 Napoli, Italy\\
$^{19}$ Centro de Astrof\'{\i}sica da Universidade do Porto, Rua das Estrelas, 4150-762 Porto, Portugal\\
$^{20}$ Instituto de Astrof\'isica e Ci\^encias do Espa\c{c}o, Universidade do Porto, CAUP, Rua das Estrelas, PT4150-762 Porto, Portugal\\
$^{21}$ INAF-IASF Milano, Via Alfonso Corti 12, I-20133 Milano, Italy\\
$^{22}$ Port d'Informaci\'{o} Cient\'{i}fica, Campus UAB, C. Albareda s/n, 08193 Bellaterra (Barcelona), Spain\\
$^{23}$ Institut de F\'{i}sica d’Altes Energies (IFAE), The Barcelona Institute of Science and Technology, Campus UAB, 08193 Bellaterra (Barcelona), Spain\\
$^{24}$ INAF-Osservatorio Astronomico di Roma, Via Frascati 33, I-00078 Monteporzio Catone, Italy\\
$^{25}$ Department of Physics "E. Pancini", University Federico II, Via Cinthia 6, I-80126, Napoli, Italy\\
$^{26}$ INFN section of Naples, Via Cinthia 6, I-80126, Napoli, Italy\\
$^{27}$ Dipartimento di Fisica e Astronomia ''Augusto Righi'' - Alma Mater Studiorum Universit\'a di Bologna, Viale Berti Pichat 6/2, I-40127 Bologna, Italy\\
$^{28}$ INAF-Osservatorio Astrofisico di Arcetri, Largo E. Fermi 5, I-50125, Firenze, Italy\\
$^{29}$ Institut national de physique nucl\'eaire et de physique des particules, 3 rue Michel-Ange, 75794 Paris C\'edex 16, France\\
$^{30}$ Centre National d'Etudes Spatiales, Toulouse, France\\
$^{31}$ Institute for Astronomy, University of Edinburgh, Royal Observatory, Blackford Hill, Edinburgh EH9 3HJ, UK\\
$^{32}$ Jodrell Bank Centre for Astrophysics, School of Physics and Astronomy, University of Manchester, Oxford Road, Manchester M13 9PL, UK\\
$^{33}$ European Space Agency/ESRIN, Largo Galileo Galilei 1, 00044 Frascati, Roma, Italy\\
$^{34}$ ESAC/ESA, Camino Bajo del Castillo, s/n., Urb. Villafranca del Castillo, 28692 Villanueva de la Ca\~nada, Madrid, Spain\\
$^{35}$ Univ Lyon, Univ Claude Bernard Lyon 1, CNRS/IN2P3, IP2I Lyon, UMR 5822, F-69622, Villeurbanne, France\\
$^{36}$ Institute of Physics, Laboratory of Astrophysics, Ecole Polytechnique F\'{e}d\'{e}rale de Lausanne (EPFL), Observatoire de Sauverny, 1290 Versoix, Switzerland\\
$^{37}$ Departamento de F\'isica, Faculdade de Ci\^encias, Universidade de Lisboa, Edif\'icio C8, Campo Grande, PT1749-016 Lisboa, Portugal\\
$^{38}$ Instituto de Astrof\'isica e Ci\^encias do Espa\c{c}o, Faculdade de Ci\^encias, Universidade de Lisboa, Campo Grande, PT-1749-016 Lisboa, Portugal\\
$^{39}$ Department of Astronomy, University of Geneva, ch. d\'Ecogia 16, CH-1290 Versoix, Switzerland\\
$^{40}$ Universit\'e Paris-Saclay, CNRS, Institut d'astrophysique spatiale, 91405, Orsay, France\\
$^{41}$ INFN-Padova, Via Marzolo 8, I-35131 Padova, Italy\\
$^{42}$ University of Lyon, UCB Lyon 1, CNRS/IN2P3, IUF, IP2I Lyon, France\\
$^{43}$ Aix-Marseille Univ, CNRS/IN2P3, CPPM, Marseille, France\\
$^{44}$ Istituto Nazionale di Astrofisica (INAF) - Osservatorio di Astrofisica e Scienza dello Spazio (OAS), Via Gobetti 93/3, I-40127 Bologna, Italy\\
$^{45}$ Istituto Nazionale di Fisica Nucleare, Sezione di Bologna, Via Irnerio 46, I-40126 Bologna, Italy\\
$^{46}$ INAF-Osservatorio Astronomico di Padova, Via dell'Osservatorio 5, I-35122 Padova, Italy\\
$^{47}$ Universit\"ats-Sternwarte M\"unchen, Fakult\"at f\"ur Physik, Ludwig-Maximilians-Universit\"at M\"unchen, Scheinerstrasse 1, 81679 M\"unchen, Germany\\
$^{48}$ Dipartimento di Fisica "Aldo Pontremoli", Universit\'a degli Studi di Milano, Via Celoria 16, I-20133 Milano, Italy\\
$^{49}$ INFN-Sezione di Milano, Via Celoria 16, I-20133 Milano, Italy\\
$^{50}$ Institute of Theoretical Astrophysics, University of Oslo, P.O. Box 1029 Blindern, N-0315 Oslo, Norway\\
$^{51}$ Jet Propulsion Laboratory, California Institute of Technology, 4800 Oak Grove Drive, Pasadena, CA, 91109, USA\\
$^{52}$ von Hoerner \& Sulger GmbH, Schlo{\ss}Platz 8, D-68723 Schwetzingen, Germany\\
$^{53}$ Max-Planck-Institut f\"ur Astronomie, K\"onigstuhl 17, D-69117 Heidelberg, Germany\\
$^{54}$ Institut d'Astrophysique de Paris, 98bis Boulevard Arago, F-75014, Paris, France\\
$^{55}$ Mullard Space Science Laboratory, University College London, Holmbury St Mary, Dorking, Surrey RH5 6NT, UK\\
$^{56}$ Department of Physics and Helsinki Institute of Physics, Gustaf H\"allstr\"omin katu 2, 00014 University of Helsinki, Finland\\
$^{57}$ NOVA optical infrared instrumentation group at ASTRON, Oude Hoogeveensedijk 4, 7991PD, Dwingeloo, The Netherlands\\
$^{58}$ INAF-Osservatorio Astronomico di Trieste, Via G. B. Tiepolo 11, I-34131 Trieste, Italy\\
$^{59}$ Argelander-Institut f\"ur Astronomie, Universit\"at Bonn, Auf dem H\"ugel 71, 53121 Bonn, Germany\\
$^{60}$ Dipartimento di Fisica e Astronomia “Augusto Righi” - Alma Mater Studiorum Università di Bologna, via Piero Gobetti 93/2, I-40129 Bologna, Italy\\
$^{61}$ INFN-Sezione di Bologna, Viale Berti Pichat 6/2, I-40127 Bologna, Italy\\
$^{62}$ Institute for Computational Cosmology, Department of Physics, Durham University, South Road, Durham, DH1 3LE, UK\\
$^{63}$ Universit\'e C\^{o}te d'Azur, Observatoire de la C\^{o}te d'Azur, CNRS, Laboratoire Lagrange, Bd de l'Observatoire, CS 34229, 06304 Nice cedex 4, France\\
$^{64}$ University of Applied Sciences and Arts of Northwestern Switzerland, School of Engineering, 5210 Windisch, Switzerland\\
$^{65}$ California institute of Technology, 1200 E California Blvd, Pasadena, CA 91125, USA\\
$^{66}$ European Space Agency/ESTEC, Keplerlaan 1, 2201 AZ Noordwijk, The Netherlands\\
$^{67}$ Department of Physics and Astronomy, University of Aarhus, Ny Munkegade 120, DK–8000 Aarhus C, Denmark\\
$^{68}$ Perimeter Institute for Theoretical Physics, Waterloo, Ontario N2L 2Y5, Canada\\
$^{69}$ Department of Physics and Astronomy, University of Waterloo, Waterloo, Ontario N2L 3G1, Canada\\
$^{70}$ Centre for Astrophysics, University of Waterloo, Waterloo, Ontario N2L 3G1, Canada\\
$^{71}$ Institute of Space Science, Bucharest, Ro-077125, Romania\\
$^{72}$ Dipartimento di Fisica e Astronomia “G.Galilei", Universit\'a di Padova, Via Marzolo 8, I-35131 Padova, Italy\\
$^{73}$ Centro de Investigaciones Energ\'eticas, Medioambientales y Tecnol\'ogicas (CIEMAT), Avenida Complutense 40, 28040 Madrid, Spain\\
$^{74}$ Instituto de Astrof\'isica e Ci\^encias do Espa\c{c}o, Faculdade de Ci\^encias, Universidade de Lisboa, Tapada da Ajuda, PT-1349-018 Lisboa, Portugal\\
$^{75}$ Universidad Polit\'ecnica de Cartagena, Departamento de Electr\'onica y Tecnolog\'ia de Computadoras, 30202 Cartagena, Spain\\
$^{76}$ Kapteyn Astronomical Institute, University of Groningen, PO Box 800, 9700 AV Groningen, The Netherlands\\
$^{77}$ Infrared Processing and Analysis Center, California Institute of Technology, Pasadena, CA 91125, USA\\
$^{78}$ INAF-Osservatorio Astronomico di Brera, Via Brera 28, I-20122 Milano, Italy\\
$^{79}$ Center for Computational Astrophysics, Flatiron Institute, 162 5th Avenue, 10010, New York, NY, USA\\
$^{80}$ School of Physics and Astronomy, Cardiff University, The Parade, Cardiff, CF24 3AA, UK\\
$^{81}$ Universit\'e de Paris, CNRS, Astroparticule et Cosmologie, F-75013 Paris, France\\
$^{82}$ IFPU, Institute for Fundamental Physics of the Universe, via Beirut 2, 34151 Trieste, Italy\\
$^{83}$ SISSA, International School for Advanced Studies, Via Bonomea 265, I-34136 Trieste TS, Italy\\
$^{84}$ INFN, Sezione di Trieste, Via Valerio 2, I-34127 Trieste TS, Italy\\
$^{85}$ Departamento de Astrof\'{i}sica, Universidad de La Laguna, E-38206, La Laguna, Tenerife, Spain\\
$^{86}$ Instituto de Astrof\'isica de Canarias, Calle V\'ia L\'actea s/n, E-38204, San Crist\'obal de La Laguna, Tenerife, Spain\\
$^{87}$ Institut de Physique Th\'eorique, CEA, CNRS, Universit\'e Paris-Saclay F-91191 Gif-sur-Yvette Cedex, France\\
$^{88}$ Dipartimento di Fisica - Sezione di Astronomia, Universit\'a di Trieste, Via Tiepolo 11, I-34131 Trieste, Italy\\
$^{89}$ INFN-Bologna, Via Irnerio 46, I-40126 Bologna, Italy\\
$^{90}$ Dipartimento di Fisica e Scienze della Terra, Universit\'a degli Studi di Ferrara, Via Giuseppe Saragat 1, I-44122 Ferrara, Italy\\
$^{91}$ INAF, Istituto di Radioastronomia, Via Piero Gobetti 101, I-40129 Bologna, Italy\\
$^{92}$ INAF-IASF Bologna, Via Piero Gobetti 101, I-40129 Bologna, Italy\\
$^{93}$ Aix-Marseille Univ, CNRS, CNES, LAM, Marseille, France\\
$^{94}$ INFN-Sezione di Genova, Via Dodecaneso 33, I-16146, Genova, Italy\\
$^{95}$ INAF-Istituto di Astrofisica e Planetologia Spaziali, via del Fosso del Cavaliere, 100, I-00100 Roma, Italy\\
$^{96}$ Department of Physics, Oxford University, Keble Road, Oxford OX1 3RH, UK\\
$^{97}$ Department of Physics, P.O. Box 64, 00014 University of Helsinki, Finland\\
$^{98}$ Department of Physics, Lancaster University, Lancaster, LA1 4YB, UK\\
$^{99}$ Universit\'e PSL, Observatoire de Paris, Sorbonne Universit\'e, CNRS, LERMA, F-75014, Paris, France\\
$^{100}$ Department of Physics and Astronomy, University College London, Gower Street, London WC1E 6BT, UK\\
$^{101}$ Helsinki Institute of Physics, Gustaf H{\"a}llstr{\"o}min katu 2, University of Helsinki, Helsinki, Finland\\
$^{102}$ Centre de Calcul de l'IN2P3, 21 avenue Pierre de Coubertin F-69627 Villeurbanne Cedex, France\\
$^{103}$ Dipartimento di Fisica, Sapienza Universit\`a di Roma, Piazzale Aldo Moro 2, I-00185 Roma, Italy\\
$^{104}$ Institut f\"ur Theoretische Physik, University of Heidelberg, Philosophenweg 16, 69120 Heidelberg, Germany\\
$^{105}$ Zentrum f\"ur Astronomie, Universit\"at Heidelberg, Philosophenweg 12, D- 69120 Heidelberg, Germany\\
$^{106}$ INFN, Sezione di Lecce, Via per Arnesano, CP-193, I-73100, Lecce, Italy\\
$^{107}$ Department of Mathematics and Physics E. De Giorgi, University of Salento, Via per Arnesano, CP-I93, I-73100, Lecce, Italy\\
$^{108}$ Department of Physics, P.O.Box 35 (YFL), 40014 University of Jyv\"askyl\"a, Finland\\
$^{109}$ Ruhr University Bochum, Faculty of Physics and Astronomy, Astronomical Institute (AIRUB), German Centre for Cosmological Lensing (GCCL), 44780 Bochum, Germany\\
}

\date{\today}

\authorrunning{Euclid Collaboration}

\titlerunning{\Euclid preparation. XIX.}
  \abstract
   {}
   {We investigate the importance of lensing magnification for estimates of galaxy clustering and its cross-correlation with shear for the photometric sample of \Euclid. Using updated specifications, we study the impact of lensing magnification on the constraints and the shift in the estimation of the best fitting cosmological parameters that we expect if this effect is neglected. 
  }
   {We follow the prescriptions of the official \Euclid Fisher matrix forecast
   for the photometric galaxy clustering analysis and the combination of photometric clustering and cosmic shear. 
   The slope of the luminosity function (local count slope), which regulates the amplitude of the lensing magnification, and the galaxy bias have been estimated from the \Euclid Flagship simulation. 
   }
   {
   We find that magnification significantly affects both the best-fit estimation of cosmological parameters and the constraints in the galaxy clustering analysis of the photometric sample. In particular, including magnification in the analysis reduces the 1$\sigma$ errors on $\Omega_{\text{m},0}, \wzero, \wa$ at the level of 20--35\%, depending on how well we will be able to independently measure the local count slope. In addition, we find that neglecting magnification in the clustering analysis leads to shifts of up to 1.6$\sigma$ in the best-fit parameters.
In the joint analysis of galaxy clustering, cosmic shear, and galaxy--galaxy lensing, magnification
does not improve precision, but it leads to an up to 6$\sigma$ bias if neglected.
Therefore, for all models considered in this work, 
magnification has to be included in the analysis of galaxy clustering and its cross-correlation with the shear signal ($3\times2$pt analysis) for an accurate parameter estimation.
}
   {}
   
   \maketitle
   
\section{Introduction}
In the past few decades, observational cosmology has undergone unprecedented advances in terms of experimental techniques. The anisotropies of the cosmic microwave background (CMB) have been mapped with stunning accuracy~\citep{Aghanim:2018eyx},
and the low-redshift window has become accessible 
with observations of the large-scale distribution of galaxies and the statistics of weak gravitational lensing \citep{Alam:2016hwk,Abbott:2017wau,Lee:2021ebd,Sevilla-Noarbe:2020jpu,Abbott:2021bzy, KiDS:2020suj,Heymans:2020gsg},
as have distance measurements from supernovae~\citep{Scolnic:2017caz}. 
This progress on the experimental side
has led to the affirmation of  $\Lambda$ cold dark matter (CDM) as the concordance model for cosmology.
Despite the remarkable success of $\Lambda$CDM, 
there are two ingredients whose nature is still unknown: dark matter and dark energy. In addition, the value of the cosmological constant corresponds to a vacuum energy in the millielectronvolt regime, which is unsatisfactory from a theoretical point of view. Furthermore, the constancy of $\Lambda$ leads to the question of why its contribution to the expansion rate of the Universe should be of the same order of magnitude as the one from the matter density only at the present time. These fine-tuning and coincidence (`why now') problems motivate researchers in the field to consider alternatives to $\Lambda$CDM, such as scalar field dark energy (quintessence, k-essence) and more general tensor-scalar gravity theories or other modifications of general relativity (see e.g. \citealt{Amendola:2016saw} for an extended discussion). 
The next generation of large-scale structure probes is expected to provide crucial information on the dark sector that will allow us to test many of these different models of dark energy and our theory of gravity  on cosmological scales. Due to the statistical power of these future surveys, 
new efforts are needed to reduce systematic uncertainties to a higher degree than previously required. Such systematic effects arise not only from observational aspects, but also from the theoretical predictions that may have to be improved as well to exploit the full power of the upcoming observations.

The \Euclid survey
\citep{Amendola:2016saw, Laureijs:2011gra} will contribute to the challenge of constraining the dark sector with the combination of two complementary probes: a) a spectroscopic sample of  about $30$ million galaxies that will be used to study the growth of structure in the redshift range $z \in [0.9, 1.8]$ \citep{Pozzetti:2016cch} and
b) a photometric catalogue of about 1.5 billion galaxy images, which will provide a direct tomographic map of the distribution of matter through measurements of cosmic shear in the redshift range $z \in [0, 2]$ \citep{Amendola:2016saw}. 

In this paper, we  focus on the photometric sample. Galaxy images and positions in this sample will be used both for extracting the galaxies' shapes and their weak lensing (WL) distortions and for galaxy clustering measurements in
photometric redshift bins. 
However, the statistics of galaxy number counts are not only determined by the local density of sources; they are also affected by gravitational lensing due to the foreground matter distribution \citep{Menard:2002vz,  Menard:2002da, Menard:2002ia, Matsubara:2004fr, Scranton:2005ci, LoVerde:2007ke, Hui:2007tm, Hildebrandt:2009,  VanWaerbeke:2010yp, Heavens:2011ei,  Bonvin:2011bg,Challinor:2011bk,Duncan:2013haa, Unruh:2019pyk, Liu:2021gbm}.
Gravitational lensing affects
the observed number count of galaxies in two ways, which have opposite signs: it modifies the observed size of the solid angle, diluting the number of galaxies per unit of solid
angle behind an overdensity, and it magnifies the apparent luminosity of galaxies behind an overdensity, enhancing the number of galaxies
above the magnitude threshold of a given survey. The second effect is survey dependent. To model it, we need to know the luminosity function and the magnitude cut of the galaxies in the sample.
The combination of these two effects is known as `lensing magnification'.

Lensing magnification has not been taken into account in the validated \Euclid forecast \citep[][EC20 in the following]{Blanchard:2019oqi}, and the aim of this work is 
to assess its impact on the analysis of the \Euclid photometric sample. There has been extensive work in investigating the relevance of magnification for future cosmological surveys (see for example \citealt{Namikawa:2011, Bruni:2011ta, Gaztanaga_2012, Duncan:2013haa, Montanari:2015rga, Eriksen:2014zua, Eriksen:2015nha, Raccanelli:2015vla, Cardona:2016qxn, DiDio:2016ykq,  Eriksen:2015hqa, Lorenz:2017iez, Villa_2018, Thiele:2019fcu, Tanidis:2019fdh, Bellomo:2020pnw, Jelic-Cizmek:2020pkh, Viljoen:2021ocx}). The consensus is that lensing should be taken into account in the analysis of photometric clustering for the following reasons: i) Including lensing will significantly improve the cosmological  constraints by breaking the degeneracy between  galaxy bias and the amplitude of  primordial perturbations. This is especially relevant for photometric samples where redshift-space distortions (RSDs) are smeared out. ii) Neglecting this effect can lead to significant shifts in the estimation of some cosmological parameters -- especially for models beyond the minimal $\Lambda$CDM~\citep{Camera:2014sba, Lorenz:2017iez, Villa_2018}. iii) Lensing magnification provides a tomographic measurement of the lensing potential that is complementary to cosmic shear analysis and can be used to test general relativity~\citep{Montanari:2015rga}. 

In this work, we study the impact of lensing magnification on the analysis of the photometric sample of \Euclid, using for the first time realistic specifications for the local count slope based on the \Euclid Flagship simulation. 
Apart from $\Lambda$CDM and massive neutrinos, we consider a simple phenomenological parametrisation of dark energy as a function of redshift, $z$, via an equation of state of the form
$$w(z) = \wzero +\wa \frac{z}{1+z} \,,$$ which is the so-called Chevallier-Polarski-Linder (CPL), or $\wzero\wa$, parametrisation~\citep{Chevallier:2000qy, Linder_2003}. While these simple models do not fully allow one to explore the additional information that lensing magnification may add to photometric galaxy clustering ($\text{GCph}$) as a cosmological probe, they are sufficient to 
assess whether we need to include lensing magnification to avoid systematically biasing our results.  An extended analysis that includes dark energy models with a stronger impact on the growth of structure is beyond the scope of this paper and is left for future work. 

The paper is structured as follows. In the next section, we introduce the theoretical, linear perturbation theory expressions for the quantities measured in the survey. In \Cref{sec:flag} we present the \Euclid specifics used in this work, and we outline how they have been extracted from the Flagship simulation. 
In \Cref{s:met} we describe the Fisher formalism used in our analysis. In \Cref{sec:res} we present the results and discuss them. In \Cref{sec:rob-tests} we show the outcome of several tests that we performed to assess the robustness of our results. We conclude in \Cref{s:con}.
In the appendix we discuss in more detail some technical aspects of our work.
\section{The photometric sample observables: Number counts and cosmic shear}
In this section we define our observables: the galaxy number counts, the shear, and their cross-correlation. We consider them as quantities on the sphere at different redshifts. We first give a brief recap on power spectra and correlation functions on the sphere for different tensorial quantities. We then discuss our specific observables in more detail.

\subsection{Angular power spectra}\label{s:powerspec}
Whenever we have a function on the sphere, such as the number counts, $\De(\bn,z)$, the lensing potential, $\psi(\bn,z)$,  or the convergence, $\ka(\bn,z)$, observed in the direction $\bn$ at fixed redshift, $z$, or integrated over a redshift bin centred at $z$, we can expand it in spherical harmonics,
\bea
\De(\bn, z) &=& \sum_{\ell m}a^{\Delta}_{\ell m}(z)Y_{\ell m}(\bn)\,, \\
\ka(\bn, z) &=& \sum_{\ell m}a^\ka_{\ell m}(z)Y_{\ell m}(\bn) \,.
\eea

Due to statistical isotropy, which we assume here\footnote{Observational evidence of statistical isotropy in the galaxy distributions has been found for example in \cite{Blake:2002gx}, \cite{Alonso:2014xca}, and \cite{Bengaly:2016amk}. However, recent measurements of the local radio dipole exhibit an anomaly when compared to the CMB dipole, which may indicate a not purely kinematic origin (see for example \citealt{Siewert:2020krp}). The presence of a large-scale anisotropy has been also found in CMB data (see \citealt{Fosalba:2021} for details).}, the $a_{\ell m}$ coefficients for different $\ell$ and $m$ values are uncorrelated, and we obtain the angular power spectra
\bea
\left\langle a^{\Delta}_{\ell m}(z)\,a^{\Delta\, *}_{\ell' m'}(z')
\right\rangle &=& C_\ell^{\Delta\Delta}(z, z')\,\de_{\ell\ell'}^{\rm K}\de_{mm'}^{\rm K}\,,\\
\left\langle a^\ka_{\ell m}(z)\,a^{\ka\, *}_{\ell' m'}(z')
\right\rangle &=& C_\ell^{\ka\ka}(z, z')\,\de_{\ell\ell'}^{\rm K}\de_{mm'}^{\rm K}\,,\\
\left\langle a^{\Delta}_{\ell m}(z)\,a^{\ka\, *}_{\ell' m'}(z')
\right\rangle &=& C_\ell^{\Delta\ka}(z, z')\,\de_{\ell\ell'}^{\rm K}\de_{mm'}^{\rm K}\,,
\eea
where the symbol $\de_{ab}^{\rm K}$ denotes the Kronecker delta and the superscripts $\Delta$ and $\kappa$ denote the number counts and the convergence field as example of functions on the sphere for which we can compute the angular power spectrum. 
For Gaussian fluctuations, these power spectra contain the full statistical information. In the presence of non-Gaussianities, reduced higher-order spectra and other statistics contain additional information.
The fact that the power spectra depend on redshift is what makes clustering surveys so useful. They contain three-dimensional information, which we exploit in this case by considering several redshifts and their cross-correlations. 

For functions on the sphere, the link between the power spectrum and the correlation function is given by
\be
\langle f(\bn, z)\,f(\bn', z')\rangle = \frac{1}{4\pi}\sum_\ell (2\ell+1)\,\,C^{ff}_\ell(z, z')\,\,P_\ell(\bn\cdot\bn')\,,
\ee
where $P_\ell$ denotes the Legendre polynomial of degree $\ell$ and $f$ is the considered function on the sphere.

The shear is not a function, but a helicity-2 object on the sphere, which has to be expanded in spin-weighted spherical harmonics (see~\citealt{Bartelmann:1999yn} for an introduction). Denoting the complex shear by $\ga = \ga_1+\mathrm{i}\ga_2$ we can write
\be
\ga(\bn, z) = \sum_{\ell m}a^\ga_{\ell m}(z)\,_{\;2\!}Y_{\ell m}(\bn) \,.
\ee
Here $_{\;2\!}Y_{\ell m}$ are the spin-2 spherical harmonics (see e.g.\ \citealt{RuthBook} for details). The correlators
\be
\left\langle a^\ga_{\ell m}(z)\,a^{\ga\, *}_{\ell' m'}(z')
\right\rangle = C_\ell^{\ga\ga}(z, z')\,\de_{\ell\ell'}^{\rm K}\de_{mm'}^{\rm K}
\ee
denote the shear power spectrum. In order to compare the shear spectrum with the convergence $\ka$, we first act on $\ga$ with the spin-lowering operator $\ds^*$ (again, see e.g.\ \citealt{RuthBook} for details). This allows us to define the function
\bea
\beta(\bn, z) = (\ds^*)^2\ga(\bn, z) =
\sum_{\ell m}\sqrt{\frac{(\ell+2)!}{(\ell-2)!}}a^\ga_{\ell m}(z)\,Y_{\ell m}(\bn) \,.
\eea
For the second equality we made use of the identity
$$
 (\ds^*)^2_{\;2\!}Y_{\ell m}(\bn) =\sqrt{\frac{(\ell+2)!}{(\ell-2)!}}Y_{\ell m}(\bn) \,.
$$
The scalar quantity $\beta$ is actually just the Laplacian of $\ka$, which implies
\be\label{e:ka-ga}
[\ell(\ell+1)]^2C_\ell^{\ka\ka} = \frac{(\ell+2)!}{(\ell-2)!}C_\ell^{\ga\ga} \,.
\ee
On small angular scales, $\ell\gg 1$, these spectra therefore agree,
\be
C_\ell^{\ka\ka} \simeq C_\ell^{\ga\ga}\, .\label{eq:conv_shear}
\ee
A similar relation can be derived for the cross-correlation of the shear and a scalar function (see \Cref{ap:shear-func-cross} for details).

Given the power spectra correlating two quantities $A$ and $B$,  $C_\ell^{AB}(z,z')$, we can compute the corresponding spectra obtained from two bins $i$ and $j$ with (normalised) galaxy distributions $n_i(z)$ and $n_j(z)$. They are simply given by
\be
C^{AB}_{\ell}(i, j) = \int \diff z \,\diff z'\,
n_i(z)\,n_j(z')\,C^{AB}_{\ell}(z, z') \,.
\ee

The observables $AB$ used in this paper are the galaxy number counts $\De\De$, the cosmic shear $\gamma\gamma$ and their cross-correlation $\De\gamma$ (galaxy--galaxy lensing). 
We discuss them in more detail in the following section. 

\subsection{Galaxy number counts}\label{s:galaxycounts}
The clustering of matter in the Universe is a very promising observable not only to determine cosmological parameters but also to test the theory of gravity, general relativity, on cosmological scales. While we cannot observe the matter density directly, it is generally assumed that on large scales the distribution of galaxies is a faithful biased tracer of the matter distribution. On large enough scales (roughly $\ell < 500$), the bias depends on redshift but not on scale~\citep[see for example][]{Fosalba:2013mra}. An important issue is, however, that we  do not observe galaxies in a three-dimensional spatial hypersurface but on our past light cone. More precisely, we measure angular positions and redshifts, which are affected  by the perturbed geometry and the peculiar motion of galaxies. While the galaxy velocities have been taken into account in galaxy number counts since the seminal paper by~\cite{Kaiser1987}, the fully relativistic perturbed light-cone projection has been considered first about a decade ago. In{~\cite{Yoo:2009}, \cite{Yoo:2010ni}, \cite{Bonvin:2011bg}, and \cite{Challinor:2011bk} these light-cone or projection effects  have been studied at first order in perturbation theory. A numerical code for the fast calculation of all relativistic effects is presented in~\cite{CLASSgal}, with vanishing curvature, and~\cite{DiDio:2016ykq}, including curvature. These codes are publicly available and included in the newer releases of {\sc class}~\citep{class2}. Attempts to go to second order in the light-cone projection have also been published~\citep{Bertacca:2014wga,Yoo:2014sfa,DiDio:2014lka}.

On small scales, $k\gg \HH/c$, where $k$ is the comoving wave number, $c$ is the speed of light, and $\HH$ denotes the comoving Hubble parameter, $\HH(z) = \frac{1}{a}\frac{\diff a}{\diff \eta}$, with $a$ the scale factor and $\eta$ conformal time, only  density, peculiar velocity (which enters through RSDs) and lensing magnification are relevant. These terms lead to the following simple formulae in angular and redshift space
\bea
\De(\bn, z) &=& b(z)\,\de[r(z)\bn,z] - \frac{1}{\HH(z)}\partial_rV_r[r(z)\bn, z] \nonumber \\
 && {+} [5s(z)-2]\,\kappa(\bn, z)\,, \qquad  \label{e:numbercount}\\  
\kappa(\bn, z) & = & \int_0^{r(z)}\hspace{-2mm}\diff r'\frac{[r(z)-r']}{2 r(z)r'}\,\De_\Omega(\Phi+\Psi)[r'\bn, z(r')]  \qquad \label{e:kappa-def1}\\
 & = & \frac{1}{2}\De_\Omega\psi(\bn, z) \, , \label{e:kappa-def}
\eea
where the unit vector $\bn$ denotes the direction of observation, $z$ is the measured redshift, $V_r = -\, \mathbf{V} \cdot \bn$ is the peculiar velocity in longitudinal gauge $\mathbf{V}$ projected along the  radial direction, and $\De_\Omega$ is the Laplace operator on the sphere.\footnote{The operator $\De_\Omega$ is defined in terms of the spin lowering and raising operators $\ds^*$ and $\ds$, that is, $\De_\Omega \equiv (\ds\,\,\ds^* + \ds^*\, \ds)/2$ {\citep[see][for details]{Bernardeau:2010}}.}
Here, $\psi$ is the lensing potential\footnote{We use the sign convention of~\cite{Bartelmann:1999yn} for the lensing potential, which is the opposite of the one in~\cite{Lewis:1999bs}.}, $b(z)$ is the galaxy bias, $r(z)$ is the comoving distance out to redshift $z$ and $z(r)$ is its inverse. $\Phi$ and $\Psi$ are the Bardeen potentials, which in $\Lambda$CDM are related to the Newtonian potential by $\Psi\simeq\Phi\simeq \Phi_{\rm Newton}/c^2$. The function $s(z)$ is the 
local count slope\footnote{In the literature this is often called the `magnification bias'.} given by the logarithmic derivative of the cumulative number density of galaxies as a function of their flux $F$ measured at the flux limit of the survey under consideration, $F_{\lim}$. More precisely,
\be
\frac{5}{2}s(z, F_{\rm lim}) \equiv - 
\frac{\partial\logten N(z,F> F_{\rm lim})}{\partial\logten F_{\rm lim}}\,.
\label{e:s_mlim}
\ee

Contrary to the bias $b(z)$, which is estimated through the clustering analysis together with the cosmological parameters, the local count slope $s(z)$ can in principle be measured directly from the luminosity function of the galaxy sample, which provides a measurement independent of the cosmological analysis.

The angular power spectrum of galaxy clustering is given by
\begin{align}
&C_\ell^{\Delta\Delta}(z,z')=C_\ell^{\rm gg}(z,z')+[5s(z')-2]\,C_\ell^{\text{g}\kappa}(z,z')\label{eq:ClDelta}\\
&+[5s(z)-2]\,C_\ell^{\kappa \rm g}(z,z')+[5s(z)-2][5s(z')-2]\,C_\ell^{\kappa\kappa}(z,z')\nonumber\\
&+C_\ell^{\rm RSD}(z,z')\nonumber\, ,
\end{align}
where the term in the last line contains the RSD-RSD correlation as well as the density-RSD and the magnification-RSD correlations.

In our analysis, we used Limber's approximation for these spectra~\citep{Limber:1954zz}, which is very good for the lensing potential and for $\ell\gtrsim 30$. 
We also made use of the Einstein constraint equation in the late Universe, where radiation can be neglected, such that
\be
P_{\Phi+\Psi}(k, z) = 9\left(\frac{H_0}{k}\right)^4\omegamatter ^2(1+z)^2\,P_{\delta\delta}(z, k)\,.
\ee

Here $P_{\delta\delta}$ is the matter power spectrum in comoving gauge, and $P_{\Phi+\Psi}$ is the power spectrum of the two Bardeen potentials (which are equal in our regime), which enters into the computation of the convergence in Eq.~\eqref{e:kappa-def}, and $\omegamatter $ is the matter density parameter.
Using Limber's approximation~\citep{Limber:1954zz}, the galaxy-magnification correlation in Eq.~\eqref{eq:ClDelta} can be written as
\bea
C_\ell^{\text{g}\ka}(z, z') &\hspace{-0.2cm}=&\hspace{-0.2cm}
\begin{cases}
\begin{aligned}[t]
6\,b(z)\omegamatter\, \left(\frac{H_0}{c}\right)^2\frac{\ell(\ell+1)}{(2\ell+1)^2} \frac{[r(z')-r(z)]}{r(z')r(z)}  
\\
\times\, (1+z)\,P_{\de\de}\left[\frac{\ell+1/2}{r(z)},z\right]\,,
\end{aligned}
& z < z' \\
 0\,, & z\ge z'\,,
\end{cases}
\label{e:ggl-limber}
\eea
and the magnification-magnification correlation becomes 
\bea
C_\ell^{\kappa\kappa}(z, z') &=& \left(\frac{2 H_0}{c}\right)^4(3\omegamatter )^2\frac{\ell^2(\ell+1)^2}{(2\ell+1)^4}  \label{e:kappa_limber}
\\  &&  \hspace*{-1.7cm}
\times\,\int_0^{r_{\min}}
\diff r \frac{[r(z)-r][r(z')-r]}{r(z)r(z')}[1+z(r)]^2\,P_{\delta\delta}\left(\frac{\ell+1/2}{r}, z\right) \,, \nonumber
\eea
where $r_{\min} = \text{min}\{r(z), r(z')\} $ (for more details on Limber's approximation, see e.g.\ \citealt{RuthBook}).

In Fig.\ \ref{fig:ncount1} we show the main contributions to the galaxy number counts for the \Euclid specifics described in \Cref{sec:flag}. We show two representative configurations: the auto-correlation at mean redshift $\bar{z}_1 = \bar{z}_2 = 0.69$, where the density contribution dominates, and the cross-correlation of two far-apart redshift bins, $\bar{z}_1 = 0.14$ and $\bar{z}_2 = 1.91$, where the entire signal consists of the cross-correlation of density at $\bar{z}_1$ and magnification at $\bar{z}_2$.

While RSDs, the second term on the first line of Eq. \eqref{e:numbercount}, are very important for spectroscopic surveys, they are smeared out in photometric surveys: their contribution to the auto-correlations is $\sim 30\%$ at $\ell \sim 10$ and drops below $1\%$ at $\ell > 90$. For this reason, they have been neglected in the official forecast presented in \citetalias{Blanchard:2019oqi}. In this paper, we focus on lensing magnification. Therefore, we neglect RSDs in the main analysis presented in this manuscript, and we test the impact of this approximation on our results in \Cref{sec:rob-tests}. 
A detailed study on the impact of RSDs on the \Euclid analysis is left to future work, as it has been pointed out in \cite{Tanidis:2019teo} that correct modelling of RSDs is crucial so as not to bias cosmological parameter estimation. 

\begin{figure}[!ht]
\centering
\begin{subfigure}[]{0.48\textwidth}
   \includegraphics[width=1\linewidth]{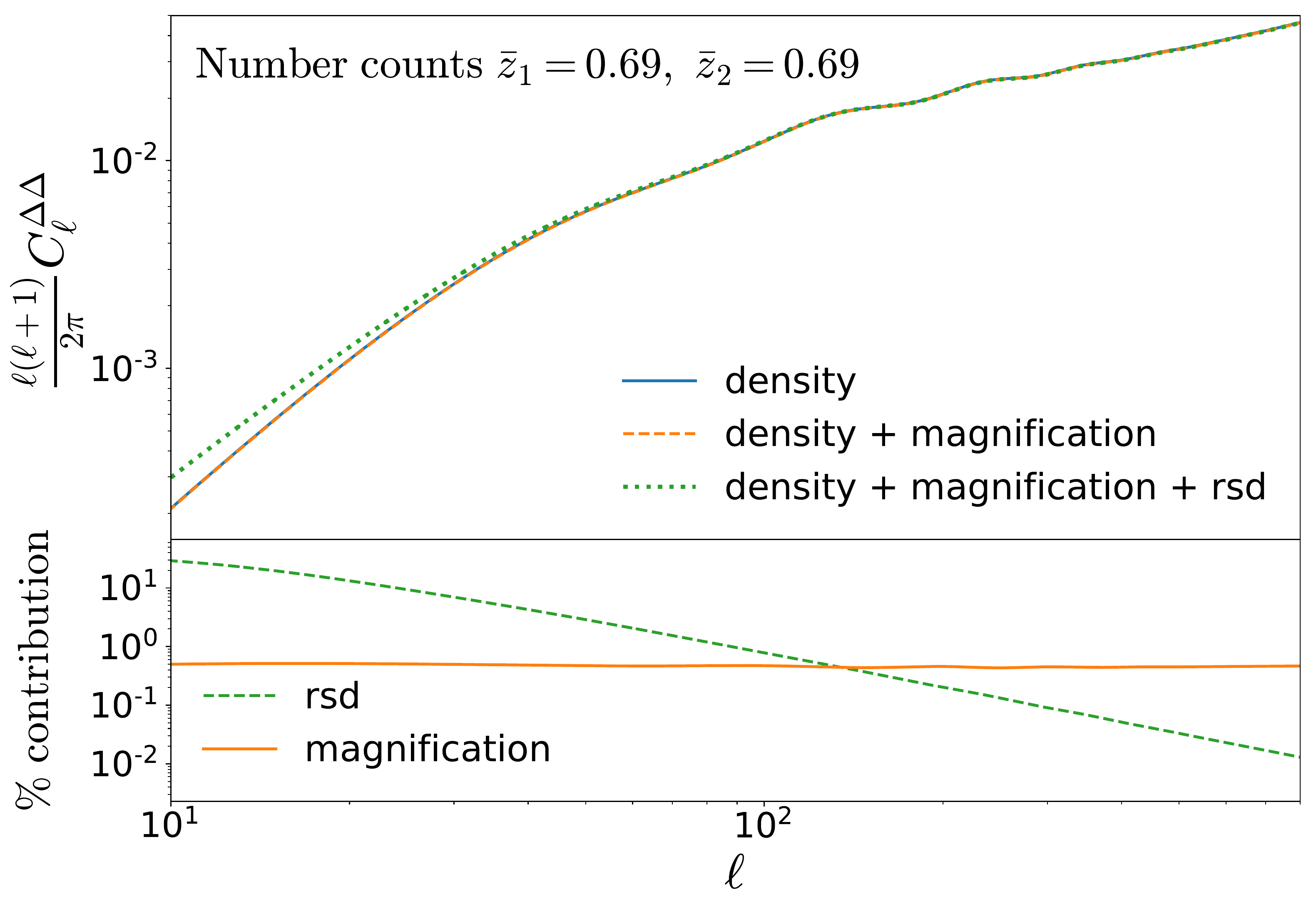}
\end{subfigure}

\begin{subfigure}[]{0.48\textwidth}
   \includegraphics[width=1\linewidth]{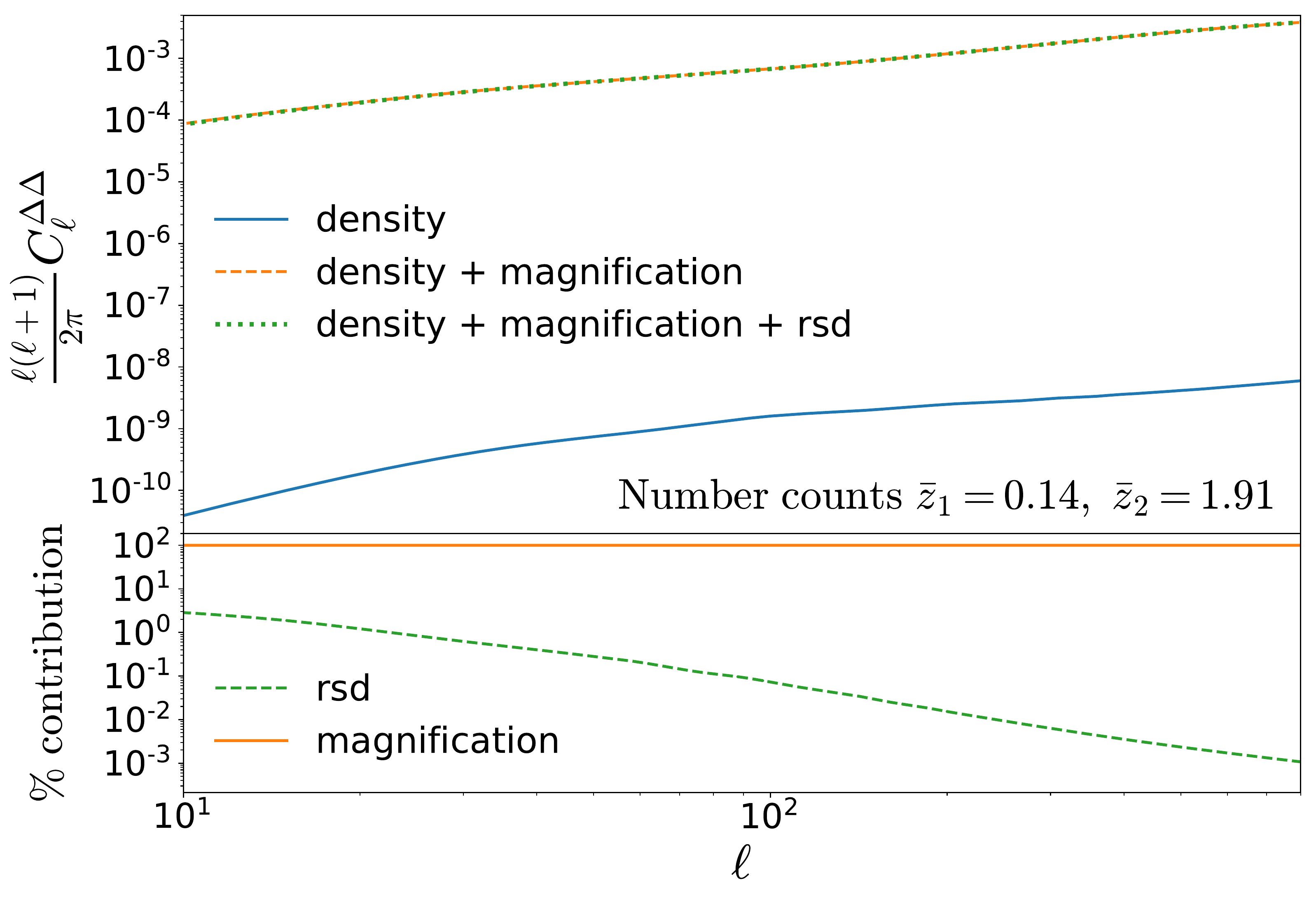}
\end{subfigure}
\caption[]{\label{fig:ncount1}
Number counts of power spectra for the \Euclid photometric sample (top panels) and percentage contributions of magnification and RSDs, $100 \times C^\text{RSD/magn}_\ell / C^{\De\De}_\ell$ (bottom panels). The contribution of magnification includes the $\ka\ka$ contribution as well as the density-$\ka$ contributions, given by the second, third, and fourth terms in Eq. \eqref{eq:ClDelta}. The contribution of RSDs, third line in Eq. \eqref{eq:ClDelta}, comprises the RSD-RSD correlation and the cross-correlation of RSDs with density and magnification. The magnification-RSD correlation is sub-dominant. The top sub-figure refers to the auto-correlation at $\bar{z}_1 = \bar{z}_2 = 0.69$. While the contribution of RSDs is $30\%$ on large scales, that is, $\ell \sim 10$, it drops below $1\%$ at $\ell > 90$. For this configuration, the contribution of magnification is at the sub-percent level on all scales (the blue line and the orange line overlie on all scales). 
The bottom sub-figure shows the cross-correlation of two bins with large redshift separation, $\bar{z}_1 = 0.14,\, \bar{z}_2 = 1.91$. The contribution of density alone and RSDs is 
negligible in this case. Magnification (and its cross-correlation with the density) constitutes the totality of the spectrum. }
\end{figure}

Even though Eq. \eqref{e:numbercount} is strictly valid only within linear perturbation theory, the density term and the magnification term are well modelled by replacing the linear power spectrum with a non-linear prescription (see e.g.~\citealt{Fosalba:2013mra, Fosalba:2013wxa, Lepori:2021lck}). This is not at all the case for RSDs, but since we do not include this effect in the analysis, the main results of this work, namely the relevance of magnification for parameter estimation, can be trusted when obtained with a non-linear prescription. At equal redshifts, the density fluctuation is usually the dominant contribution to the number counts, while at unequal redshifts, the lensing terms $\de\ka$ and $\ka\ka$ dominate, as can be seen in \Cref{fig:ncount1}.

\subsection{Cosmic shear}
\label{sec:cosmic shear}

The paths followed by photons coming from distant galaxies are deflected due to the large-scale structure of the Universe. These deflections introduce distortions in the images of these galaxies. We can decompose these distortions (at the linear level and locally) into convergence given by $\kappa$ and complex shear $\gamma$. The former is related to the magnification of the images, while the latter is linked to the shape distortion of the images. More specifically, these two effects correspond to the trace and trace-free part of the Jacobian  of the lens map given by
\bea
\bn &\mapsto & \bn -\bal(\bn, z)\,, \\
\bal(\bn, z) &=& \gradientsphere\psi(\bn, z) \,,
\eea
where $\gradientsphere$ denotes the gradient on the sphere.

Although cosmological information can be extracted from the convergence \citep[see e.g.][]{Alsing2015}, we focus here on the cosmological signal that can be obtained from the shear field. Under the assumption of homogeneity and isotropy of our Universe, the mean of the shear field vanishes. However, its angular power spectrum $C_{\ell}^{\gamma\gamma}$ contains cosmological information sensitive to both the expansion and the growth of structures.

Linking the shear field to observations, the ellipticity of a given galaxy, at linear order, can be expressed as
\begin{equation}\label{eq:epsilon}
    \epsilon=\gamma+\epsilon^{\rm I}\,,
\end{equation}
where $\epsilon^{\rm I}$ stands for the intrinsic ellipticity of the object. Under the assumption that galaxies are randomly oriented, the ellipticity provides an unbiased estimator of the complex shear. However, in practice tidal interactions during the formation of galaxies or other astrophysical effects may induce an intrinisic alignment of galaxies \citep[see e.g.][]{Joachimi:2015mma}, resulting in one of the major systematic effects in WL analyses.

Considering the angular power spectra of \Cref{eq:epsilon}, we can express the ellipticity angular power spectrum as
\begin{equation}\label{eq:ClsWL}
    C^{\epsilon\epsilon}_{\ell}=C^{\gamma\gamma}_{\ell}+C^{\text{I}\gamma}_{\ell}+C^{\gamma\rm I}_{\ell}+C^{\rm II}_{\ell}\,,
\end{equation}
where the two indexes represent two tomographic redshift bins. Therefore, the cosmic shear angular power spectra are contaminated by the correlations between background shear and foreground intrinsic ellipticity, $C^{\text{I}\gamma}_{\ell}$, the correlations between background and foreground intrinsic ellipticity, $C^{\rm II}_{\ell}$, and the correlations between background intrinsic ellipticity and foreground shear, $C^{\gamma \rm I}_{\ell}$. We note that $C^{\gamma \rm I}_{\ell}$ should be equal to zero because foreground shear should not be correlated with a background ellipticity except if galaxies are misplaced due to the photometric redshift uncertainty.
Using \Cref{eq:conv_shear}, the cosmic shear (without intrinsic alignments) angular power spectra, $C_\ell^{\gamma\gamma}$, is directly given by Eq.~\eqref{e:kappa_limber} within Limber's approximation.

In this work, we model the remaining terms in \Cref{eq:ClsWL}, using the extended non-linear alignment model for intrinsic alignments presented in \citetalias{Blanchard:2019oqi}. In this model, the three-dimensional matter-intrinsic and intrinsic-intrinsic power spectra can be expressed as

\begin{align}
    P_{\delta \rm I}(k,z)&=-\mathcal{A}_{\rm IA}\mathcal{C}_{\rm IA}\Omega_{\rm m,0}\frac{\mathcal{F}_{\rm IA}(z)}{D(z)}P_{\delta\delta}(k,z)\,,\\
    P_{\rm II}(k,z)&=\left[\mathcal{A}_{\rm IA}\mathcal{C}_{\rm IA}\Omega_{\rm m,0}\frac{\mathcal{F}_{\rm IA}(z)}{D(z)}\right]^2P_{\delta\delta}(k,z)\,,
\end{align}
with
\begin{equation}
    \mathcal{F}_{\rm IA}(z)=(1+z)^{\eta_{\rm IA}}\left[\frac{\braket{L}(z)}{L_*(z)}\right]^{\beta_{\rm IA}}\,,
\end{equation}
where $\mathcal{A}_{\rm IA}, \eta_{\rm IA}, \beta_{\rm IA}$ are nuisance parameters controlling the intrinsic alignment amplitude, redshift dependence, and luminosity dependence, respectively. Following the standard convention in the literature to model the intrinsic alignments (see e.g. \citealt{Joachimi:2020abi}), the constant $\mathcal{C}_{\rm IA}$ is set to a fixed value of 0.0134 as it is fully degenerate with $\mathcal{A}_{\rm IA}$. The $\braket{L}(z)$ and $L_*(z)$ stand for the redshift-dependent mean and the characteristic luminosity of source galaxies. We refer the reader to \citetalias{Blanchard:2019oqi} for more details on this model.

Given these three-dimensional power spectra, again using Limber's approximation, we can express the full ellipticity angular power spectra as
\bea
C^{\epsilon\epsilon}_{\ell}=C^{\gamma\gamma}_{\ell}+C^{\text{I}\gamma}_{\ell}+C^{\rm II}_{\ell}\,,
\eea
where $C^{\text{I}\gamma}_{\ell}$ and $C^{\rm II}_{\ell}$ are given by
\bea
C_\ell^{\rm II}(z, z')  &=& \delta_{\rm D}(z-z')\,\frac{H(z)}{c\,r(z)^2}\,P_{\text{II}}\left[\frac{\ell+1/2}{r(z)},z\right] \,, \\
C_\ell^{\rm I\gamma}(z, z')&=&
\begin{cases}
\begin{aligned}[b]
6\,\omegamatter \left(\frac{H_0}{c}\right)^2\frac{\ell(\ell+1)}{(2\ell+1)^2}
\frac{[r(z')-r(z)]}{r(z')r(z)}\\
\times\,(1+z)\,P_{\de \rm I}\left[\frac{\ell+1/2}{r(z)},z\right]\,,
\end{aligned}
&z \,<\, z'\,,\\
 0\,, &z\ge z'\,.
\end{cases}
\eea

Considering photometric redshift bins $i$ and $j$, even if the mean redshift $\bar z_i >\bar z_j$ we have to include not only $C_\ell^{\rm I\gamma}(j,i)$ but also $C_\ell^{\rm I\gamma}(i,j)=C_\ell^{\gamma\rm I}(j,i)$ in $C_\ell^{\epsilon\epsilon}(i,j)$ due to the significant overlap of photometric redshift bins.

It is important to mention that relativistic effects are also present in the source sample and therefore in cosmic shear analyses. For example, magnification effects can also change the number of sources in a magnitude-limited survey. However, these effects are of second order and the inclusion of magnification effects in cosmic shear requires the modelling of the matter bispectrum. Furthermore, its overall impact is significantly smaller than for galaxy number counts \citep[see e.g.][]{Duncan:2013haa,Deshpande:2019sdl}. Because of this, and the fact that the impact of magnification effects in cosmic shear has already been studied in \citet{Deshpande:2019sdl} in the context of \Euclid, we do not consider this effect (and other relativistic effects that appear at second order) in the cosmic shear part of our analysis.

\subsection{Galaxy--galaxy lensing}
In the photometric survey of \Euclid, we measure both galaxy number counts and cosmic shear, and we will also cross-correlate these measurements \citep[see e.g.][]{Tutusaus2020}. For purely scalar perturbations, the correlation function between the tangential shear and number counts is given by \Cref{e:cor-fk}:
\be
\langle\De(\bn, z)\ga_{\rm t}(\bn', z')\rangle = -\frac{1}{4\pi}\sum_\ell\frac{2\ell+1}{\ell(\ell+1)}\,P_{\ell 2}(\bn\cdot\bn')\,C_\ell^{\De\ka}(z, z')\,,
\ee
where $P_{\ell\,2}$ is the modified Legendre function,
of degree $\ell$ and index $m=2$ (see~\cite{Abram}).
Here,  $C_\ell^{\De\ka}(z,z')$
is the angular
correlation spectrum between the number counts $\De$ and the convergence $\ka$ (see \Cref{s:powerspec}).

As before, for a photometric survey, we can neglect RSD and large-scale relativistic contributions, so that
\be
C_\ell^{\De\ka}(z, z') \simeq C_\ell^{\text{g}\ka}(z, z') 
+[5s(z)-2]\,C_\ell^{\ka\ka}(z, z') \,. \label{eq:crossCl}
\ee

Using Limber's approximation, the two contributions in Eq.~\eqref{eq:crossCl} are given by Eqs.~\eqref{e:ggl-limber} and~\eqref{e:kappa_limber}, respectively.
For $z'>z$ the dominant term is $C_\ell^{\text{g}\ka}(z,z')$
since the foreground density fluctuations contribute to the integral $\ka$ (see Eq.~\eqref{e:kappa-def1}) This correlation has been measured\ by, for example, the Dark Energy Survey~\citep[DES;][]{Abbott:2017wau}. For $z>z'$, this term (nearly) vanishes and the correlation is dominated by the $C_\ell^{\ka\ka}(z,z')$ term. This term has also been recently measured~\citep{Liu:2021gbm}.
Considering distributions $n_i(z)$ for galaxy number counts and $n_j(z)$ for the shear measurements in bins $i$ and $j$, respectively, one obtains in Limber's approximation~\citep[see e.g.][]{Ghosh:2018nsm}:
\bea
\left\langle\De^{(i)}\ga^{(j)}_{\rm t}\right\rangle\,(\theta) = \hspace{5cm}&&
\nonumber\\
\int_0^\infty \diff z\, n_i(z)\int _0^\infty \diff z'\, n_j(z')\int_0^\infty \frac{\ell \diff \ell}{2\pi} J_2(\ell \theta)C_\ell^{\De\ka}(z,z')\,. &&
\eea

In Fig.~\ref{fig:ncount2} we show 
two representative configurations of these spectra for the \Euclid specifics. For $\bar{z}_1 < \bar{z}_2$ the density term in the number counts is the largest contribution to the cross-correlation; vice versa, the configuration with $\bar{z}_1 > \bar{z}_2$ is dominated by the cross-correlation of magnification and lensing. 
It should be noted that RSDs have an effect of $< 3\%$ on both configurations. 

\begin{figure}
\centering
\begin{subfigure}[]{0.48\textwidth}
   \includegraphics[width=1\linewidth]{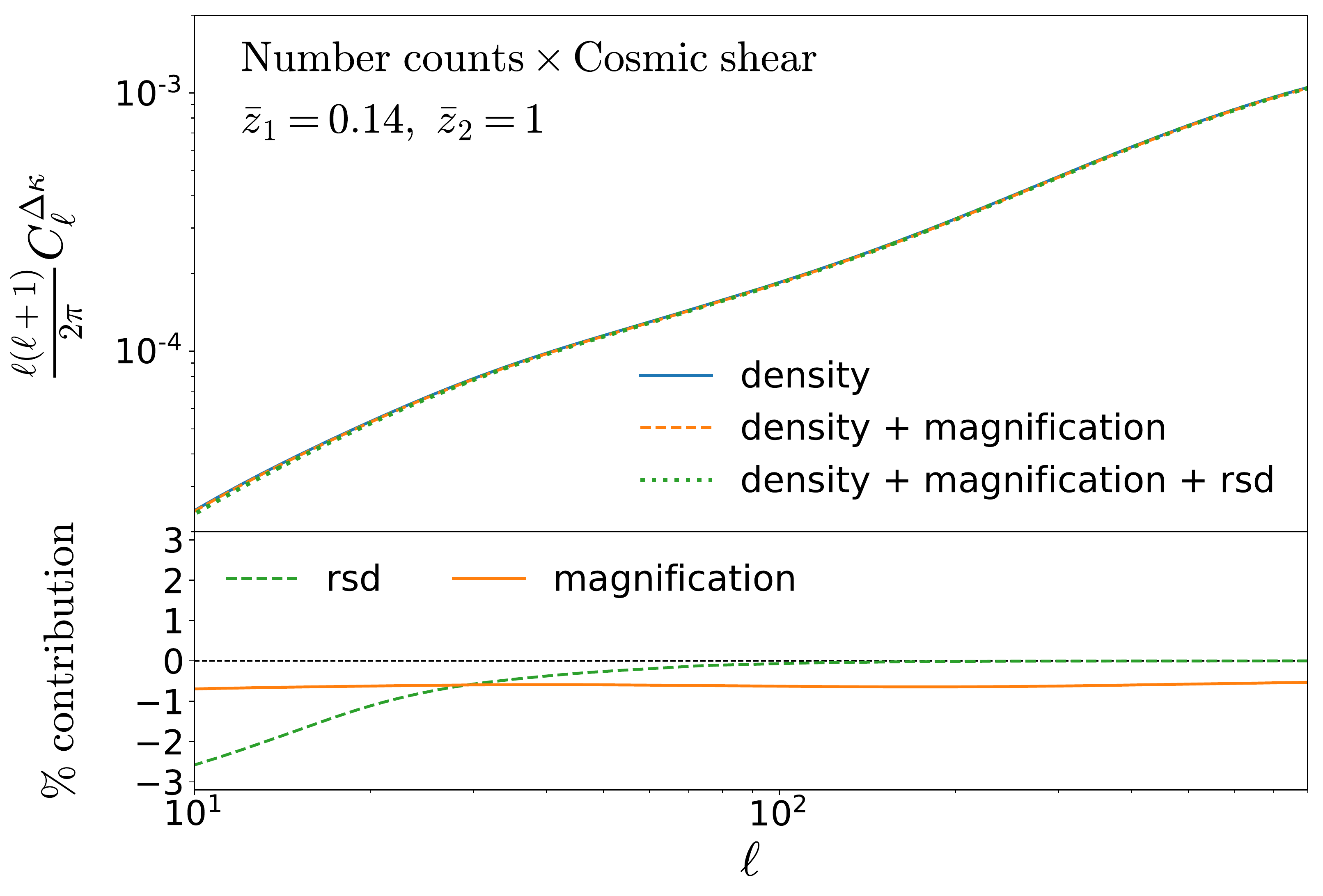}
\end{subfigure}

\begin{subfigure}[]{0.48\textwidth}
   \includegraphics[width=1\linewidth]{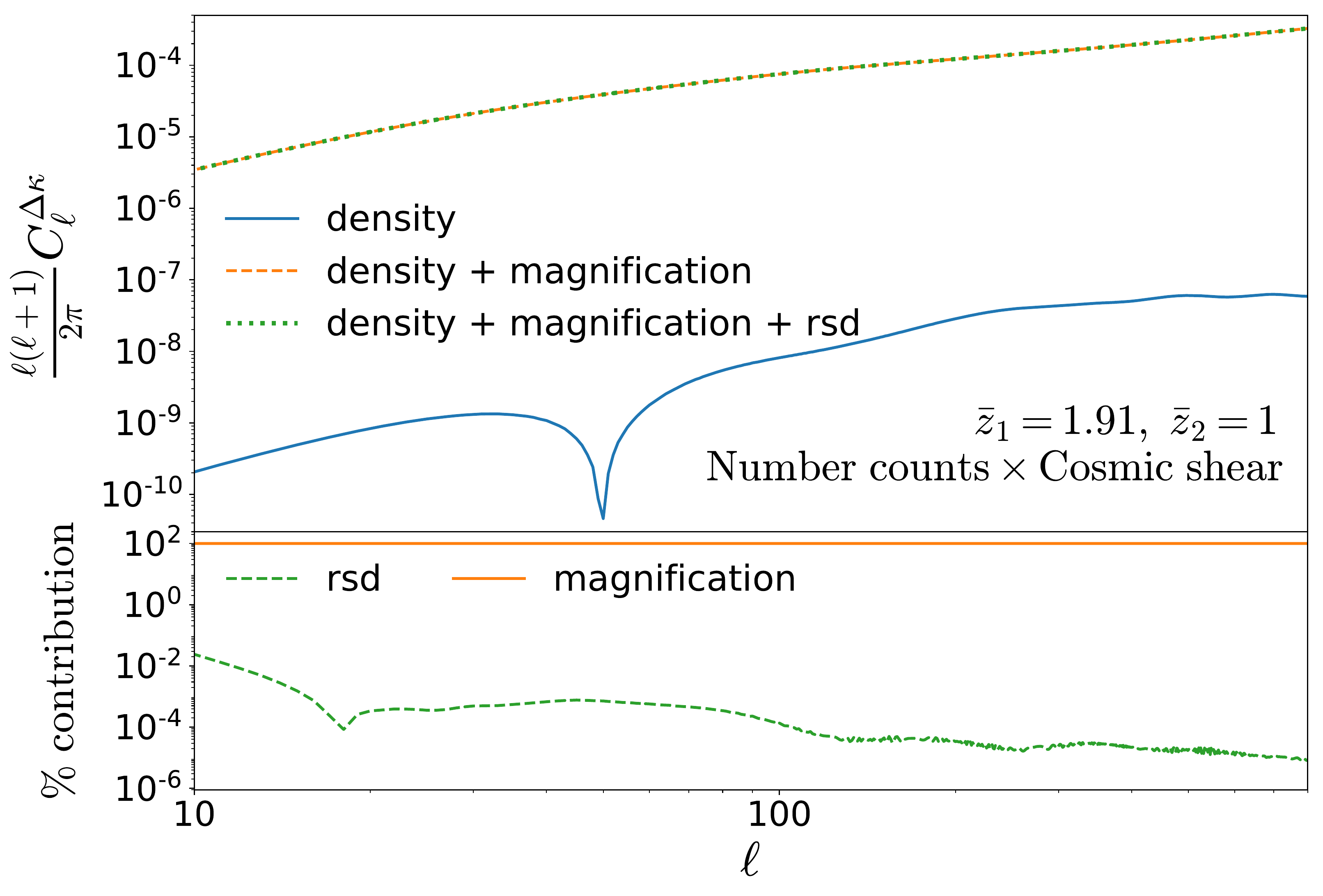}
\end{subfigure}
\caption[]{\label{fig:ncount2}
Angular power spectra of the GGL cross-correlation for the \Euclid photometric sample (top panels) and percentage contributions of magnification and RSDs, $100 \times C^{\De\ka, \text{RSD/magn}}_\ell / C^{\De\ka}_\ell$ (bottom panels). 
The contribution of magnification is the second term in Eq. \eqref{eq:crossCl}. The contribution of RSDs, which is neglected in Eq. \eqref{eq:crossCl}, is given by the cross-correlation RSD-$\ka$.
The top sub-figure refers to the configuration $\bar{z}_1 < \bar{z}_2$, that is, we correlate galaxies at low redshift with the background lensing. The contribution of RSDs is $3\%$ on large scales and drops below the percent level at $\ell \approx 30$, while the contribution of magnification is at the sub-percent level on all scales. 
The bottom sub-figure shows the configuration $\bar{z}_1 > \bar{z}_2$, that is, we correlate number counts at high redshift with foreground lensing. 
The contribution of density alone and RSDs is negligible in this case: we observe the correlation of magnification with the foreground cosmic shear. The small contribution of density alone, the blue curve in the top panel, changes sign at $\ell \sim 50$: it is negative on small scales and positive on large scales. 
}
\end{figure}

\section{\Euclid specifics from the Flagship simulation}
\label{sec:flag}
In this section, we briefly describe the Flagship galaxy catalogue and the ingredients extracted from this simulation to obtain realistic input for our forecasts.

We use the Flagship galaxy mock catalogue of the Euclid Consortium adapted the photometric sample (Euclid Collaboration, in preparation). 
The catalogue uses the Flagship $N$-body dark matter simulation \citep{Potter_2017}. 
The cosmological model assumed in the simulation is a flat $\Lambda$CDM model with fiducial values
$\Omega_\text{m,0} = 0.319$, $\Omega_\text{b,0}=0.049$, $\Omega_{\Lambda}=0.681$, $\sigma_{8}=0.83$, $n_{\rm s}=0.96$, $h=0.67$. The $N$-body simulation ran in a $3.78$ \hGpc{} box with particle mass $m_{\rm p}=2.398\times10^{9}$ \hsolarmass.
Dark matter halos are identified using the code `Robust Overdensity Calculation using K-Space Topologically Adaptive Refinement', known as {\sc Rockstar}  \citep{Behroozi_2013}, and are retained down to a mass of $2.4\times10^{10}$ \hsolarmass, which corresponds to ten particles. Galaxies are assigned to dark matter halos using the halo abundance matching (HAM) and halo occupation distribution (HOD) techniques, closely following \citet{Carretero:2015}. 
The galaxy mock generated has been calibrated using local observational constraints, such as the luminosity function from \citet{Blanton_2003} and \citet{Blanton_2005} for the faintest galaxies, the galaxy clustering measurements as a function of luminosity and colour from \citet{Zehavi_2011}, and the colour-magnitude diagram as observed in the New York University Value Added Galaxy Catalog \citep{Blanton_2005_b}. The mock calibration is automated and reproducible thanks to a novel and efficient minimisation technique that works in the presence of stochastic noise inherent to the galaxy mock construction (Tutusaus et al, in preparation). The catalogue contains about $3.4$ billion galaxies over $5000$ deg$^{2}$ and extends up to redshift $z = 2.3$. 

Given this galaxy catalogue, we extract three different quantities to adapt our forecasts to \Euclid specifications: the galaxy distributions as a function of redshift, $n(z)$, the galaxy bias, and the local count slope.
The Flagship mock galaxy catalogue is complete for magnitude limits below $25.5-26$ in the \Euclid VIS band. The specifics for the \Euclid photometric sample used in this work have been extracted applying a magnitude cut of $24.5$ in the VIS band, which is well within the completeness limit. 

\paragraph*{Number density distributions:}
    
The different galaxy distributions used in this analysis correspond to the fiducial selection presented in \citet{Pocino2021}. In this reference, the authors generated photometric redshift estimates for all objects in an area of 400 square degrees of the Flagship catalogue. Using the directional neighbourhood fitting \citep[DNF;][]{DNFref} training-based algorithm, two different redshift estimates were provided for each object. The DNF algorithm estimates the photometric redshifts based on the closeness in colour and magnitude space of the galaxies with unknown redshift to reference galaxies with known redshifts (training sample). The average of the redshifts from the neighbourhood in colour and magnitude space is one of the estimates, denoted $z_{\rm mean}$. But DNF can also provide a second estimate consisting of a Monte Carlo draw from the nearest neighbour, denoted as $z_{\rm mc}$. This estimate can be understood as a one-point sampling of the photometric redshift probability density function. In this work, we consider the fiducial settings from \citet{Pocino2021}, which were selected to optimise the constraining power of galaxy clustering and galaxy--galaxy lensing (GGL) with the \Euclid photometric sample. Such settings imply that DNF was trained with an incomplete spectroscopic  training sample to mimic the expected lack of spectroscopic information at very faint magnitudes. We consider the optimistic magnitude limits for all photometric bands shown in Table 1 of \citet{Pocino2021}. Given these two photometric redshifts estimates per galaxy, and following \citet{Pocino2021}, we select all Flagship galaxies with $z_{\rm mean}$ between 0 and 2, and split the sample into 13 bins with equal redshift width. We then obtain the final $n(z)$ used in our predictions by computing the histogram of $z_{\rm mc}$ of all the galaxies within each one of these bins. For these photometric bins, the fraction of outliers is $2.2\%$ (see Table 3 in \citealt{Pocino2021}). 
In \Cref{fig:photo-bins} we represent the 13 normalised $n(z)$ distributions obtained by binning in $z_{\rm mean}$ and computing the histogram of $z_{\rm mc}$, while the vertical grey lines show the mean redshift for each sample, $\bar{z}$. We note that it should not be confused with the $z_{\rm mean}$ estimate provided by DNF for each object. Moreover, although the bins were selected with equal width in $z_{\rm mean}$, given the non-Gaussianity of the $z_{\rm mc}$ distributions, their mean redshift $\bar{z}$ is not equispaced, as can be seen in \Cref{tab:flag}. The number density for each 
of the bins is also provided in the same table.

\paragraph*{Galaxy bias:}
    
The linear galaxy bias is calculated as the square-root ratio between the angular galaxy-galaxy power spectrum, $C_\ell^\text{gg}$, from the different $n(z)$ samples and the angular matter-matter power spectrum, $C_\ell^{\delta \delta}$. The $C_\ell^\text{gg}$ is obtained from the maps of the fractional overdensity of galaxies, generated using the HEALPix framework \citep{Gorski:2004by}. The maps have 
a resolution of $N_\text{side}=4096$ (that is 0.85 arcmin/pixel). 
We estimated the angular power spectra using \textsc{PolSpice}\,\footnote{\url{www2.iap.fr/users/hivon/software/PolSpice}} \citep{Szapudi:2000xj, Chon:2003gx}.
Mask effects for the 400 square degrees photo-$z$ region are also accounted for in this harmonic space analysis. The resulting $C_\ell$ values are corrected for shot noise using $C_\ell^\text{corr} = C_\ell - 4\pi f_\text{sky}/n_\text{gal}$, where $f_\text{sky}$ is the fraction of the sky covered by the photo-$z$ sample and $n_\text{gal}$ is the number of galaxies in the sample. The $C_\ell^{\delta \delta}$ is modelled with the public code Core Cosmology Library\,\footnote{\url{ccl.readthedocs.io/en/latest}} \citep[CCL;][]{Chisari:2018vrw} using the fiducial cosmology of
the Flagship simulation.
We used Limber's approximation for every multipole since CCL does not yet allow a non-Limber framework to be used. We note that the (linear) galaxy bias is calculated as the mean value across the multipole range $\ell \in [50, 500]$ to avoid non-linear (or higher-order) bias effects.
    
\paragraph*{Local count slope:}
    
As described in \Cref{s:galaxycounts}, the local count slope can be calculated from Eq. \eqref{e:s_mlim}.  We use the observed magnitude in the \Euclid VIS band with error realisation, assuming a $10\sigma$ magnitude limit of 24.6. For our analysis, we use a magnitude cut of 24.5. A binned magnitude cumulative function is calculated for the photo-$z$ sample at the different redshifts, and the corresponding slope is calculated at the magnitude cut using bins centred at 24.45 and 24.55.

The results for $n(z)$, $b(z)$, and $s(z)$ are shown in Table~\ref{tab:flag}  and Fig.~\ref{fig:specifics-photo}. 
    
\begin{table}[!ht]
\caption{Summary of \Euclid specifics from Flagship.
}
\begin{center}
\begin{tabular}{ c c c c }
 \toprule
 $\bar{z}$ & $n_\text{gal}(\bar{z}) [\text{gal}/\text{bin}/\text{arcmin}^2]$ & $b(\bar{z})$ & $s(\bar{z})$ \\
 \midrule
 $0.14$ & $0.758$  & $0.624$   & $0.023$\\
 $0.26$ & $2.607$ & $0.921$   & $0.135$ \\
 $0.39$ & $4.117$ & $1.116$   & $0.248$  \\
 $0.53$ & $3.837$ & $1.350$   & $0.253$  \\
 $0.69$ & $3.861$ & $1.539$   &  $0.227$  \\
 $0.84$ & $3.730$ & $1.597$   & $0.280$  \\
 $1.0$  & $3.000 $ & $1.836$   &  $0.392$  \\
 $1.14$ & $2.827$ & $1.854$   & $0.481$   \\
 $1.3$  & $1.800$ & $2.096$   &  $0.603$  \\
 $1.44$ & $1.078$ & $2.270$ &  $0.787$  \\
 $1.62$ & $0.522$ & $2.481$    & $1.057$   \\
 $1.78$ & $0.360$ & $2.193$  & $1.138$   \\
 $1.91$ & $0.251$ & $2.160$    &  $1.094$  \\
 \bottomrule
\end{tabular}
\end{center}
Number density (in units of gal/bin/$\text{arcmin}^2$), galaxy bias, and local count slope used in each photometric bin. Values are extracted from the Flagship simulation. A simple fit for $b(z)$ and $s(z)$ can be found in~\Cref{ap:fits}.
\label{tab:flag}
\end{table}

\begin{figure}
  \includegraphics[width=\linewidth]{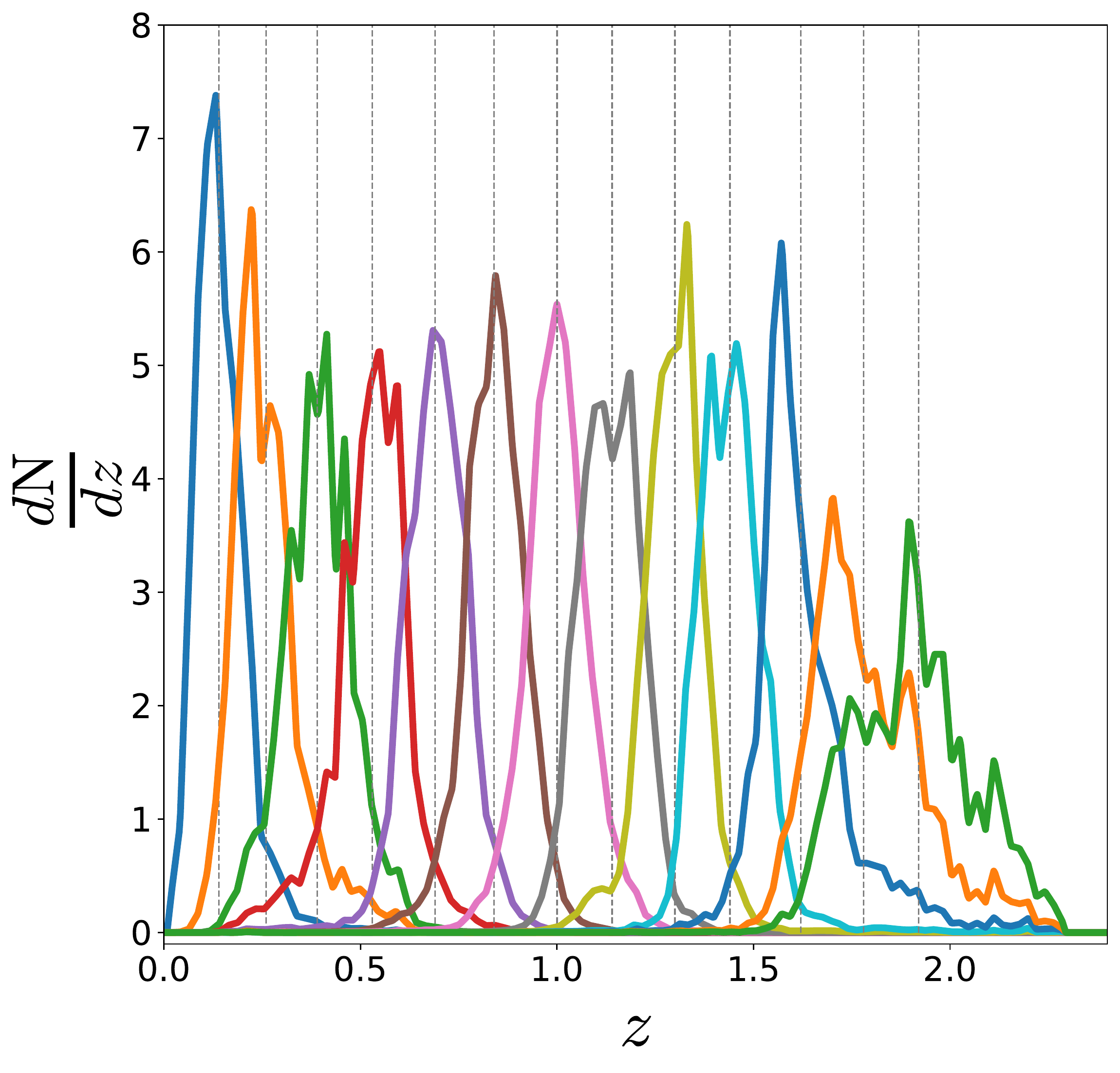}
  \caption{Normalised number of galaxies in the photometric redshift bins of \Euclid, as inferred from the Flagship simulation. 
  The sample is split into 13 equally spaced redshift bins defined by $z_{\rm mean}$. The redshift distribution of the galaxies inside the bins is estimated computing the histogram of the redshift defined by $z_{\rm mc}$.
  The vertical lines indicate the mean redshifts of the bins, $\bar{z}$.
  We note that this is the fiducial setting from \citet{Pocino2021}, selected to optimise the constraining power of galaxy clustering and GGL with the \Euclid photometric sample.
  }
  \label{fig:photo-bins}
\end{figure}

\begin{figure}
  \includegraphics[width=\linewidth]{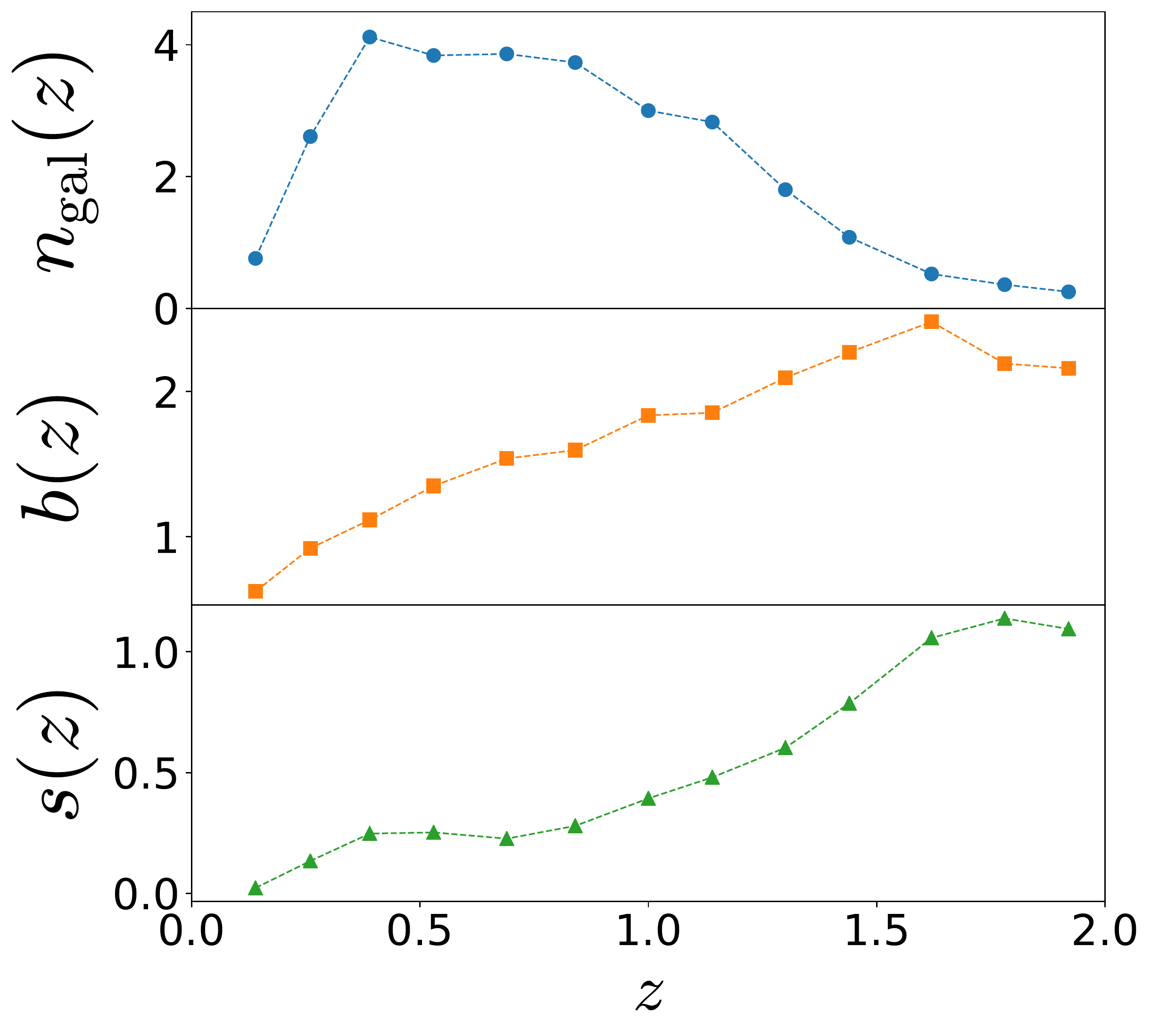}
  \caption{Galaxy number density in units of gal/bin/$\text{arcmin}^2$ (top panel), galaxy bias (middle panel), and local count slope (bottom panel) as a function of redshift. These results are obtained from the Flagship simulation. We note that at $z=1$ we have $s\simeq 0.4,$ so $2-5s(z=1)\simeq 0$. Hence, the lensing term exactly cancels at this redshift.
  A simple fit for $b(z)$ and $s(z)$ is found in~\Cref{ap:fits}.
  }
  \label{fig:specifics-photo}
\end{figure}

\section{Method}\label{s:met}

\subsection{The Fisher matrix formalism}
\label{fisher-form}

In this work, we follow \citetalias{Blanchard:2019oqi} in estimating the uncertainties on the cosmological parameters using a Fisher matrix formalism.
We used the Fisher matrix code \texttt{FisherCLASS}, based on version \texttt{v2.9.4} of the {\sc class} code \citep{class2, CLASSgal}, adapted to the prescription described in the previous section. The code has been validated against \citetalias{Blanchard:2019oqi}. More details on the code and its validations are presented in  \Cref{ap:code-valid}. 

We should recall that the Fisher matrix is defined as the expectation value of the second derivative with respect to the model parameters of the logarithm of the likelihood function of the data~\citep{Tegmark:1997rp},
\begin{equation}
F_{\alpha\beta}=\Braket{-\frac{\partial^2\text{ln}L}{\partial \theta_{\alpha}\partial \theta_{\beta}}}\,,
\end{equation}
where $\alpha$ and $\beta$ label the parameters of interest $\theta_{\alpha}$ and $\theta_{\beta}$.

Under the assumption of a Gaussian likelihood for the data, the Fisher matrix can be written as
\begin{equation}\label{eq:fisher}
    F_{\alpha\beta}=\frac{1}{2}\text{tr}\left[\frac{\partial \covariance}{\partial \theta_{\alpha}}\covariance^{-1}\frac{\partial \covariance}{\partial \theta_{\beta}}\covariance^{-1}\right]+\sum_{pq}\frac{\partial \mu_p}{\partial \theta_{\alpha}}\left(\covariance^{-1}\right)_{pq}\frac{\partial \mu_q}{\partial \theta_{\beta}}\,,
\end{equation}
where $\mu$ is the mean of the data vector and $\covariance$ is the 
covariance matrix of the data. The trace and sum over $p$ or $q$ stand for summations over the components of the data vector. It is important to note
that, in practice, we consider the angular power spectra as observables, which follow a Wishart distribution if the fluctuations are Gaussian. As shown for example in \citet{refId0,Bellomo:2020pnw}, the Fisher matrix for such distributions is given by \Cref{eq:fisher} but without the first term. Therefore, in the following, we only consider the second term when computing the Fisher matrix.

Once the Fisher matrix is constructed, we estimate the expected covariance matrix of the cosmological parameters as the inverse of the Fisher matrix:
\begin{equation}
    C_{\alpha\beta}=\left(\fisher^{-1}\right)_{\alpha\beta}\,. \label{e:Cor-exp}
\end{equation}

The Fisher matrix formalism is a powerful tool to quickly forecast the constraining power of future surveys. 
The main limitation of this approach is the Gaussian approximation, which results in optimistic cosmological constraints. \cite{Wolz:2009} and \cite{Takada:2008fn} show that, for observables that trace structure formation, such as WL and tomographic galaxy clustering analysis, the $1\sigma$ errors are typically underestimated by $10\%$ to $20\%$. 
The purpose of our analysis is assessing the impact of magnification. Therefore, when we compare the cosmological constraints with and without magnification we do not expect the Gaussian approximation to change significantly our results because both constraints with and without magnification are affected in the same way.

Another limitation of the Fisher approach
is that it only provides the uncertainties for a fiducial model. Therefore, it cannot quantify the bias in the posterior distributions if a wrong model is used to forecast the data vector and its covariance. This can be fixed using extensions of the Fisher matrix formalism, as explained at the end of this section.

We consider analyses of $\text{GCph}$, WL, and their cross-correlation terms. In the case of a joint analysis, a joint covariance matrix is required. In this work, since we consider the angular power spectra as observables \citepalias[see e.g.][for the equations when using the spherical harmonic coefficients as observables]{Blanchard:2019oqi}, we use the fourth-order Gaussian covariance given by
\begin{equation}
\begin{aligned}
\covariance &\left[C_{\ell}^{AB}(i,j),C_{\ell'}^{A'B'}(k,l)\right]=\frac{\delta_{\ell\ell'}^{\rm K}}{(2\ell+1)f_{\rm sky}\Delta \ell}\\
&\times\bigg\{\left[C_{\ell}^{AA'}(i,k)+N_{\ell}^{AA'}(i,k)\right]\left[C_{\ell'}^{BB'}(j,l)+N_{\ell'}^{BB'}(j,l)\right]\\
    &+\left[C_{\ell}^{AB'}(i,l)+N_{\ell}^{AB'}(i,l)\right]\left[C_{\ell'}^{BA'}(j,k)+N_{\ell'}^{BA'}(j,k)\right]\bigg\}\,,
\end{aligned}
\end{equation}
where $A,B,A',B'$ run over WL and galaxy clustering, and $i,j,k,l$ run over all tomographic bins. The noise terms $N_{\ell}^{XY'}$ are given by $\sigma_{\epsilon}^2/\bar{n}_i\delta_{ij}^{\rm K}$, $\delta_{ij}^{\rm K}/\bar{n}_i$, and $0$ for WL, galaxy clustering, and the cross-correlation terms, respectively. $\sigma_{\epsilon}^2$ is the variance of the ellipticity measurement (equal to $0.3^2$ in \citetalias{Blanchard:2019oqi} and in this work), and $\bar{n}_i$ is the number density in the corresponding tomographic bin.

With this covariance matrix, we can compute the final joint Fisher matrix as
\begin{equation}
F_{\alpha\beta}=\!\!\sum_{\ell=\ell_{\rm min}}^{\ell_{\rm max}}\sum_{\substack{{ABCD} \\{ij,mn}}}\hspace{-0.2cm}\frac{\partial C_{\ell}^{AB}(i,j)}{\partial \theta_{\alpha}}\covariance^{-1}\left[C_{\ell}^{AB}(i,j),C_{\ell}^{CD}(m,n)\right]\frac{\partial C_{\ell}^{CD}(m,n)}{\partial \theta_{\beta}}\,,
\end{equation}
where $A,B,C,D$ run over the different probes. The indices $ij$ and $mn$ run over all unique pairs of tomographic bins $(i\leq j, m \leq n)$ for WL and galaxy clustering, while they run over all pairs of tomographic bins for the cross-correlation terms. 

Throughout this study, we consider the pessimistic scenario presented in \citetalias{Blanchard:2019oqi} as a conservative choice for the lensing effects. We include all multipoles from $\ell=10$ up to $\ell=1500$ for WL and all multipoles from $\ell=10$ up to $\ell=750$ for galaxy clustering and the cross-correlation terms. These maximum $\ell$ values have been determined in \citetalias{Blanchard:2019oqi} by mapping the signal-to-noise ratio (S/N) between an analysis with and without the super-sample covariance contribution. In more detail, such $\ell$ values correspond to the values providing the same S/N in an analysis considering a Gaussian covariance 
 and in an analysis going to very non-linear scales ($\ell_{\rm max}=5000$ for WL and $\ell_{\rm max}=3000$ for galaxy clustering and the cross-correlation terms) but accounting for the super-sample covariance. We note that the maximum multipole considered for galaxy clustering and the cross-correlation terms is significantly smaller than the maximum multipole considered for WL. The main reason behind this choice is that galaxy clustering (and cross-correlations) is more sensitive to non-linearities, and their relevance appears sooner than in the WL case when including small scales. Given the fact that we consider a linear galaxy bias model, we prefer to be more conservative when selecting the scale cuts for galaxy clustering and the cross-correlation terms.

\subsection{Beyond the Fisher matrix formalism}
\label{subsec:fisher-shifts}
In this analysis, beyond providing the expected constraints on the cosmological parameters, we want to quantify the amount of information that is misinterpreted in an analysis that neglects magnification and how this affects the estimation of cosmological parameters. 
This is a model comparison problem, where the two models have a common set of cosmological parameters, and they differ by an extra model parameter, which is fixed in both models, but to a different value \citep[see for example,][]{Taylor:2006aw}.
We can generically express our theoretical model for the
angular power spectra as
\begin{align}
C^{\Delta\Delta}_\ell(i, j) &= C^{\text{g}\text{g}}_\ell(i, j) + \epsilon_\text{L}  C^{\Delta\Delta, \text{magn}}_\ell(i,j) \,, \label{eq-mod-1} \\
C^{\Delta\kappa}_\ell(i, j) &= C^{\text{g}\kappa}_\ell(i,j) + \epsilon_\text{L}  C^{\Delta\kappa, \text{magn}}_\ell(i,j)\,, \label{eq-mod-2}
\end{align}
where $\epsilon_\text{L}$ is the extra model parameter, fixed to $\epsilon_\text{L} = 1$ in the \emph{correct} model and to $\epsilon_\text{L} = 0$ in the {wrong} model.
We note that in \Cref{eq-mod-1} the magnification contribution $C^{\Delta\Delta, \text{magn}}_\ell(i,j)$ includes both the density-magnification cross-correlation and the magnification-magnification auto-correlation, while in \Cref{eq-mod-2} $C^{\Delta\kappa, \text{magn}}_\ell(i,j)$
is the cross-correlation between magnification and $\kappa$.

The shift in the fixed parameter in the wrong model leads to a shift 
in the maximum of the likelihood and, therefore, to a bias in the estimation of the common set of cosmological parameters.
A first-order Taylor expansion of the likelihood around the wrong model leads to the following expression for the shift in the best fit of common parameters $\{\theta_\alpha\}$:
\begin{equation}
\Delta \theta_\alpha = \sum_{\beta} \left(\fisher^{-1}\right)_{\alpha\beta} B_\beta\,, \label{eq:shift}
\end{equation}
where
\begin{equation}
B_\beta = \hspace{-0.1cm}\sum_{\ell=\ell_{\rm min}}^{\ell_{\rm max}}\sum_{\substack{{ABCD} \\ {ij,mn}}}\hspace{-0.2cm}\frac{\partial C_{\ell}^{AB}(i,j)}{\partial \theta_{\beta}}\covariance^{-1}\left[C_{\ell}^{AB}(i,j),C_{\ell}^{CD}(m,n)\right] \frac{\partial C_{\ell}^{CD}(m,n)}{\partial \epsilon_\text{L}}\,. 
\label{eq:lens-correl}
\end{equation}

We note that since we are expanding the likelihood around the wrong model, the Fisher matrix in \Cref{eq:shift} must
be computed neglecting magnification. This difference is of course of second order, but since we neglect other second-order terms, this is the more consistent approach.
This formalism provides a fast and straightforward method
to test the accuracy of our analysis if a known systematic effect is neglected. However, it is important to keep in mind the implicit assumptions behind the formula: since 
we are Taylor-expanding our likelihood around the incorrect model, we are assuming that the neglected systematic effect 
is small and, therefore, this formula can be quantitatively trusted only for small values of the shifts. 
If this assumption is violated, the computation of the shifts with this formalism gives a clear indication that the systematic effect is important for a precise parameter estimation.

Our analysis aims to assess whether magnification must be modelled for the analysis of the photometric sample of \Euclid or if the effect can be neglected. Therefore, for the purpose of our paper, a Fisher matrix analysis is a reliable tool to qualitatively study the impact of neglecting magnification. A quantitative determination of the parameter shifts is beyond the scope of this work and would require a full Markov chain Monte Carlo (MCMC) analysis to be run.

\section{Results}
\label{sec:res}

We investigate the impact of magnification for the primary cosmological probes in the photometric sample of \Euclid: the $\text{GCph}$ and the probe combination of
galaxy clustering, WL and GGL ($\text{GCph} + \text{WL} + \text{GGL}$). 

The fiducial cosmology adopted in our analysis is a flat $\Lambda$CDM model with one massive neutrino species.
The set of parameters considered in the analysis comprises: the present matter and baryon critical density parameters, respectively $\Omega_{\text{m}, 0}$ and $\Omega_{\text{b}, 0}$; the dimensionless Hubble parameter $h$; the amplitude of the linear density fluctuations within a sphere of radius 8 \hMpc, $\sigma_8$;
the spectral index of the primordial matter power spectrum $n_\text{s}$; the equation of state for the dark energy component $\{\wzero, \wa\}$; and the sum of the neutrino masses $\sum m_\nu$.

The fiducial values of the cosmological parameters are reported in Table \ref{tab:fid}. They correspond to the $\Lambda$CDM best-fit parameters from the 2015 Planck release \citep{Ade:2015xua}. This choice is consistent with the baseline cosmology adopted in \citepalias{Blanchard:2019oqi}.

\begingroup
\setlength{\tabcolsep}{0.5em}
\begin{table}[!ht]
\caption{Fiducial values of the cosmological parameters.  
}
\begin{center}
\begin{tabular}{ c c c c c c c c}
\toprule
$\Omega_{\text{m}, 0}$& $\Omega_{\text{b}, 0}$ & $\wzero$ & $\wa$& $h$  & $\ns$ & $\sigma_8$ & $\sum m_\nu\,[\text{eV}]$\\
\midrule
0.32& 0.05& $-$1.0& 0.0& 0.67& 0.96& 0.8156 & 0.06 \\
\bottomrule
\end{tabular}
\end{center}
\label{tab:fid}
\end{table}
\endgroup

In addition to these cosmological parameters, we introduce nuisance parameters and marginalise over them. For galaxy clustering the bias in each redshift bin, $\{b_i\}, \ i = 1, ..., N_\text{bins}$, are included as nuisance parameters. We modelled them as constant within each redshift bin, and we estimated their fiducial values in the Flagship simulation, as described in Sect. \ref{sec:flag} (see values in 
Table \ref{tab:flag}). For WL, the nuisance parameters are the ones used to model the intrinsic alignment 
contamination to cosmic shear, as defined in Sect.~\ref{sec:cosmic shear}: $\{\mathcal{A}_\text{IA}, \eta_\text{IA}, \beta_\text{IA}\}$. We note that since $\mathcal{C}_\text{IA}$ is fully degenerate with $\mathcal{A}_\text{IA}$, it is kept fixed in the Fisher analysis. Their fiducial values are given by: $\mathcal{A}_\text{IA} = 1.72 $, $\eta_\text{IA} = -0.41$, $\beta_\text{IA} = 2.17$,
and $\mathcal{C}_\text{IA} = 0.0134$. We note that these fiducial values correspond to the values considered in \citetalias{Blanchard:2019oqi}. However, the amplitude $\mathcal{A}_{\rm IA}$ might be smaller in practice\,\citep[see][for a discussion on the intrinsic alignment amplitude for different types of galaxies]{MAFortuna}.

The impact of magnification on the cosmological parameters may depend on the model chosen to describe our Universe. We therefore ran our analysis for four different cosmological models and comment on the difference between the results when relevant. We considered: 1) a minimal $\Lambda$CDM model, with five free parameters $\{\Omega_{\text{m}, 0}, \Omega_{\text{b}, 0}, h, \ns, \sigma_8\} \ plus $ nuisance parameters; 2) a $\Lambda$CDM model plus the sum of the neutrino masses as an additional free parameter: $\{\Omega_{\text{m}, 0}, \Omega_{\text{b}, 0}, h, \ns, \sigma_8, \sum m_\nu\}\ $ plus  nuisance parameters; 3) dynamical dark energy with seven free parameters
$\{\Omega_{\text{m}, 0}, \Omega_{\text{b}, 0}, \wzero, \wa, h, \ns, \sigma_8\}\  $ plus nuisance parameters; and 4) dynamical dark energy plus the sum of the neutrino masses as an additional free parameter: 
$\{\Omega_{\text{m}, 0}, \Omega_{\text{b}, 0}, \wzero, \wa, h, \ns, \sigma_8, \sum m_\nu\} \  $ plus nuisance parameters.

Although we ran our analysis for the four models described above, some results and tests that we performed will be reported only for model 3 
that we consider as our baseline analysis. 
In the baseline model, we did not vary the sum of the neutrino masses because its likelihood is highly non-Gaussian due to a physically forbidden region: it cannot be negative. Since the Fisher approach assumes Gaussian statistics, it is not accurate for computing constraints on the neutrino mass. The results reported for models 2
and 4 are therefore less accurate than the ones for models 1 and 3. 
An MCMC analysis that does not rely on Gaussianity for the effect of lensing magnification in the estimated neutrino mass is presented in \cite{Cardona:2016qxn}.

\subsection{Magnification information in the photometric sample} 
\label{sec:sn}

\begin{figure*}
\begin{center}
\begin{subfigure}[]{0.455\textwidth}
   \includegraphics[width=1\linewidth]{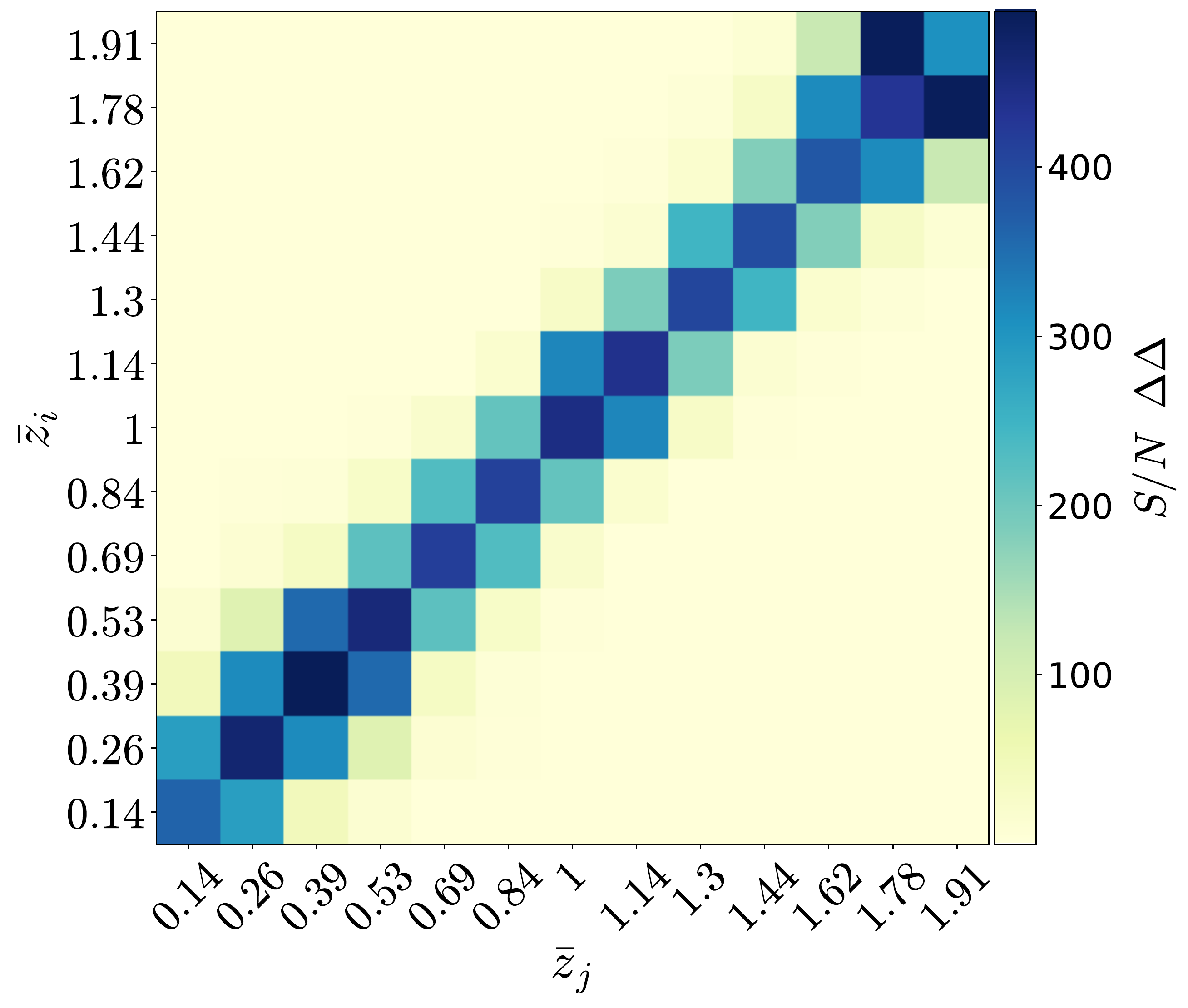}
\end{subfigure}
\begin{subfigure}[]{0.455\textwidth}
   \includegraphics[width=1\linewidth]{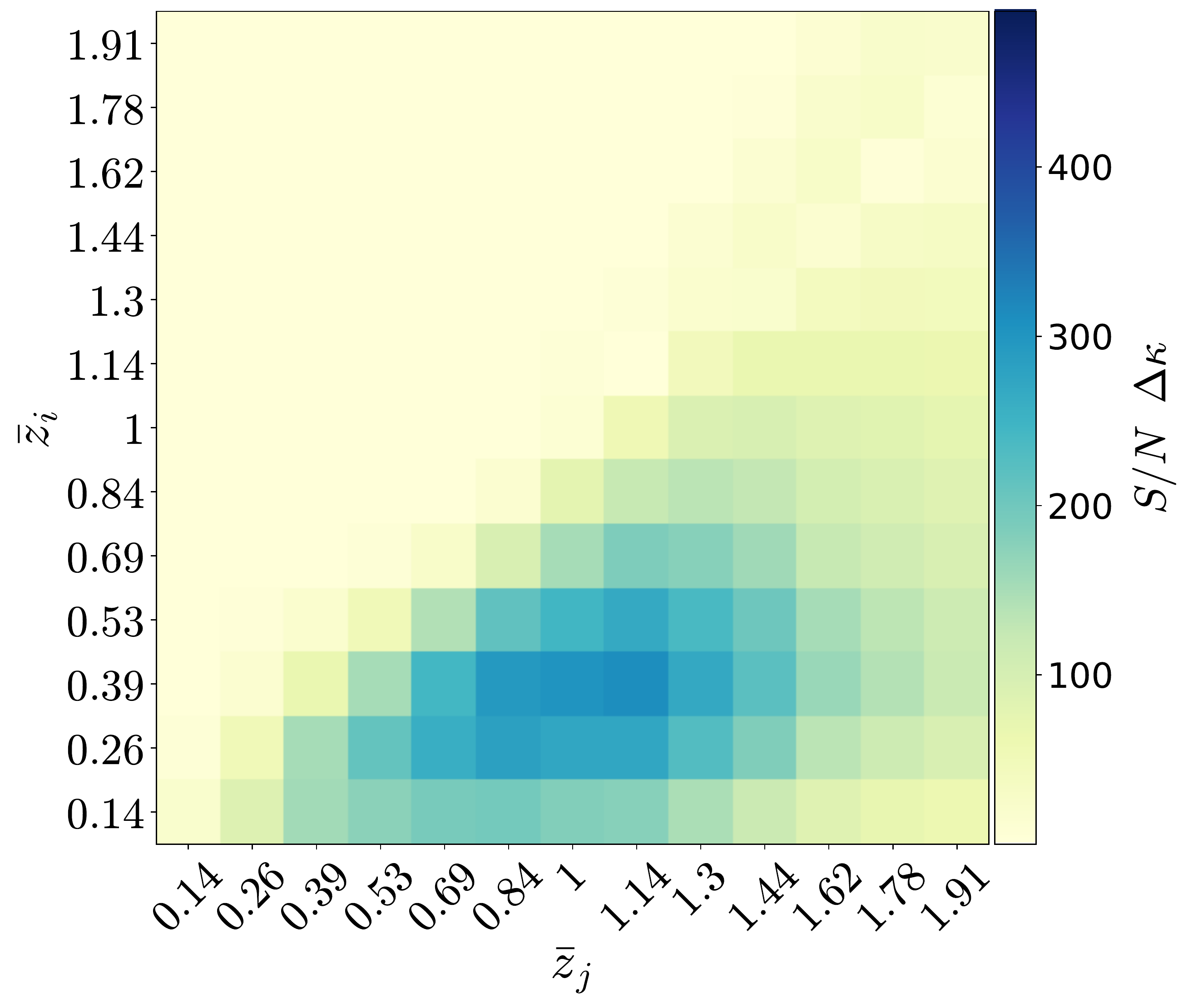}
\end{subfigure}
\begin{subfigure}[]{0.455\textwidth}
   \includegraphics[width=1\linewidth]{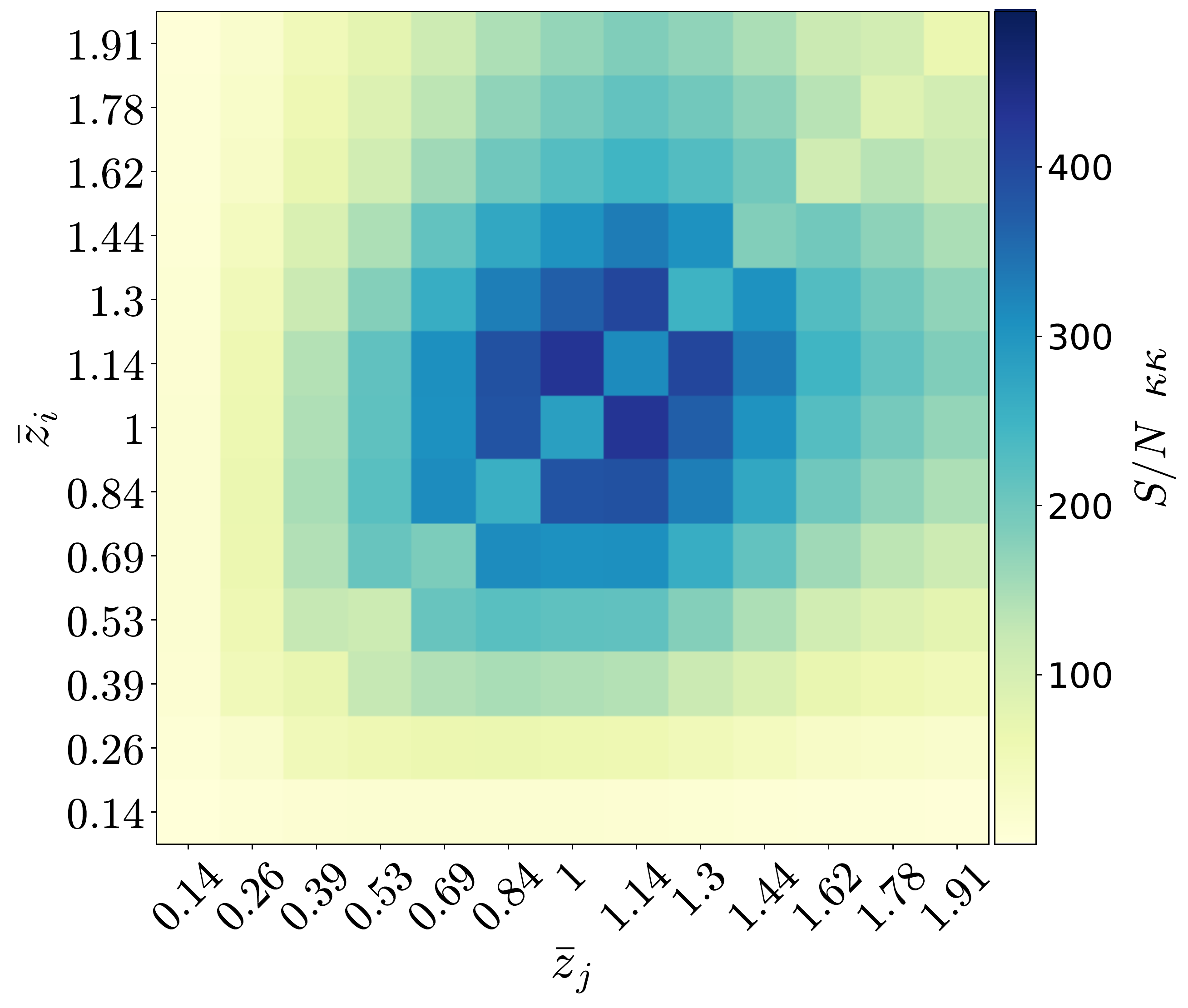}
\end{subfigure}
\caption{\label{fig:sntot} S/N per bin neglecting lensing magnification for the observables: GCph (top left), GGL (top right), and WL (bottom). The index $i$ refers to the $i$th redshift bin defined in Table~\ref{tab:flag}. The S/N is computed from Eq. \eqref{eq:SNR-obs}.
}
\vspace{0.7cm}
\begin{subfigure}[]{0.455\textwidth}
   \includegraphics[width=1\linewidth]{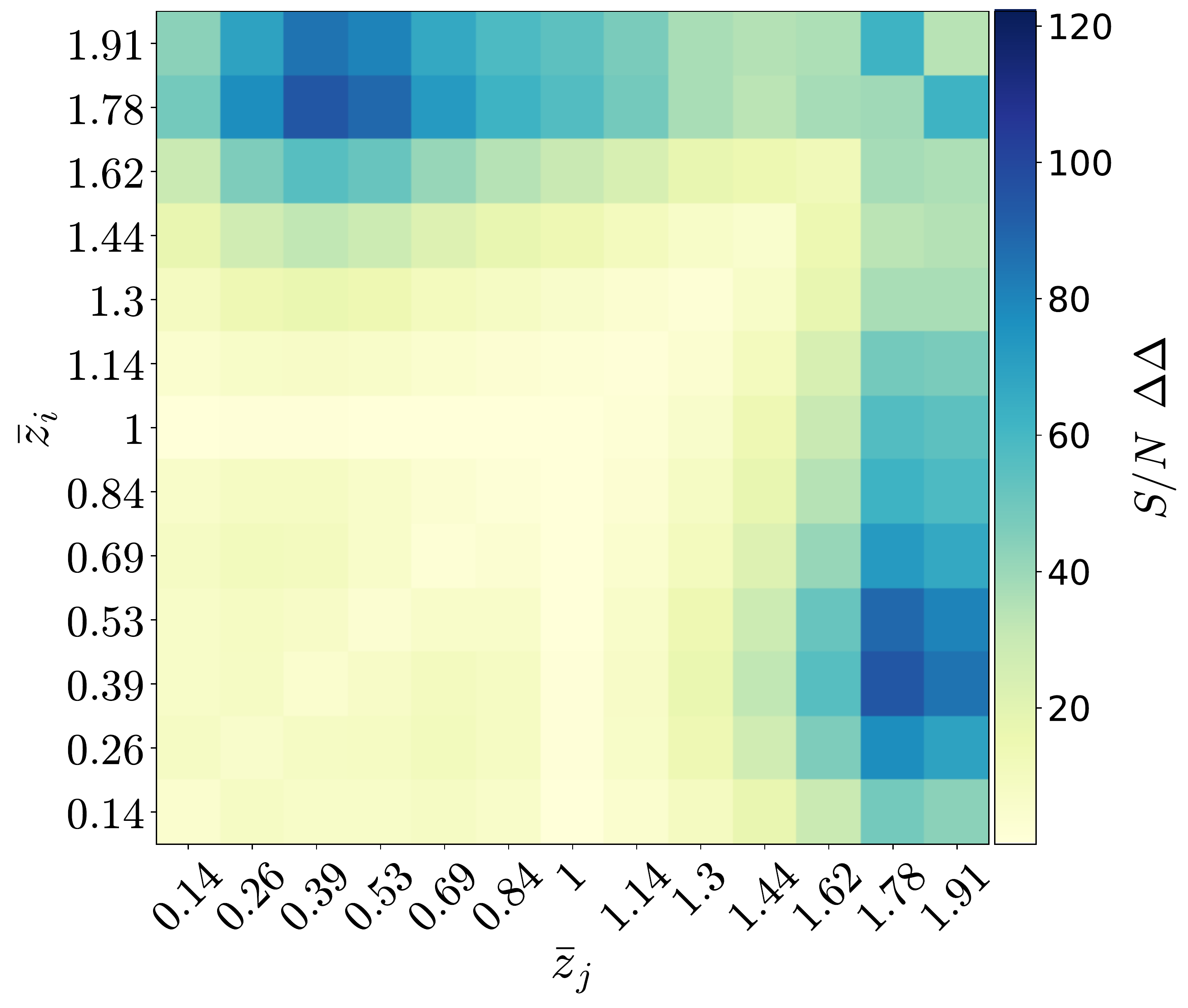}
\end{subfigure}
\begin{subfigure}[]{0.455\textwidth}
   \includegraphics[width=1\linewidth]{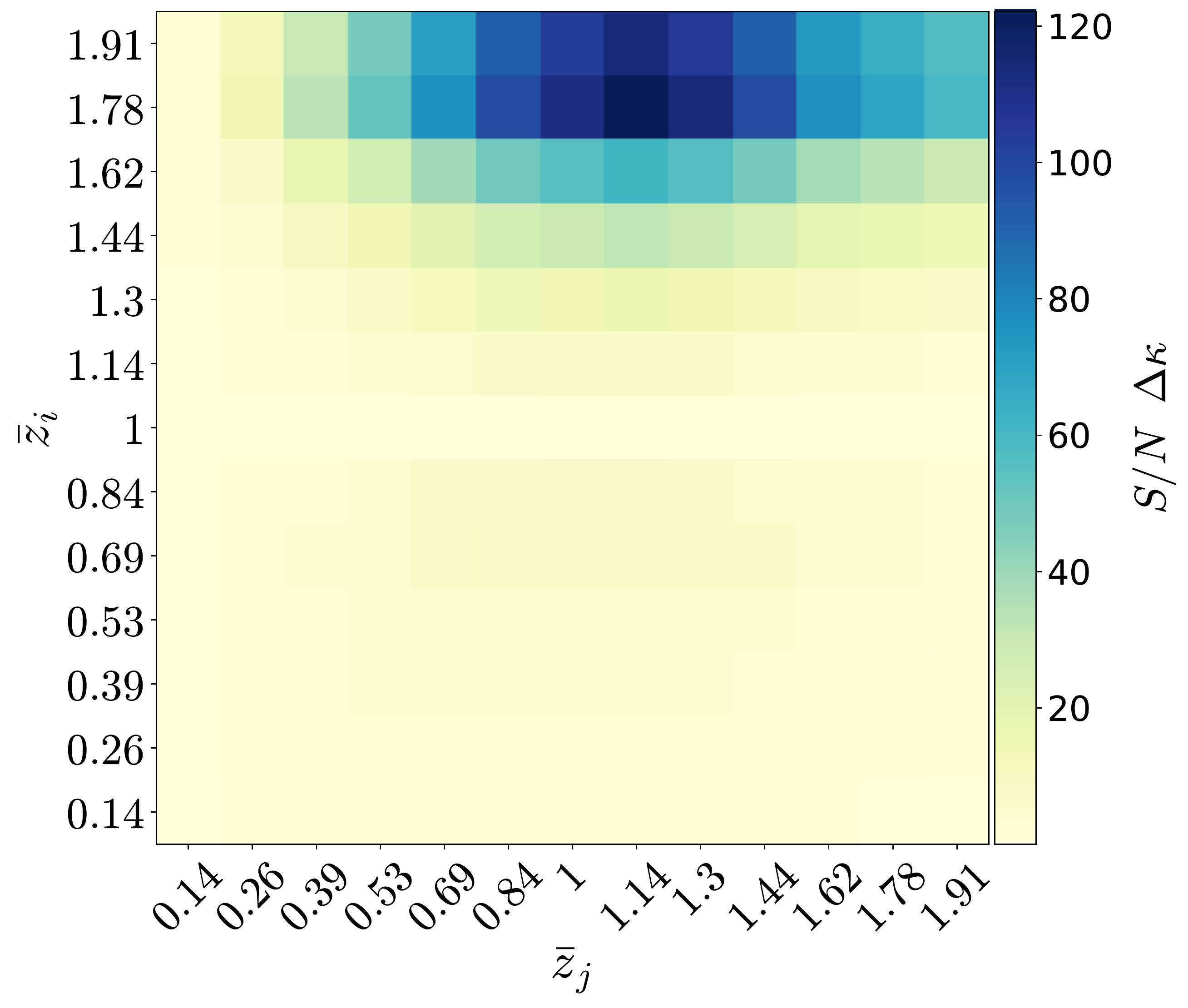}
\end{subfigure}
\caption{\label{fig:snmag} S/N per bin from lensing magnification in the GCph (left panel) and GGL analysis (right panel).The index $i$ refers to the $i$th redshift bin defined in Table~\ref{tab:flag}. The S/N is computed from Eq. \eqref{eq:SNRmag}.}
\end{center}
\end{figure*}

As discussed in the introduction, neglecting magnification in the modelling of the clustering signal will have two effects on the results of the \Euclid analysis: first, it will lead to incorrect estimations of the error bars on cosmological parameters, and second, it will lead to wrong estimations of the best-fit values of the cosmological parameters. The importance of these two effects is directly related to the S/N of the observables, compared to the S/N of magnification. We therefore start by computing these various S/N. Since we are interested in the redshift dependence of the S/N, we did not sum over all redshift bins, but rather computed the S/N for each pair of redshift bins $(z_i, z_j)$ separately. The S/N for our observables is given by 
\begin{equation}
\left(\frac{S}{N}\right)^{AB}_{ij} =\sqrt{ \sum_{\ell=\ell_{\rm min}}^{\ell_{\rm max}}
C^{AB}_{\ell}(i,j)
\,\,\covariance^{-1}\left[C^{AB}_{\ell}(i,j),C^{AB}_{\ell}(i,j)\right]\, C^{AB}_{\ell}(i,j)}\,,
\label{eq:SNR-obs}
\end{equation}
where $\{AB\} = \{\Delta \Delta\}, \{\Delta \kappa\}, \{\kappa \kappa\}$ for GCph, GGL, and WL, respectively, and $(i,j)$ refers to the pair of redshift bins. The S/N for the magnification contribution in GCph and GGL is given by
\begin{equation}
\left(\frac{S}{N}\right)^{\kappa AB}_{ij} \hspace{-0.1cm}=\sqrt{ \sum_{\ell=\ell_{\rm min}}^{\ell_{\rm max}}
 \Delta C^{AB}_{\ell}(i,j)
\covariance^{-1}\left[C^{AB}_{\ell}(i,j),C^{AB}_{\ell}(i,j)\right] \Delta C^{AB}_{\ell}(i,j)}\,,\label{eq:SNRmag}
\end{equation}
where $\Delta C^{AB}_{\ell}(i,j)$ denotes the contribution of magnification to the angular power spectrum ${AB}$. We note that in Eq.~\eqref{eq:SNRmag} only the magnification is included in the signal, but the covariance is that of the full observable.

In Fig.~\ref{fig:sntot} we show the S/N for GCph, GGL, and WL (without magnification) for each pair of redshift bins (the index $i$ refers to the $i$th redshift bin defined in Table~\ref{tab:flag}). We see that the GCph signal is most significant in the auto-correlations and in the cross-correlation of nearby bins. The S/N is slightly larger at low redshift (it peaks for bins 2 and 3). Interestingly, the S/N of the GCph signal in the cross-correlations of bins 12 and 13 is larger than the one in the corresponding auto-correlations. There are two reasons for this: on the one hand, these bins have a very significant overlap, as can be seen from Fig.~\ref{fig:photo-bins}; and on the other hand, correlations of different bins have no shot noise, which is the dominant
source of noise in high-redshift bins. The GGL S/N is prominent in the cross-correlations of cosmic shear at intermediate redshift ($z \sim$ 0.7--1.3) and the galaxy density at low $z$ ($z \sim$ 0.25--0.55). Finally, the S/N of WL is found to be prominent in the cross-correlation of nearby bins in
the redshift range $z \sim$ 0.7--1.5, reaching a maximum for the configuration $i = 7$ ($\bar{z} = 1$), $j = 8$ ($\bar{z} = 1.14$). 
The peak of the WL S/N per bin is comparable to the peak of the GCph S/N and to the peak of the GGL S/N.

The S/N of magnification is shown in Fig. \ref{fig:snmag} for the GCph alone analysis and for the GGL alone analysis (the WL analysis is not affected by magnification).
In the GCph analysis, we find that the S/N
of magnification is largest for the cross-correlation of widely separated redshift bins, reaching a maximum in
the cross-correlation of $i = 3$ and $j = 12$. For these pairs, the contribution of magnification is dominated by the cross-correlation of 
density at low $z$ and magnification at high $z$. 
We also note that the minimum S/N is found for the auto-correlation of the bin $i = 7$ and its cross-correlations with other bins. This is due to the value of the local count slope, close to the critical value $s = 0.4$ for these configurations. 
In fact, for $s = 0.4$ the effect of magnification on the apparent luminosity of the observed galaxies compensates exactly 
for the change in the observed solid angle due to lensing, and therefore, the magnification contribution to the number counts is exactly zero for this critical value (see Eq.~\eqref{e:numbercount}). Comparing with Fig.~\ref{fig:sntot}, we see that the maximum S/N for magnification is roughly four times smaller than the maximum S/N for GCph (due to density). 

In the GGL observable,
the magnification signal is given by the cross-correlation of the magnification contribution to the number count and cosmic shear. The largest S/N is found cross-correlating the magnification at high redshift $(z > 1.7)$ and cosmic shear at intermediate and high redshift $(z \in [0.8, 1.5])$. For these configurations the contributions of density to the galaxy counts is very small: the background density field is (almost) uncorrelated with the lensing signal in the foreground and the small correlations that we estimate
are due to the overlap between the redshift distribution of the sources in the bins. Comparing with Fig.~\ref{fig:sntot}, we see that the maximum S/N for magnification in the GGL observable (which is due to the magnification-shear correlation) is roughly 2.5 times smaller than the maximum S/N for GGL (which comes from the density-shear correlation). 

In general, comparing Fig.~\ref{fig:sntot} with Fig.~\ref{fig:snmag} we see that the contamination due to magnification is maximal for the bins in which the S/N of the corresponding observable is minimal. This will somewhat mitigate the impact of magnification on the analysis, but as we will see in Sects.~\ref{sec:shift} and~\ref{sec:shift_combined} it is not enough to make magnification negligible.

\subsection{Impact of magnification on the galaxy clustering analysis} 
\label{sec:gch-alone}

We now compute the impact of magnification on the constraints and on the best-fit values of the cosmological parameters. We first consider an analysis based on galaxy clustering alone.

\subsubsection{Cosmological constraints} 
\label{sec:constraint}
In order to quantify the amount of cosmological information encoded in the magnification signal, for each cosmological model we ran two Fisher matrix analyses: 
a) one that includes only the density contribution to the galaxy clustering observable and covariance, and b) one that also takes into account lensing magnification, both in the theoretical signal and in the covariance. We then compared the constraints in both cases.

\begin{table*}[!ht]
\caption{Constraints on cosmological parameters for GCph alone.
}
\begin{center}
\begin{tabular}{ cccccccccc}
 \toprule
 model & & $\Omega_{\text{m},0}$ & $\Omega_{\text{b},0}$
      & $\wzero$ & $\wa$ & $h$
      & $n_\text{s}$ & $\sigma_8$ & $\sum m_\nu$\\
 \midrule
\multirow{2}{*}{$\Lambda$CDM}& only density & 1.4 & 4.2 & --&  -- &  2.8 & 1.2 & 0.80 & --\\ 
    &+ magnification & 1.1  & 4.2  & --&  -- &  2.8  & 1.1  & 0.57 & --\\ 
\midrule
 \multirow{2}{*}{$\Lambda$CDM + $\sum m_\nu$}& only density & 1.9 & 4.2 & --&  -- &  2.9 & 1.4 & 1.2 & 140 \\ 
    &+ magnification & 1.6  & 4.2 & --&  -- &  2.9 & 1.2 & 1.0 & 130 \\
 \midrule
 \multirow{2}{*}{$\wzero\,\wa$CDM}& only density & 7.3 & 9.1 & 25 &  84 &  3.7 & 1.8 & 1.9 & --\\
    &+ magnification & 4.7 & 6.9 & 16 &  54 &  3.2 & 1.2  & 1.6 & --\\ 
 \midrule
  \multirow{2}{*}{$\wzero\,\wa$CDM + $\sum m_\nu$}& only density & 7.4 & 9.6 & 25 &  84 &  3.7 & 1.9 & 1.9 & 160 \\ 
    &+ magnification & 4.7 & 7.2 & 16.5 &  54 &  3.2  & 1.3  & 1.6 & 150 \\ 
 \bottomrule
\end{tabular} 
\end{center}
The 1$\sigma$ constraints on cosmological parameters are relative to their corresponding fiducial values (in \%), without and with magnification. For the parameter $\wa$, we report the absolute error times 100.
We have marginalised over the galaxy bias parameters, and the values of the local count slope are kept fixed in the computation of the constraints with magnification. We report the results for four cosmological models: a minimal $\Lambda$CDM model with one massive neutrino species and fixed neutrino mass, an analogue model that includes dynamical dark energy, denoted as $\wzero\,\wa$CDM, and  two extensions of these models where the sum of the neutrino masses is a free parameter.
\label{tab:gc-const}
\end{table*}

\begin{table*}[!ht]
\caption{Improvement in the constraints for GCph alone, including magnification. 
}
\begin{center}
\begin{tabular}{ ccccccccc}
 \toprule
 model & $\Omega_{\text{m},0}$ & $\Omega_{\text{b},0}$
      & $\wzero$ & $\wa$ & $h$
      & $n_\text{s}$ & $\sigma_8$ & $\sum m_\nu$\\
 \midrule
$\Lambda$CDM & $21\%$ & $0.3\%$ & --&  -- &  $1\%$ & $12\%$ & $28\%$ & --\\
$\Lambda$CDM + $\sum m_\nu$ & $11\%$ & $0.5\%$ & --&  -- &  $1.65\%$ & $13\%$ & $16\%$ & $3\%$ \\
$\wzero\,\wa$CDM & $36\%$ & $24\%$ & $34\%$ &  $35\%$ &  $14\%$ & $32\%$ & $18\%$ & --\\
$\wzero\,\wa$CDM + $\sum m_\nu$ & $37\%$ & $25\%$ & $35\%$ &  $35\%$ &  $15\%$ & $30\%$ & $15\%$ & $4\%$ \\ 
 \bottomrule
\end{tabular} 
\end{center}
Shown is the improvement in the constraints (given by $1-\sigma_\text{magn}/\sigma_\text{dens}$, in \%), including magnification. We report the results for the same models as in Table \ref{tab:gc-const}, and, in the same way, we marginalise over the galaxy bias parameters. The values of the local count slope are fixed, and thus we assume a perfect knowledge of $s(z)$ in each redshift bin.
\label{tab:gc-const-improv}
\end{table*}

The impact of magnification strongly depends on the value of the local count slope $s(z)$. As we see from Eq.~\eqref{e:numbercount}, if $s(z)=0.4$, magnification has no effect in the corresponding bin. For \Euclid's photometric survey, this is nearly the case for redshift bin~7 around $z=1$ (see Table~\ref{tab:flag}). As a first step, we assumed that we know the value of the local count slope $s(z)$ exactly in each redshift bin. This local count slope can indeed be measured directly from the distribution of galaxies as a function of luminosity. In Table \ref{tab:gc-const} we
report the constraints obtained for the two analyses.  In Table \ref{tab:gc-const-improv} we show the relative difference between the 1$\sigma$ constraints obtained in the two cases.

Including magnification significantly improves the constraints on cosmological parameters. For a $\Lambda$CDM model, magnification provides additional information on $\Omega_{\text{m}, 0}$ and $\sigma_8$, improving their constraints at the level of $21\%$ and $28\%$. This can be understood by the fact that the density contribution is proportional to the bias, which is a free parameter (over which we marginalise). In the linear regime, there is therefore a strong degeneracy between the amplitude of perturbations $\sigma_8$ and the bias, both of which control the amplitude of the density term. The non-linear evolution of the density field breaks this degeneracy. However, since we restrict the analysis to mildly non-linear scales, the degeneracy is only partially broken. Including magnification then significantly improves the constraints on $\sigma_8$, since it helps to break the degeneracy further. Looking at the magnification contribution to GCph we see that it contains two terms: one that depends linearly on the bias (from the correlation between density and lensing) and one that is independent of bias (from the lensing-lensing correlation). These two terms break the degeneracy between $\sigma_8$ and the bias, leading to a significant improvement in the constraints. We verified that this improvement is even stronger when we use a smaller $\ell_{\rm max}$ since in this case non-linearities are less relevant and are therefore not able to break the degeneracy: for example, for $\ell_{\rm max}=300$, the constraint on $\sigma_8$ is improved by 50\%. Adding magnification also improves the constraints on $\Omega_{\rm m,0}$, which is not surprising since $\Omega_{\rm m,0}$ is itself also degenerate with $\sigma_8$: it determines the redshift of matter-radiation equality 
where density perturbations start to grow. This degeneracy is evident in Fig.~\ref{fig:contour-plot-smarg}. Breaking the degeneracy between the bias and $\sigma_8$ therefore automatically leads to better constraints on $\Omega_{\rm m,0}$. 

For our baseline model with dynamical dark energy, we have a large improvement for all the parameters, up to roughly $35\%$ for $\Omega_{\text{m}, 0}$ and $\{\wzero, \wa\}$. From Table~\ref{tab:gc-const}, we see that adding $\{\wzero, \wa\}$ as free parameters strongly degrades the constraints on $\Omega_{\rm m, 0}$. This is due to the fact that these quantities are degenerate, as can be seen from Fig.~\ref{fig:contour-plot-smarg}: changing $\Omega_{\rm m,0}$ means changing $\Omega_\text{DE,0}$, which can be partially counterbalanced by a change of the equation of state. When only density is included in the analysis, this degeneracy is worsened by the fact that the bias is free and can be adjusted at each redshift. However, when magnification is included, it tightens the constraints since the lensing-lensing contribution is independent of  bias. This leads to a significant improvement in the constraints on $\Omega_{\text{m}, 0}$ and $\{\wzero, \wa\}$.

Finally, adding the sum of the neutrino mass as a free parameter degrades the constraints with respect to the $\Lambda$CDM case, especially for $\Omega_{\rm m,0}$ and $\sigma_8$. Adding magnification mitigates this degradation, again due to the fact that magnification has a contribution that is bias independent.

\begin{table}[!ht]
\caption{Uncertainty in the local count slope, GCph alone. }
\begin{center}
\begin{tabular}{ cccc }
\toprule
parameter & $s(z_i)$ fixed & $s(z_i)$ marg & + $10 \%$ prior on $s(z_i)$\\
 \midrule
$\Omega_{\text{m},0}$ & $36\%$ & $17\%$& $23\%$\\
$\Omega_{\text{b},0}$ & $24 \%$& $13\%$ & $16\%$\\
$\wzero$ & $34\%$& $14\%$& $21\%$\\
$\wa$ & $35\%$& $17\%$ & $20\%$\\
$h$ & $14\%$& $-8\%$& $13\%$\\
$n_\text{s}$ & $32\%$& $-22\%$& $9\%$\\
$\sigma_8$ & $18\%$& $-21\%$& $18\%$\\
 \bottomrule
\end{tabular} 
\end{center}
The relative difference, $1-  \sigma_\text{magn}/\sigma_\text{dens}$, in percentage, is shown for
three cases: a) an optimistic scenario, when the local count slope is measured with high accuracy and thus $s(z)$ is kept fixed in the analysis (Col. 2), b) a pessimistic scenario where the local count slope cannot be constrained by an independent measurement, and therefore we marginalise over its values (Col. 3), and c) a realistic scenario such that the local count slope is assumed to be measured independently with a $10\%$ precision (Col. 4).
The results reported here refer to our baseline cosmology, the $\wzero\,\wa\textrm{CDM}$ model.
\label{tab:prior}
\end{table}

\begin{figure*}[!ht]
  \includegraphics[width=\linewidth]{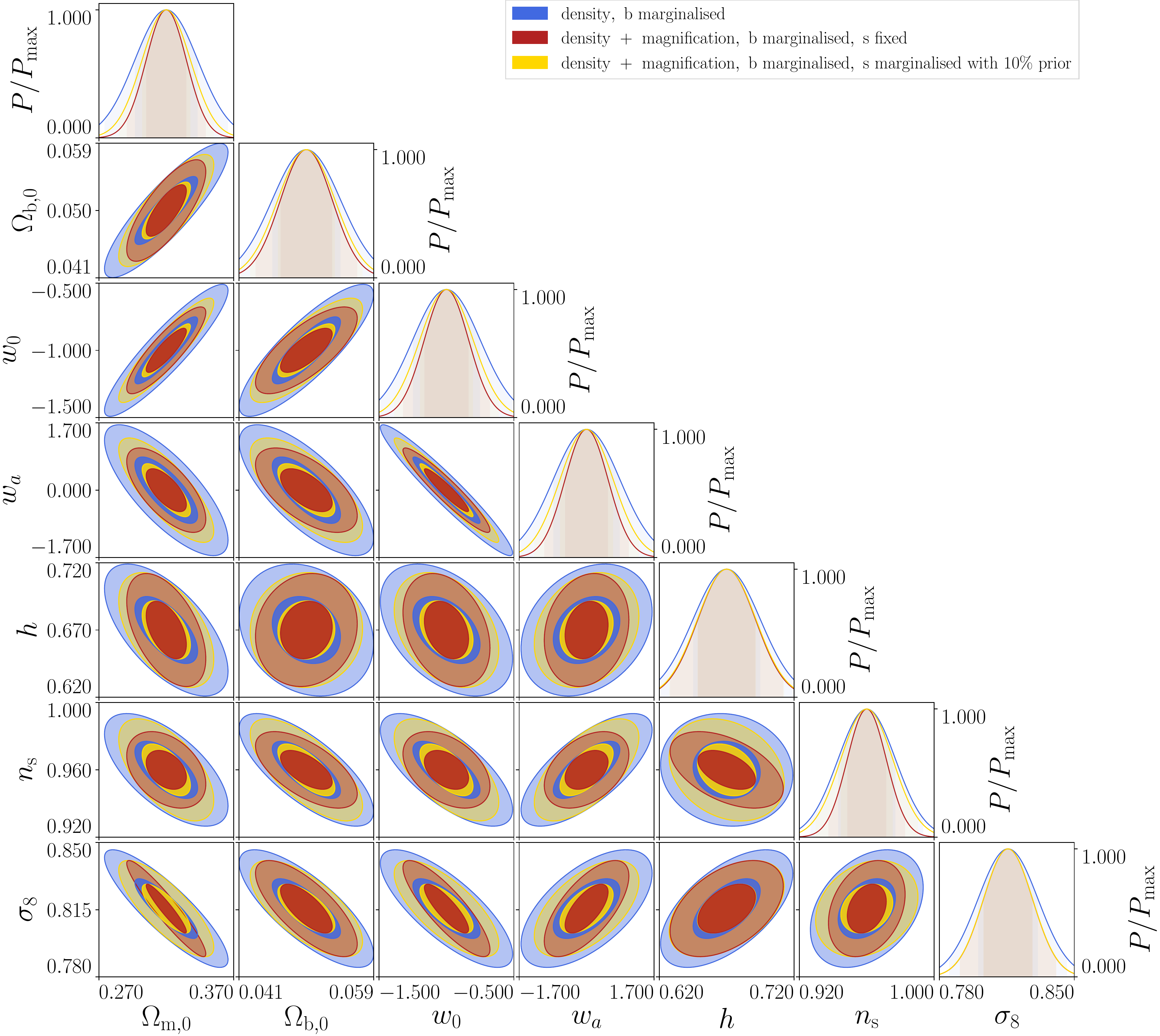}
  \caption{Cosmological constraints from the GCph analysis neglecting magnification (blue contours), including magnification and assuming a perfect knowledge of the local count slope (red contours), and marginalising over the local count slope parameters with a $10\%$ prior (yellow contours). The results reported here refer to our baseline cosmology, that is, the $\wzero\,\wa\textrm{CDM}$ model.
  The contour plot was generated using the Python library CosmicFish \citep{Raveri:2016leq}. Dark and light contours refer to the 1$\sigma$ and 2$\sigma$ confidence level, respectively. 
  }
  \label{fig:contour-plot-smarg}
\end{figure*}

As already mentioned, all these results were obtained assuming perfect knowledge of the local count slope, $s(z)$. However, in a realistic scenario, the local count slope will not be exactly known: it will be measured with some uncertainty. In order to take this into account, we compared the optimistic analysis previously discussed to a pessimistic case and a realistic case. In the pessimistic case, we assumed no prior knowledge of local count slope, and we treated it in the same way as the galaxy bias: we marginalised over the local count slope parameters in each redshift bin. In the realistic case, we still marginalised over the local count slope, but we included a uniform $10\%$ prior on the $N_\text{bins}$ extra parameters.

The prior information $\sigma_{s_i} = 0.1 \times s_i$ on the local count slope in the $i = 1, ..., N_\text{bins}$ bins is included adding to our Fisher matrix a diagonal prior information matrix, whose entries are
\begin{equation}
F^\text{prior}_{\alpha\beta} = \delta^{\rm K}_{\alpha\beta}\times
\begin{cases}
0 \hspace{1cm} \mbox{for} \hspace{0.3cm} \alpha \ne s_i\, , \\
\sigma^{-2}_{s_i} \hspace{0.65cm} \mbox{for} \hspace{0.3cm} \alpha = s_i\, . 
\end{cases}
\end{equation}

In Table \ref{tab:prior} we report the 
percent improvement due to magnification for the optimistic (second column), pessimistic (third column), and realistic (fourth column) scenario, for our baseline model of dynamical dark energy. 
In the pessimistic scenario, that is, assuming no prior knowledge of the local count slopes, we partially lose the information encoded in the magnification signal when constraining $\Omega_{\rm m, 0}, \Omega_{\rm b, 0}, \wzero$, and $w_{a}$. More worryingly, $h, n_\text{s}$, and $\sigma_8$ will be measured with larger errors compared to an analysis including only density. We would like to emphasise that this does not imply that an analysis without magnification is preferable for measuring these parameters: as we show in the next section, neglecting magnification generates a shift in the best-fit values of the parameters. Such an analysis would therefore be more precise but less accurate, which is not a viable option.

Finally, in the realistic scenario where we assumed that we can measure $s(z)$ with a $10\%$ precision, we see from Table~\ref{tab:prior} that magnification improves the constraints on all cosmological parameters. The improvement is smaller than in the optimistic scenario, but it still reaches $\sim 20\%$ for 
$\Omega_{\text{m},0}$ and the dark energy equation of state.
This test suggests that an independent precise measurement of the local count slope is crucial for an optimal analysis of the photometric galaxy number counts.
There are several difficulties associated with this measurement. In particular, systematic effects such as noise, colour selection, and dust extinction can have a significant impact (see e.g. ~\citealt{Hildebrandt:2015kcb}). Furthermore, galaxy samples are in general not purely flux-limited. A novel method for estimating the local count slope for a complex selection function has been developed for the Kilo-Degree Survey (KiDS; see \citealt{vonWietersheim-Kramsta:2021lid}). Assessing whether this method will be accurate enough for \Euclid, that is, whether it can be used to estimate the local count slope within a $10\%$ uncertainty, requires further investigation. 

\subsubsection{Shift in the best fit} 
\label{sec:shift}

In an optimal cosmological analysis, we aim to estimate the parameters of our models in a precise and accurate way. 
In this section, we study the impact of magnification on the accuracy of the analysis, that is, we calculate the shift induced on the best-fit values of the parameters due to neglecting magnification in the theoretical modelling of the clustering signal.

\begin{table*}[!ht]
\caption{Shift in
best-fit parameters for GCph alone. 
}
\begin{center}
\begin{tabular}{ ccccccccc}
 \toprule
 model & $\Omega_{\text{m},0}$ & $\Omega_{\text{b},0}$
      & $\wzero$ & $\wa$ & $h$
      & $n_\text{s}$ & $\sigma_8$ & $\sum m_\nu$\\
 \midrule
$\Lambda$CDM & $-0.18$ & $0.004$ & --&  -- &  $-0.02$ & $0.33$ & $-0.57$ & --\\ 
$\Lambda$CDM + $\sum m_\nu$ & $0.96$ &  $-0.15$ & --&  -- &   $-0.42$ & $0.98$ &  $-1.62$ & 1.64 \\
$\wzero\,\wa$CDM&  $-0.65$ &  $-0.64$ & $-1.02$ &  $1.20$ & $0.05$ & $1.04$ & $0.17$ & --\\ 
$\wzero\,\wa$CDM + $\sum m_\nu$&  $-0.90$ & $-1.12$ &  $-1.27$ &  $1.21$ & $0.22$ & $1.59$ & $-0.13$ & $1.62$ \\ 
 \bottomrule
\end{tabular} 
\end{center}
The shift in
best-fit parameters are in units of $1\sigma$. 
We report the results for the same models as in Tables~\ref{tab:gc-const} and  \ref{tab:gc-const-improv}. The shifts are estimated with the formalism described in Sect.~\ref{subsec:fisher-shifts}. The values of shifts that are larger than $1\sigma$ cannot be trusted but indicate that the shift is large. We marginalise over the galaxy bias parameters, and the values of the local count slope are fixed to their fiducial values.
\label{tab:gc-shift}
\end{table*}

 As discussed in Sect.~\ref{fisher-form}, the estimation of the shift is based on a Taylor expansion of the likelihood around the correct model and, therefore, it can be trusted quantitatively only when the shifts $\Delta{\theta}$ are much smaller than the $1\sigma$ error. The results of our analysis should therefore be regarded as a diagnostic to determine whether magnification can be neglected or not: if we find small values for the shifts $\Delta \theta \ll \sigma$, the Taylor expansion is valid, and we can confidently conclude that it is safe to neglect magnification in the theoretical modelling. On the other hand, if large values $\Delta \theta \gtrsim\sigma$ are found, we cannot quantitatively trust the value of the shift, but we can conclude that the shifts are large and that, consequently, magnification cannot be neglected in the theoretical modelling.

In Table~\ref{tab:gc-shift} we report the shift in the best-fit estimation of our parameters for the four models under consideration. 
For a five-parameter $\Lambda$CDM model, all parameter shifts in the best-fit estimation are below $1\sigma$. The measurement of $\sigma_8$ is the most affected by magnification ($\Delta {\sigma_8} \sim 0.6\sigma$). The shifts are negative for $\Omega_{\rm m,0}$ and $\sigma_8$, which means that the magnification contamination decreases the clustering signal. The sign of the magnification contamination depends on the sign of $5s-2$ and on the relative importance of the density-magnification correlation (which is proportional to $5s-2$ and therefore changes sign at $z\simeq 1$), and the magnification-magnification correlation (which is proportional to $(5s-2)^2$ and is therefore always positive). To understand the sign of the shifts, we performed the following test: we ran an analysis where we remove the magnification from the signal for $z>1$, that is, we pretended that magnification contaminates only the redshifts $z\leq 1$. We found that the shifts on all parameters remain almost the same in this case\footnote{The only parameters for which the shift decreases are the bias parameters governing the bias evolution at high redshift.}. This shows that the shifts are not due to the high magnification contamination (S/N $\sim 80$) at high redshift ($z\geq 1.62$ in Fig.~\ref{fig:snmag}) but rather to the (relatively) small contamination (S/N $\sim 10-20$)  at $z\leq 1$. At those redshifts, the factor $5s-2$ is negative. From Fig.~\ref{fig:sntot} we see that the GCph signal peaks for the auto-correlations of redshift bins. We expect therefore the constraints, and consequently the shifts, to come mainly from these auto-correlations. Since the bins are relatively wide, both the density-magnification and the magnification-magnification contribute to the auto-correlations, and we checked that the density-magnification always dominates at $z\leq 1$. As a consequence, the magnification contamination is negative for the bins that contribute most to the constraints, leading to a decrease in $\Omega_{\rm m,0}$ and $\sigma_8$.

For all the models beyond $\Lambda$CDM, we find shifts above $1\sigma$. The parameters that are mostly affected are the parameters beyond the $\Lambda$CDM minimal model: the neutrino mass and the dynamical dark energy parameters $\{\wzero, \wa\}$.
This can be understood by looking at Fig.~\ref{fig:sntot}, where we see that the S/N for GCph peaks at low redshift: $z\in [0.26,0.39]$, which corresponds to bins $i=2,3$. For $\Lambda$CDM, we expect the constraints to be driven by these bins. For models beyond $\Lambda$CDM, however, the evolution with redshift becomes relevant: the sum of the neutrino mass and the dark energy equation of state modify indeed the redshift evolution of perturbations. More redshift bins contribute therefore to the constraints, which increases proportionally the impact of magnification and leads to a larger shift. Since the impact of dark energy and neutrino mass decreases with redshift, we expect however the highest-redshift bins to be irrelevant for the constraints. As before, to check this, we ran an analysis without the magnification contamination at $z>1$ and we found that the shifts on all parameters remain almost the same. This again means that the shifts do not come from the high-redshift bins where the magnification contamination is the largest, but rather from the low-redshift bins. A direct consequence of this is that any alternative model that would be constrained by the highest-redshift bins of \Euclid, would be significantly more biased when neglecting magnification. We note that these results are in agreement with previous analyses on this subject (see e.g.\ \citealt{Cardona:2016qxn,Lorenz:2017iez,Villa_2018}).

Looking at the sign of the shifts of $\Omega_{\rm m,0}$ and $\sigma_8$ for models beyond $\Lambda$CDM, we see from Table~\ref{tab:gc-shift} that when the neutrino mass is included the shift in $\Omega_{\rm m,0}$ becomes positive, whereas in the dynamical dark energy model the shift in $\sigma_8$ becomes positive. However, the overall amplitude is still decreased by magnification, since the negative shifts are always larger than the positive ones.

For our calculation of the shifts, we used the fiducial values of the local count slope measured in the Flagship simulation. We did not consider the local count slope as a free parameter in this part of the analysis, since our goal was to determine the shifts induced  on the other cosmological parameters by a magnification signal of a given fixed amplitude. However, we tested
the stability of our results by repeating the analysis with different fiducial values of the local count slope. We found that,
in the range $s_i = (1 \pm 0.1) s_i^\text{fid}$, the values of the shifts do not change significantly. Therefore, our results are robust with respect to the fiducial $s_i$ used in the analysis.

\subsection{Impact of magnification on the probe combination analysis}

In this section, we present the same analysis described in Sect.~\ref{sec:gch-alone}, but this time for the joint data $\text{GCph} + \text{WL} + \text{GGL}$. 
We note that magnification contributes to the galaxy clustering observable and to the cross-correlation GGL, while in our analysis it does not affect cosmic shear.

\subsubsection{Constraints on cosmological parameters}
\begin{figure}[!ht]
  \includegraphics[width=\linewidth,trim=0 70 0 0,clip]{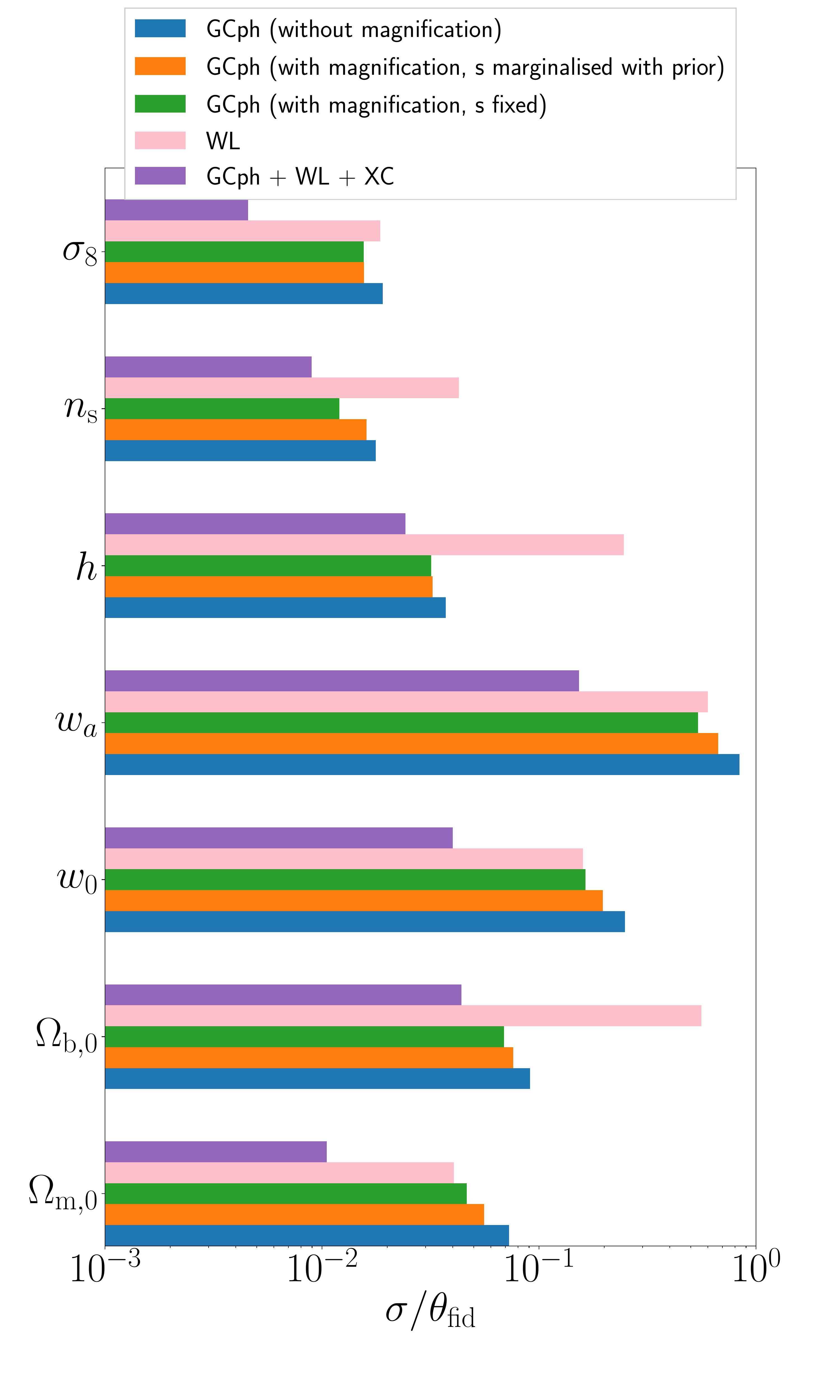}
  \caption{Marginalised 1$\sigma$ errors on cosmological parameters, relative to their corresponding fiducial values for the baseline model of dynamical dark energy.
  The error bars for $\wa$ represent the absolute error, $\sigma$, for this parameter since a relative error cannot be computed for a fiducial value of $0$.
  Each histogram refers to a different cosmological analysis or observational probe. We show in blue a GCph analysis that neglects magnification, in orange a GCph analysis that includes magnification and assumes a $10\%$ prior on the measurement of the local count slope (realistic scenario), and in green a GCph analysis that models magnification assuming a perfect knowledge of the local count slope (optimistic scenario). For comparison, we show in pink the constraints from the WL analysis and in violet the one obtained from the probe combination GCph + WL + GGL.  
  }
  \label{fig:constraints-bar-comparison}
\end{figure}

\begin{figure}
  \includegraphics[width=\linewidth,trim=0 40 0 0,clip]{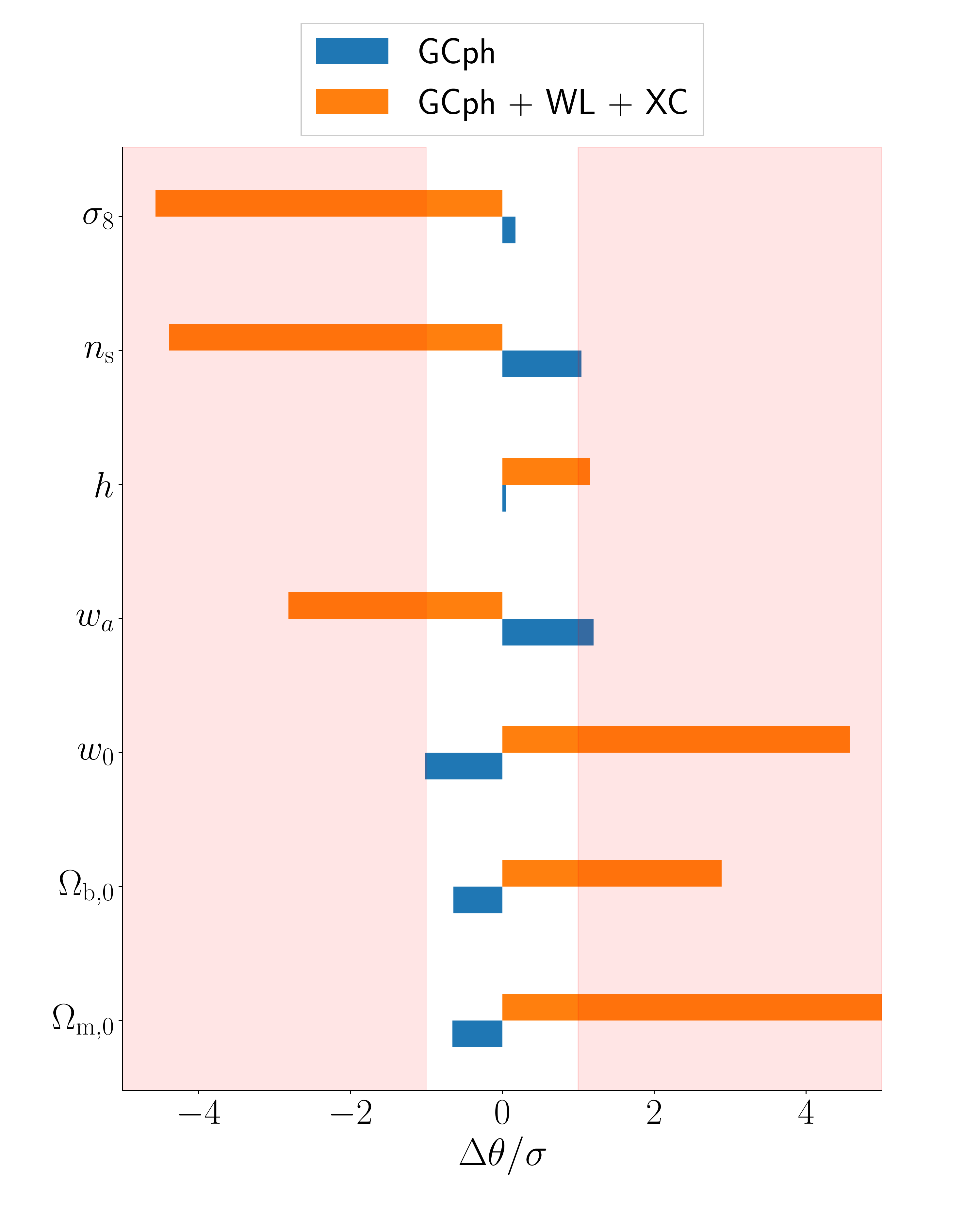}
  \caption{Shift in the best-fit estimation of cosmological parameters induced by neglecting magnification in our theoretical model. 
  The values of the shift are expressed in units of the marginalised 1$\sigma$ constraints. 
  The blue histogram refers to the parameters estimated from the GCph alone analysis, and the orange histogram represents the shifts for the $3\times2$pt analysis GCph + WL + GGL.
  The red regions highlight shifts above 1$\sigma$ in absolute value. The values of the shifts computed with the Fisher formalism cannot be trusted quantitatively in this region.}
  \label{fig:bias-bar-comparison}
\end{figure}

Similar to the discussion in the previous section, we studied the impact of magnification on the constraints on cosmological parameters by comparing a Fisher matrix analysis for the probe combination that neglects this effect and an analysis that consistently includes it. As before, we considered an optimistic case where we assume that the local count slope is exactly known, a pessimistic case where the local count slope is considered as a free parameter, and a realistic case where we include a 10$\%$ prior on the local count slope.

In the optimistic case, that is, assuming a perfect knowledge of the local counts slope, we found that the improvement on the constraints due to magnification is negligibly small, that is, smaller than $3\%$ for all cosmological parameters and all models under consideration. This is due to the fact that the information encoded in magnification is the same as the one in the cosmic shear. As a consequence, adding magnification does not help to break degeneracies between parameters anymore, since these degeneracies are already broken by the inclusion of cosmic shear. This can be seen by looking at Table \ref{tab:wl-gc-xc-const-lens}, where we report the 1$\sigma$ constraints for the joint analysis. Comparing with Table~\ref{tab:gc-const}, we see for example that the constraints on $\Omega_{\rm m,0}$ for our baseline dynamical dark energy model are four times better in the joint analysis, and the constraints on $\sigma_8$ are three times better. This reflects the fact that cosmic shear breaks the degeneracy between the amplitude of perturbations and the bias, and since its S/N is significantly higher than that of magnification (as can be seen from Figs.~\ref{fig:sntot} and~\ref{fig:snmag}), adding magnification does not help anymore. This also becomes clear by looking at Fig.~\ref{fig:constraints-bar-comparison}, which compares the constraints from galaxy clustering alone, with the ones from the joint analysis for our baseline dynamical dark energy model: we see that adding cosmic shear brings a much larger improvement in the constraints than including magnification in the clustering signal.

\begin{table*}[!ht]
\caption{Constraints on cosmological parameters for $\text{GCph} + \text{WL} + \text{GGL}$. 
}
\begin{center}
\begin{tabular}{ ccccccccc}
 \toprule
 model & $\Omega_{\text{m},0}$ & $\Omega_{\text{b},0}$
      & $\wzero$ & $\wa$ & $h$
      & $n_\text{s}$ & $\sigma_8$ & $\sum m_\nu$\\
 \midrule
$\Lambda$CDM & 0.75 & 3.4 & --&  -- &  2.2 & 0.76 & 0.37 & --\\ 
$\Lambda$CDM + $\sum m_\nu$ & 0.91 & 4.0 & --&  -- &  2.3 & 0.76 & 0.60 & 100 \\ 
$\wzero\,\wa$CDM & 1.1 & 4.4 & 4.0 &  15 &  2.4 & 0.89 & 0.46 & --\\ 
$\wzero\,\wa$CDM + $\sum m_\nu$ & 1.2 & 4.5 & 4.0 &  16 &  2.4 & 1 & 0.83 & 140 \\ 
 \bottomrule
\end{tabular} 
\end{center}
1$\sigma$ constraints are relative to their corresponding fiducial values, including magnification (in \%). For the parameter $\wa$, we report the absolute error times 100.
We have marginalised over the galaxy bias and the intrinsic alignment parameters, and the values of the local count slope are kept fixed. We report the results for four cosmological models: a minimal $\Lambda$CDM with one massive neutrino species and fixed neutrino mass, an analogue model that includes dynamical dark energy, denoted as $\wzero\,\wa$CDM, and their extensions where the sum of the neutrino masses is also a free parameter. The constraints obtained when neglecting magnification differ from the values reported here by less than $3\%$ for all cosmological parameters and all models considered.
\label{tab:wl-gc-xc-const-lens}
\end{table*}

\begin{table}[!ht]
\caption{Uncertainty in the local count slope, $\text{GCph} + \text{WL} + \text{GGL}$.
}
\begin{center}
\begin{tabular}{ cccc}
 \toprule
parameter & $s(z_i)$ fixed & $s(z_i)$ marg & + $10 \%$ prior on $s(z_i)$\\
 \midrule
$\Omega_{\text{m},0}$ & $1\%$ & $-23\%$& $-3\%$\\
$\Omega_{\text{b},0}$ & $< 1\%$& $-3\%$ & $<1\%$\\
$\wzero$ & $2\%$& $-16\%$& $<1\%$\\
$\wa$ & $2\%$& $-11\%$ & $2\%$\\
$h$ & $<1\%$& $<1\%$& $<1\%$\\
$n_\text{s}$ & $<1\%$& $-4\%$& $-2\%$\\
$\sigma_8$ & $1\%$& $-14\%$& $<1\%$\\
 \bottomrule
\end{tabular} 
\end{center}
The relative difference, $1 - \sigma_\text{magn}/\sigma_\text{dens}$, is in percentage. As in Table \ref{tab:prior}, we report the results for three scenarios: a) an optimistic scenario, when the local count slope is measured with high accuracy and thus $s(z)$ is kept fixed in the analysis (Col. 2), b) a pessimistic scenario where the local count slope cannot be constrained by an independent measurement, and therefore we marginalise over its values (Col. 3), and c) a realistic scenario such that the local count slope is assumed to be measured independently with a $10\%$ precision (Col. 4).
The results reported here refer to our baseline cosmology, the $\wzero\,\wa\textrm{CDM}$ model.
\label{tab:joint-prior}
\end{table}

These constraints refer to the optimistic scenario. In Table \ref{tab:joint-prior} we compare this with the pessimistic scenario (second column) and the realistic scenario (last column). In the pessimistic scenario, the constraints are degraded at the level of 10--20\%. This degradation, especially in $\sigma_8$ and $\Omega_\text{m,0}$, is due to the fact that we no longer have a precise measure of the density fluctuation amplitude if the amplitude of lensing magnification is completely unknown. In a realistic scenario we are able to recover the same information as in the optimistic case. 

To conclude, including magnification has a negligible impact on the constraints for the joint analysis, provided that the local count slope will be measured independently with a $10\%$ uncertainty. If we do not have independent  measurements of the local count slope, an analysis with no magnification will provide constraints that are up to 10--20\% too optimistic. 

\subsubsection{Shift in the best fit} 
\label{sec:shift_combined}
\begin{table*}[!ht]
\caption{Shift in
best-fit parameters for $\text{GCph} + \text{WL} + \text{GGL}$. 
}
\begin{center}
\begin{tabular}{ ccccccccc}
 \toprule
 model & $\Omega_{\text{m},0}$ & $\Omega_{\text{b},0}$
      & $\wzero$ & $\wa$ & $h$
      & $n_\text{s}$ & $\sigma_8$ & $\sum m_\nu$\\
 \midrule
$\Lambda$CDM & $4.73$ & $0.41$ & --&  -- &  $-0.56$ &  $-1.76$ &  $-2.88$ & --\\ 
$\Lambda$CDM + $\sum m_\nu$ & $5.64$ & $0.65$ & --&  -- &  $0.07$ &  $-1.51$ &  $-4.21$ & $3.08$ \\ 
$\wzero\,\wa$CDM & $6.90$ & $2.89$ & $4.58$ &  $-2.82$ & $1.16$ & $-4.39$ & $-4.56$ & --\\ 
$\wzero\,\wa$CDM + $\sum m_\nu$& $6.21$ & $2.71$ & $4.57$ &   $-2.82$ & $1.09$ & $-3.60$ &  $-2.91$ & $0.51$ \\ 
 \bottomrule
\end{tabular} 
\end{center}
We report the shift in the values of the
best fitting parameters, in units of $1\sigma$, for the same models as in Table \ref{tab:wl-gc-xc-const-lens}. The shifts are computed with the formalism described in Sect.~\ref{subsec:fisher-shifts}, and therefore, the values of shifts that are larger than $1\sigma$ cannot be quantitatively trusted but indicate that the shift is large. We marginalise over the galaxy bias and intrinsic alignment parameters, and the values of the local count slope are fixed to their fiducial values.
\label{tab:wl-gc-xc-shift}
\end{table*}

The fact that magnification has little impact on the constraints on cosmological parameters extracted from the joint analysis does not mean that an analysis that neglects this effect is accurate in terms of parameter estimation.
Applying the Fisher formalism to our model comparison problem, we compute the shift 
in the best-fit estimation for an analysis 
that assumes the incorrect model with no magnification. 

The values of the shifts are reported in Table \ref{tab:wl-gc-xc-shift}. For all four cosmological models under consideration we find large deviations, that is, above $1\sigma$. 
Although the Fisher formalism that we use cannot be trusted quantitatively in this case, we can conclude that an analysis that neglects magnification does not provide an accurate estimation of cosmological parameters.
This important result agrees with previous studies (see \citealt{Duncan:2013haa}): although magnification has little impact on the precision of the cosmological constraints in the $3\times2$pt analysis, inferred cosmological parameter values are highly biased when the effect is neglected. 
Comparing the above with the shifts obtained from galaxy clustering alone (see Table~\ref{tab:gc-shift}), we see that the shifts (in units of $\sigma$) are significantly larger in the joint analysis, especially for $\Omega_{\rm m, 0}$ where it lies between 5 and 7$\sigma$, depending on the model, and for $\sigma_8$ where it is between 3 and 4.5$\sigma$. This is only partially due to the fact that now the $1\sigma$ errors are smaller, as is seen in Fig.~\ref{fig:constraints-bar-comparison}. More importantly, the shear measurements provide a precise estimation of the gravitational potential so that number counts are no longer well-fitted without lensing magnification.

Looking at the sign of the shifts in Table~\ref{tab:wl-gc-xc-shift}, we see that the shifts in $\sigma_8$ are negative for all models, whereas the shifts in $\omegamatter$ are always positive. Moreover, we checked the shifts of the best-fit galaxy bias parameters and found that most of them are negative. In Fig.~\ref{fig:bias-bar-comparison}, we directly compare the shifts for our baseline dynamical dark energy model in the GCph analysis and in the combined analysis. The shifts are systematically of opposite sign. We already know that in the GCph signal, the magnification contamination is negative in the pairs of redshift bins that contribute most to the constraints. In the GGL signal, the magnification contamination is proportional to $5s-2$, which is negative at $z<1$ and positive at $z>1$. The sign of the shifts will therefore depend on which range of redshift contributes most to the constraints. As before, we ran an analysis removing the magnification contamination in GCph and in GGL at $z>1$. We found that the shifts decrease slightly in amplitude but remain of the same sign: for example, the shift in $\sigma_8$ decreases from $-4.6\sigma$ to $-2.3\sigma$, whereas the shift in $\omegamatter$ decreases from 6.9$\sigma$ to $4.4\sigma$. This means that the constraints are mainly driven by $z<1$, where the magnification contamination is negative in both GCph and GGL. Indeed, if the magnification contamination at $z>1$ were to be the main driver of the shifts, we would expect the shifts to change sign when we remove the $z>1$ contamination, since at $z=1$ the contamination in GGL changes sign.
This test shows that removing from the analysis the bin configurations at high redshift, which are dominated by magnification, does not reduce the bias in the best-fit estimation due to neglecting magnification, as already pointed out in \cite{Thiele:2019fcu}. 

We then performed another test, where we fixed the value of $\omegamatter$ and computed the shifts in the other parameters for our baseline dynamical dark energy model. We found that, in this case, the shift in $\sigma_8$ becomes positive, whereas the shifts in the bias parameters become significantly more negative. This shows that there is a strong interplay between the impact of $\sigma_8, \omegamatter$, and the bias on the amplitude of the GCph signal and the GGL signal, and that there are therefore various ways of decreasing the overall amplitude of these signals. When only GCph is included, one can decrease the amplitude of the density signal by decreasing $\sigma_8, \omegamatter$, or the bias. Depending on the model, different solutions might mimic better the magnification contamination. In the joint analysis on the other hand, the problem is much more constrained: since the WL (shear-shear correlation) is not contaminated, this part of the signal has to remain unchanged. Any negative shift in $\sigma_8$ needs therefore to be compensated for by a positive shift in $\omegamatter$ to keep $S_8=\sigma_8(\omegamatter/0.3)^{0.5}$ almost constant. This explains why in all models the shift in $\sigma_8$ and the shift in $\omegamatter$ have opposite sign (see Table~\ref{tab:wl-gc-xc-shift}). In particular, for the dynamical dark energy model, we have that the positive shift $\Delta \Omega_{\rm m,0}/\Omega_{\rm m,0}=7\%$ and the negative shift $\Delta \sigma_8/\sigma_8=-2\%$ partially compensate to give a small positive shift $\Delta S_8/S_8=1\%$. Moreover, the shifts must be adjusted to decrease at the same time the GCph signal, which is proportional to $b^2\langle\delta\delta\rangle$, and the GGL signal, which is proportional to $b\langle\delta\kappa\rangle$. From Table~\ref{tab:wl-gc-xc-shift} and Fig.~\ref{fig:bias-bar-comparison} we see that all this leads to shifts that are systematically larger in the joint analysis than in the GCph analysis. This shows that including magnification in the theoretical model is absolutely crucial for the joint analysis of the photometric sample.

\section{Robustness tests}

\label{sec:rob-tests}
The results presented in the previous sections are a natural extension of the \Euclid forecast presented in \citetalias{Blanchard:2019oqi} to include magnification 
in the analysis of the photometric sample. 
We adopted three underlying simplifications: 1) non-linearities are modelled with the \texttt{Halofit} prescription \citep{Smith:2002dz}, including the Bird and Takahashi corrections; 2) the RSD contribution to the galaxy count is neglected in the analysis; and 3) both the signal and covariance are computed using Limber's approximation.
In what follows, we test the robustness of our results 
with respect to these three assumptions. 

\subsection{Non-linear prescription}

\citet{Martinelli:2020yto} investigate in
detail the impact of different non-linear prescriptions on parameter estimation for the WL analysis of \Euclid. 
In this work, we do not aim to compare the parameter estimation analysis itself for different non-linear models. Instead, we want to verify whether the impact of magnification on the analysis strongly depends on our non-linear recipe. 

With this objective in mind, 
we compared the analysis 
presented in Sect. \ref{sec:res} for three
non-linear prescriptions. The first is \texttt{Halofit} \citep{Smith:2002dz, Bird2012,Takahashi:2012em}, a model for the non-linear matter power spectrum inspired by the halo model \citep{Cooray:2002dia}. This is our reference recipe, and it is the implementation
    adopted in the forecast validation project for \Euclid \citepalias{Blanchard:2019oqi}. The second is \texttt{Halofit+Pk-equal} \citep{Casarini:2016ysv}, which is an extension to the \texttt{Halofit} fitting formula to models with a redshift-dependent equation of state for the dark energy component.
The third is \texttt{HMCODE} \citep{Mead:2016zqy}, an alternative parametrisation for the total matter power spectrum that is based on the halo model but with physically motivated free parameters. Although this model can account for baryonic feedback,  in this test we used the model fitted to the Cosmic Emulator dark-matter-only simulation \citep{Heitmann:2013bra}. The three models considered here are all implemented in version \texttt{v2.9.4} of {\sc class}~\citep{class2} and, therefore, applying our analysis to 
different recipes is straightforward. 

We performed this test on our baseline cosmology, and we assumed the optimistic scenario for the local count slope, that is, we assumed that $s(z)$ is exactly known. Therefore, its value was fixed in the analysis.

\begin{table}[!ht]
\caption{Impact of non-linear prescription on the constraints for GCph alone.}
\begin{center}
\begin{tabular}{ cccc}
 \toprule
parameter & Halofit & Halofit + Pk-equal & HMCODE\\
 \midrule
$\Omega_{\text{m},0}$ & $36\%$ & $24\%$& $31\%$\\
$\Omega_{\text{b},0}$ & $24\%$& $15\%$ & $27\%$\\
$\wzero$ & $34\%$& $22\%$& $20\%$\\
$\wa$ & $35\%$& $25\%$ & $23\%$\\
$h$ & $14\%$& $13\%$& $6\%$\\
$n_\text{s}$ & $32\%$& $18\%$& $42\%$\\
$\sigma_8$ & $18\%$& $14\%$& $11\%$\\
 \bottomrule
\end{tabular} 
\end{center}
We compare the relative difference $1-  \sigma_\text{dens+magn}/\sigma_\text{dens}$, expressed as a percentage, obtained when using three different non-linear prescriptions, as described in the text. The results reported here refer to our baseline cosmology, that is, the $\wzero\,\wa\textrm{CDM}$ model. 
\label{tab:nl-gc-constraints}
\end{table}

\begin{table}[!ht]
\caption{Impact of non-linear prescription on the shifts for GCph alone.
}
\begin{center}
\begin{tabular}{ cccc}
 \toprule
parameter & Halofit & Halofit + Pk-equal & HMCODE\\
 \midrule
$\Omega_{\text{m},0}$ &$-0.65$ & $-1.08$ & $-1.34$\\
$\Omega_{\text{b},0}$ &$-0.64$  & $-1.00$& $ -1.42$\\
$\wzero$ & $-1.02$& $-1.62$& $-1.82$\\
$\wa$ &$1.20$& $-1.84$& $2.06$\\
$h$ &$0.05$ & $0.53$& $-0.26$\\
$n_\text{s}$ &$1.04$ &$1.03$& $1.33$\\
$\sigma_8$ & $0.17$ & $0.72$& $0.67$\\
 \bottomrule
\end{tabular} 
\end{center}
We compare the shift in the
best-fit parameters, in units of $1\sigma$,
 obtained using three different non-linear prescriptions, as described in the text. The results reported here refer to our baseline cosmology, the $\wzero\,\wa\textrm{CDM}$ model.
\label{tab:nl-gc-shift}
\end{table}

In Table \ref{tab:nl-gc-constraints} we compare the improvement in terms of constraining power for the non-linear models considered here, for a GCph alone analysis. The maximum percentage improvement of the 1$\sigma$ errors between an analysis with magnification and an analysis that neglects this effect
varies between $25\%$ (\texttt{Pk-equal}) and $42\%$ (\texttt{HMCODE}). 

Table \ref{tab:nl-gc-shift}
shows a comparison of the shifts in the best-fit parameters for GCph alone analysis.
For all the non-linear prescriptions considered here, neglecting magnification can introduce a shift larger than  1$\sigma$ for several model parameters. 
Therefore, we find that magnification should not be neglected in the galaxy clustering analysis of the photometric sample of \Euclid, independently of the non-linear modelling. 

\begin{table}[!ht]
\caption{Impact of non-linear prescription on the shifts for $\text{GCph} + \text{WL} + \text{GGL}$.}
\begin{center}
\begin{tabular}{ cccc}
 \toprule
parameter & Halofit & Halofit + Pk-equal & HMCODE\\
 \midrule
$\Omega_{\text{m},0}$ &$6.90$ & $6.4$ & $6.35$\\
$\Omega_{\text{b},0}$ &$2.89$  & $2.99$& $ 2.83$\\
$\wzero$ & $4.58$& $4.58$& $4.37$\\
$\wa$ &$-2.82$& $-3.07$& $-3.31$\\
$h$ &$1.16$ & $0.71$& $0.72$\\
$n_\text{s}$ &$-4.39$ &$-4.17$& $-2.79$\\
$\sigma_8$ & $-4.56$ & $-4.73$& $-4.49$\\
 \bottomrule
\end{tabular} 
\end{center}
 We compare the shift in the
best-fit parameters, in units of $1\sigma$,
 obtained using three different non-linear prescriptions, as described in the text. The results reported here refer to our baseline cosmology, the $\wzero\,\wa\textrm{CDM}$ model.
\label{tab:nl-gc-wl-xc-shifts}
\end{table}

We repeated the same analysis for the probe combination $\text{GCph} + \text{WL} + \text{GGL}$. 
We find that the impact of magnification on the constraints is negligible ($< 3\%$) for all non-linear prescriptions considered here. 
In Table \ref{tab:nl-gc-wl-xc-shifts} we report the shifts in the best-fit estimation due to neglecting magnification in the joint analysis. The shifts do not
strongly depend on the way we model non-linearities, and they show that magnification should not be neglected in the analysis. 

In conclusion, we have shown that the results that we present in the main body of this manuscript are valid independent of the non-linear modelling. 
Although this test assumes the $\wzero\wa$ parametrisation, it has been shown in, for example, \cite{Lorenz:2017iez} and  \cite{Villa_2018} that the impact of magnification on galaxy clustering is enhanced for several modified gravity models. Therefore, we do not expect this picture to change significantly for other cosmological models. 

\subsection{Redshift-space distortions}

Redshift-space distortions are currently neglected in the \Euclid forecast for the photometric sample. 
The reason is twofold. First, in photometric redshift bins radial correlations are washed out due to poor redshift resolution and, therefore, the information encoded in the RSD contribution is highly suppressed. Second, the non-linear modelling of RSDs is a challenging task: the several prescriptions proposed to include the finger-of-god effects into our theoretical model have been proven to be inaccurate for modelling RSD contribution to the angular power spectrum \citep{Jalilvand:2019brk} and it has also been shown that finger-of-god effects change the RSD harmonic-space spectrum on all scales \citep{Gebhardt:2020imr}. Although a comprehensive study on the impact of RSDs in the analysis of the \Euclid photometric sample would require an accurate modelling of RSDs, which is beyond the scope of this work, we are interested in studying whether including 
the RSD signal could significantly affect our conclusions on the impact of magnification for the \Euclid photometric sample. 

For this purpose, we repeated the analysis presented in Sect.~\ref{sec:res}, including RSD contributions to galaxy clustering. The non-linear RSD is naively modelled using the Kaiser formula, that is, finger-of-god effects are neglected. This approximation overestimates the contribution from RSDs  to the galaxy clustering analysis, and should therefore give a first indication of whether the effect is important or not. 

\begingroup
\setlength{\tabcolsep}{0.82em}
\begin{table*}[!ht]
\caption{Impact of RSDs.
}
\begin{center}
\begin{tabular}{ ccccccccc}
 \toprule
 & & $\Omega_{\text{m},0}$ & $\Omega_{\text{b},0}$
      & $\wzero$ & $\wa$ & $h$
      & $n_\text{s}$ & $\sigma_8$\\
 \midrule
\multirow{2}{*}{$1-\frac{\sigma_\text{magn}}{\sigma_\text{dens}}\,[\%]$ (GCph)}& no RSD & $36\%$ & $24\%$ & $34\%$ &  $35\%$ &  $14\%$ & $32\%$ & $18\%$ \\ 
    & with RSD & $32\%$ & $20\%$ & $29\%$ &  $29\%$ &  $13\%$ & $27\%$ & $17\%$ \\ 
 \midrule
 \multirow{2}{*}{$\Delta\theta/\sigma_\theta$ (GCph)}& no RSD &  $-0.65$ &  $-0.64$ &  $-1.02$&  $1.20$ & $0.05$ & $1.04$ & $0.17$ \\ 
 & with RSD &   $-0.75$ &  $-0.63$ &  $-0.83$ &  $0.86$ & $0.25$ & $1.05$ & $0.09$ \\
 \midrule
\multirow{2}{*}{$1-\frac{\sigma_\text{magn}}{\sigma_\text{dens}}\,[\%]$ ($\text{GCph} + \text{WL} + \text{GGL}$)}& no RSD & $1\%$ & $<1\%$ & $2\%$  & $2\%$   & $< 1\%$ & $< 1\%$  &  $1\%$  \\ 
    & with RSD & $1\%$ &  $< 1\%$  & $2\%$  & $2\%$   & $< 1\%$  &  $< 1\%$  &  $1\%$   \\ 
 \midrule
 \multirow{2}{*}{$\Delta\theta/\sigma_\theta$ ($\text{GCph} + \text{WL} + \text{GGL}$)}& no RSD & $6.90$ & $2.89$ & $4.58$ &  $-2.82$ & $1.16$ & $-4.39$ & $-4.56$ \\ 
    & with RSD & $6.95$ & $2.82$ & $4.62$ & $-2.98$ & $1.07$ & $-4.22$  & $-4.73$  \\
 \bottomrule
\end{tabular} 
\end{center}
Impact of magnification in the GCph and $\text{GCph} + \text{WL} + \text{GGL}$ analyses, including or neglecting RSDs, for our baseline cosmology. In the results labelled as `with RSD', we add the contribution of RSDs both in the Fisher analysis that includes magnification and the one that neglects it. Vice versa, the results denoted as `no RSD'
completely neglect RSDs, and they correspond to the analysis presented in Sect.~\ref{sec:res}. 
\label{tab:rsd}
\end{table*}
\endgroup

In Table \ref{tab:rsd} we compare the impact of lensing on the constraints and the shift in the best fit induced by neglecting magnification, with and without RSDs. We stress that the lines denoted `with RSD' include the RSD signal, both in the Fisher analysis that includes magnification and the one that neglects it. Moreover, in the shift analysis, we are comparing an incorrect model that includes density and RSDs to a correct model that accounts for density, RSDs, and magnification. 
For both the GCph alone analysis and the joint analysis, including RSDs 
does not significantly change the improvement in the constraints driven by magnification and the shift in the best-fit estimation induced by neglecting this effect. Therefore, our conclusions on the impact of magnification do not depend on the RSD contribution. 
However, we stress that this result does not imply that RSDs can be neglected in the analysis. In fact, an analysis without RSDs could still provide an inaccurate estimate of cosmological parameters. This aspect will be addressed in a future work. 

\subsection{Limber's approximation}

\begingroup
\setlength{\tabcolsep}{0.75em}
\begin{table*}[!ht]
\caption{Impact of Limber's approximation.
}
\begin{center}
\begin{tabular}{ ccccccccc}
 \toprule
 & & $\Omega_{\text{m},0}$ & $\Omega_{\text{b},0}$
      & $\wzero$ & $\wa$ & $h$
      & $n_\text{s}$ & $\sigma_8$\\
 \midrule
\multirow{2}{*}{$1-\frac{\sigma_\text{magn}}{\sigma_\text{dens}}\,[\%]$ (GCph)}& Limber & $36\%$ & $24\%$ & $34\%$ &  $35\%$ &  $14\%$ & $32\%$ & $18\%$ \\ 
    & no Limber & $47\%$ & $36\%$ & $47\%$ &  $48\%$ &  $18\%$ & $43\%$ & $27\%$ \\ 
 \midrule
 \multirow{2}{*}{$\Delta\theta/\sigma_\theta$ (GCph)}& Limber &  $-0.65$ &  $-0.64$ &  $-1.02$ &  $1.20$ & $0.05$ & $1.04$ & $0.17$ \\ 
 & no Limber &   $-1.77$ &  $-1.77$ &  $-2.30$ &  $2.53$ & $0.47$ & $2.18$ & $1.20$ \\
 \midrule
\multirow{2}{*}{$1-\frac{\sigma_\text{magn}}{\sigma_\text{dens}}\,[\%]$ (GCph + WL + GGL)}& Limber & $1\%$ & $<1\%$ & $2\%$  & $2\%$   & $< 1\%$ & $< 1\%$  &  $1\%$  \\ 
    & no Limber & $1\%$ &  $< 1\%$  & $2\%$  & $2\%$   & $< 1\%$  &  $< 1\%$  &  $1\%$   \\ 
 \midrule
 \multirow{2}{*}{$\Delta\theta/\sigma_\theta$ ($\text{GCph} + \text{WL} + \text{GGL}$)}& Limber & $6.90$ & $2.89$ & $4.58$ &  $-2.82$ & $1.16$ & $-4.39$ & $-4.56$ \\ 
    & no Limber & $6.82$ & $2.87$ & $4.45$ & $-2.65$ & $1.13$ & $-4.37$  & $-4.46$  \\
 \bottomrule
\end{tabular} 
\end{center}
Impact of magnification in the GCph and $\text{GCph} + \text{WL} + \text{GGL}$ analyses for our baseline cosmology. We compare the results obtained within Limber's approximation to an analysis that does not use Limber at low $\ell$, as described in the text.
\label{tab:limber}
\end{table*}
\endgroup

An exact computation of the angular power spectra for the galaxy clustering and WL analysis requires
the estimation of double integrals in redshift (or comoving distance) of spherical Bessel functions and their derivatives, which is a numerical challenge for data-analysis pipelines due to the oscillatory behaviour of the Bessel functions. 
The computational time can be drastically reduced when making use of Limber's approximation \citep{Limber:1953,Limber:1954zz, LoVerde:2008re},
which assumes small angular scales and that the other function that appears in the radial integral varies much more slowly than the spherical Bessel functions. Effectively, this implies that we can approximate the spherical Bessel functions with a Dirac-delta function,
\begin{equation*}
j_\ell(x) \simeq \sqrt{\frac{\pi}{2\ell + 1}}  \delta_{\rm D}  \left(\ell + \frac{1}{2} - x\right)\,.
\end{equation*}

The accuracy of Limber's approximation
depends on the selection functions of the tracers and the scales that we are probing (see for example \citealt{Simon:2006gm, Eriksen:2014wda, Kitching:2016zkn, Kilbinger:2017lvu, Lemos:2017arq, Fang:2019xat, Matthewson:2020rdt}). 
For tracers with a broad kernel, such as cosmic shear, Limber's prescription has a relatively small impact on the estimation of cosmological parameters \citep{Kilbinger:2017lvu, Lemos:2017arq}. On the other hand, the approximation is inaccurate for the density and RSD contributions to the number count, especially for selection functions with a narrow radial width \citep{Eriksen:2014wda, Fang:2019xat, Matthewson:2020rdt}.
 
Since a brute-force computation of the
angular power spectra is not doable 
for a full MCMC analysis, Limber's approximation has been widely adopted in the literature \citepalias{Blanchard:2019oqi}, and we adopted the same approximation in the analysis presented in the previous sections of this paper. In this section, we study the impact of the approximation on the analysis.
For this purpose, we ran the Fisher analysis presented
in Sect. \ref{sec:res} using a brute-force integration for estimating the angular spectra on large scales, that is, for $\ell < \ell_\text{Limb}$, and turning on Limber's scheme only for sufficiently large multipoles, where the approximated spectra are accurate enough. 
In order to perform this test, we 
used the recipe implemented in the {\sc class} code \citep{CLASSgal}, where two parameters regulate the multipoles threshold at which Limber's approximation is active: 1)  \texttt{l\_switch\_limber\_for\_nc\_local\_over\_z}, which regulates the threshold at which the density contributions to the galaxy clustering power spectra are computed using Limber, that is, $\ell_\text{Limb} = \texttt{l\_switch\_limber\_for\_nc\_local\_over\_z} \times z_\text{m}$ for the density selection function, where $z_\text{m}$ is the mean redshift of the bin; and 2) \texttt{l\_switch\_limber\_for\_nc\_los\_over\_z}, which similarly defines the multipoles' threshold at which the lensing and magnification contributions to the power spectra are computed using Limber. 
We note that, in the {\sc class} implementation, Limber's threshold is redshift dependent, as the approximation
is more accurate at low $z$. 

For the purpose of our analysis, this test is \emph{de facto} equivalent to a brute-force analysis that does not employ Limber at all. We compared this setting to the less conservative  $\texttt{l\_switch\_limber\_for\_nc\_local\_over\_z} = 300$, $\texttt{l\_switch\_limber\_for\_nc\_local\_over\_z} = 40$ and we verified that the constraints differ by a few percent at most in the two cases.

In Table \ref{tab:limber} we
quantify the impact of Limber's approximation on our results. For a galaxy clustering analysis alone, 
Limber's approximation has a non-negligible effect, and in the most accurate analysis, which does not rely on Limber at low $\ell$, we find that magnification has a larger impact, both in terms of constraints on cosmological parameters, and the accuracy of the best-fit estimation. The improvement in constraining power when magnification is included reaches $ 48\%$ for $\Omega_{\text{m},0}, \wzero, \wa$, while the shifts are roughly twice as large, in absolute value.
The large impact of Limber on this analysis can be understood as follows: Limber mostly affects the analysis without magnification, degrading the constraints at the 30--40\% level. 
The effect of Limber on an analysis that includes magnification is smaller, that is, 
constraints are affected by Limber at the $10\%$ level. 
The overall effect on the constraints is that the impact of magnification is underestimated when Limber is employed on all scales.  
On the other hand, we find that the impact of Limber's approximation is marginal for the probe combination analysis.

Our results show that not using Limber's approximation does not substantially
modify the take-home message of our work, that is, that magnification needs to be taken into account in the analysis of the photometric sample of \Euclid.
However, they also point out that Limber's approximation may not be sufficiently accurate for modelling the two-point angular statistics 
of galaxy clustering. Finding a scheme that accommodates both the required accuracy and speed of the cosmological analysis would certainly be welcome and would require a specific investigation. Recent developments in this direction can be found\ in, for example, \citet{Fang:2019xat} and \citet{Matthewson:2020rdt}.
\FloatBarrier
\section{Conclusions}\label{s:con}
In this work, we have studied the effect of lensing magnification on galaxy number counts in the photometric survey of \Euclid. We have investigated the pure photometric number counts and the correlation of number counts with the tangential shear. While magnification also affects the shear power spectrum, 
we have neglected this effect in our analysis as it is of second order, and we expect it to have a smaller impact  on the probe combination analysis than the first-order magnification term in the number counts. The effect of this correction on the WL analysis has been investigated in \cite{Deshpande:2019sdl}.

In previous forecasts of the capabilities of \Euclid's photometric survey, lensing magnification was neglected. We have studied its effect for $\Lambda$CDM and a dynamical dark energy model, with and without varying neutrino
masses. We have determined the changes in error bars that are obtained by including lensing magnification in the analysis as well as the shift in the best-fit cosmological parameters due to neglecting magnification in the theoretical modelling of the signal.

When considering the galaxy clustering signal alone, lensing magnification significantly reduces the error bars on cosmological parameters (especially $\sigma_8$,
$\ns$, and $\omegamatter$), assuming a perfect knowledge of the local count slope; neglecting it leads to significant shifts in the best-fit parameters. The reduction in errors comes mainly from the fact that magnification information breaks the degeneracy between the amplitude of density fluctuations, $\sigma_8$, and galaxy bias. 

Once we also include shear and cross-correlation data, including
magnification no longer has a significant effect on the error bars, that is, on the precision of the analysis.
However, neglecting magnification leads to very significant shifts in the best-fit parameters of up to six standard deviations. In fact, all the parameters of the dynamical dark energy model are shifted by more than one standard deviation. Hence, the {accuracy} of
modelling is drastically improved by including lensing magnification.

Even though shifts of more than $1\sigma$ cannot be taken at face value in our Fisher matrix approach (since the shifts are determined at first order in
$\De\theta/\sigma$), a shift of order one or more standard deviations robustly indicates that the analysis is significantly biased. To obtain a good estimate for the value of the shift, we would have to perform an MCMC analysis as\ in, for example, \cite{Cardona:2016qxn}.

In our forecast we have adopted realistic specifications: the galaxy number density, the galaxy bias, and the local count slope have been extracted from the Flagship galaxy catalogue. The galaxy sample is split into 13 photometric bins, which optimises the figure of merit for galaxy clustering and GGL \citep{Pocino2021}. The impact of magnification is expected to depend on the number of redshift bins, as discussed in \cite{Villa_2018}. We expect magnification to have a bigger impact on the cosmological constraints for a few wider redshift bins. This is due to the facts that the radial correlations are suppressed in wider bins and the cosmological constraints come primarily from transverse correlations, which are more affected by magnification.
On the other hand, a larger number of thinner redshift bins will not increase the number of modes induced by magnification; only the modes dominated by density and RSDs will be boosted in this configuration. We thus expect that the overall impact of magnification will be smaller for this configuration. Nevertheless, in this paper we do not aim to propose an optimal binning to enhance or minimise the importance of magnification. For configurations similar to the optimal choice, that is, a number of photometric bins in the redshift range 10--15, we do not expect the overall conclusions of our paper to significantly change.

We have also tested the robustness of our predictions with respect to the most relevant approximations used in the analysis. We have compared three prescriptions for  including  non-linearities in the matter power spectrum and found that their impact on the shifts is not substantial. Moreover, we have found that we obtain similar results whether or not we include RSDs in our analysis. The use of Limber's approximation, however, has an impact on our results in the analysis of galaxy clustering alone. 
Using Limber actually leads us to underestimate both the improvements brought by magnification on the cosmological constraints and the shifts induced on the best-fit values. However, in the combined analysis, which includes the shear and cross-correlations, this difference disappears. This finding confirms similar results by~\cite{Fang:2019xat} for the Vera C. Rubin Observatory's Legacy Survey of Space and Time and the DES, where it is also found that while Limber's approximation is quite inaccurate in a clustering-only analysis, it performs significantly better in a combined analysis.  

This work presents the minimal extension of the \Euclid forecast in \citetalias{Blanchard:2019oqi} to include lensing magnification in galaxy number counts. 
The effect is included at leading order in the magnification expansion. Second-order effects, discussed for example in \cite{Menard:2002ia}, are neglected. 
Moreover, as pointed out in \cite{Monaco_2019}, galaxy bias depends on luminosity, so a modulation of survey depth on the sky (due to systematics in that paper, while here it is due to lensing) couples with galaxy density to give a contribution that is of opposite sign of the magnification bias (higher magnification will give observational access to less luminous galaxies, which are less biased). This contribution could be significant for bright galaxies, whose bias is more strongly dependent on luminosity.

We have not considered the direct estimation of magnification via flux measurements. Therefore, systematic effects such as blending and obscuration are not included in the analysis. Their correct modelling will be needed in order to optimise direct magnification measurements \citep{Menard:2010, Hildebrandt:2015kcb, Gaztanaga:2020lag}.
The final main conclusion is simply that for an accurate estimation of cosmological parameters, lensing magnification needs to be included in the analysis of the photometric survey of \Euclid.  Failing to do so would lead to an incorrect interpretation of the results of the photometric survey. In particular, using a theoretical modelling without lensing magnification could mistakenly lead us to believe that we have detected deviations from $\Lambda$CDM or even a modification of general relativity.

\begin{acknowledgements}
CB, RD, GJ, MK, JA and FL acknowledge support from the Swiss National Science Foundation. CB acknowledges funding from the European Research Council (ERC) under the European Union’s Horizon 2020 research and innovation programme (Grant agreement No. 863929; project title ”Testing the law of gravity with novel large-scale structure observables”). IT acknowledges support from the Spanish Ministry of Science, Innovation and Universities through grant ESP2017-89838, and the H2020 programme of the European Commission through grant 776247. SC acknowledges support from the `Departments of Excellence 2018-2022' Grant (L.\ 232/2016) awarded by the Italian Ministry of University and Research (\textsc{mur}).
\AckEC
\end{acknowledgements}

\bibliographystyle{aa}
\bibliography{biblio}

\begin{appendix}
\label{ap:all}
\section{Shear correlation function}
\label{ap:shear-func-cross}
Since it is not commonly discussed in the literature, we summarise here the expression for the shear and number count tangential shear correlation functions, which can be expressed in terms of the corresponding power spectra. As the shear is a helicity-2 quantity this relation is not simply given by the Legendre polynomials, but it is~\citep{Stebbins:1996wx}
\be
\langle\ga(\bn, z)\ga(\bn', z')\rangle =\hspace{-0.03cm} \sum_\ell \frac{(2\ell+1)(\ell-2)!}{\pi(\ell+2)!}\,C_\ell^{\ga\ga}(z, z')\,G_{\ell\, 2}(\bn\cdot\bn') \,.
\ee

Here the function $G_{\ell\, 2}$ is given by
\be
G_{\ell\, 2}(\mu) =  \left(\frac{4\!-\!\ell}{1\!-\!\mu^2}-\frac{1}{2}\ell(\ell-1)\right)P_{\ell\,2}(\mu)+  (\ell+2)\frac{\mu}{1-\mu^2}P_{\ell-1,\,2}(\mu)\,, 
\ee
and $P_{\ell\,2}$ is the modified Legendre function, of degree $\ell$ and index $m=2$ (see~\citealt{Abram}).

Furthermore,  the correlation spectrum between some scalar function, $f$, and a helicity-2 tensor, $\ga_{ab}$, is determined by the `tangential' component, $\ga_{\rm t}=\ga_{ab}e^ae^b$, where $\boldsymbol{e}=(e^1,e^2)$ is the vector pointing from the point $\bn$ to $\bn'$ on the sphere. A function is only correlated to the scalar part of the traceless tensor $\ga_{ab}$, which is the traceless second (angular) derivative of a potential $\psi$,
\bea
\ga_{a b}(\bn', z') &=& \left(\nabla_a\nabla_b-\frac{1}{2}\de_{ab}\laplaciansphere\right)\psi \,,
\eea
where $\laplaciansphere$ denotes the Laplacian on the sphere.
In the case, of interest to us, $\psi$ is the lensing potential.
For the correlation function of a scalar quantity $f$ and the tangential part of a helicity-2 field derived from a potential $\psi$, one obtains the following expression~\citep[see e.g.][]{Ghosh:2018nsm}:
\begin{align}
\langle f(\bn,z)\ga_{\rm t}(\bn',z')\rangle =& - \frac{1}{8\pi}\sum_\ell (2\ell+1)\,C_\ell^{f\psi}(z,z')\,P_{\ell\,2}(\bn\cdot\bn') \\
=&\frac{1}{4\pi}\sum_\ell C_\ell^{f\ka}(z,z')\frac{2\ell+1}{\ell(\ell+1)}P_{\ell\,2}(\bn\cdot\bn')\,,  \label{e:cor-fk}
\end{align}
where $\ka = \De\psi/2$. 
The angular dependence via $P_{\ell\,2}$ is a consequence of the fact that $\ga_{\rm t}(\bn')$ behaves as a helicity-2 quantity under rotations around $\bn'$.
Setting
\bea
\langle f(\bn, z)\ga_{\rm t}(\bn', z')\rangle =
\frac{1}{4\pi}\sum_\ell C_\ell^{f\ga_{\rm t}}(z, z')\,\frac{2\ell+1}{\ell(\ell+1)}\,P_{\ell\,2}(\bn\cdot\bn')\,  \label{e:cor-fgt}
\eea
implies that the correlation spectra of $f$ with $\ga_{\rm t}$ and $\ka$ agree,
\be\label{e:cfga}
C_\ell^{f\ga_{\rm t}}(z, z')=C_\ell^{f\ka}(z, z')  \,.
\ee

\section{Code validation}
\label{ap:code-valid}
The analysis presented in this work 
was carried out with the Fisher matrix code \texttt{FisherCLASS}. This code runs in two steps.

The first is computation of the angular power spectra. A Python script repeatedly calls a customised version of the code {\sc class} \texttt{v2.9.4} \citep{class2, CLASSgal} and computes all the angular power spectra needed for the analysis. The spectra are ideally computed in parallel (the script submits a job to a cluster queue for each setting required). The angular power spectra are computed using the number count feature \citep{CLASSgal} and the lensing potential feature in {\sc class}. A few modifications to the public version of the {\sc class} code have been implemented for the purpose of this paper:
i) generic and thus non-Gaussian redshift bins: the redshift distribution of the lenses and sources
can be read for each redshift bin individually; ii) galaxy bias can be redshift dependent within each bin; and iii) if the lensing potential feature is turned on, the output spectra are the shear angular power spectra, and they can 
include an intrinsic alignment systematic 
effect, modelled through the {extended} non-linear alignment model \citepalias{Blanchard:2019oqi}.

The second step is the Fisher matrix analysis. A \texttt{Jupyter Notebook} reads the angular power spectra output from step 1) and estimates the full covariance, the derivative with respect to a chosen set of parameters, and the full Fisher matrix. 
The notebook computes in addition the Fisher matrices for individual probes: GCph, WL, and the GGL terms.

The advantage of this code is that it relies on the well-maintained and tested number count feature in {\sc class}, which allows the relativistic effects to be included in the clustering observables. The code has been validated against the results in \citetalias{Blanchard:2019oqi}. For this purpose, we compared the cosmological forecast obtained with \texttt{FisherCLASS} to the
forecast computed with \texttt{CosmoSIS}\,\footnote{\url{https://bitbucket.org/joezuntz/cosmosis/wiki/Home}}\citep{Zuntz:2014csq}. 

The baseline setting used for this code comparison is the same as the one adopted in \citetalias{Blanchard:2019oqi} for the $\text{GCph} + \text{WL} + \text{GGL}^\text{(GCph, WL)}$ joint analysis. In summary: The cosmological parameter space is $\theta = \{\Omega_{\text{m}, 0}$, $\Omega_{\text{b}, 0}$, $\wzero$, $\wa$, $h$, $n_\text{s}$, $\sigma_8\}$, that is, a flat cosmology with dynamical dark energy.
The galaxy sample is split into ten equally  populated redshift bins, with 
    galaxy number density $n_\text{gal} = 30\,\text{galaxies/arcmin}^{2}$.
 We included as nuisance parameters ten galaxy bias parameters and three parameters for the intrinsic alignment contribution to the WL observable. Finally, the $\ell$-modes included in the analysis range from $\ell_\text{min} = 10$ to $\ell_\text{max, GCph} = 750$ and 
    $\ell_\text{max, WL} = 1500$ for GCph and WL, respectively.

However, we note that the specifications used in the analysis presented in this work are the ones summarised in \Cref{{sec:flag}}. This includes using the redshift distributions shown in Fig.\ \ref{fig:photo-bins}.

In \Cref{fig:code-valid-1}, we present the code comparison for the joint analysis. We show the percentage difference between the constraints obtained with the two codes and the mean values of the two results. The top panel refers to $1\sigma$ marginalised constraints, while the bottom panel shows the comparison for the un-marginalised constraints. The largest discrepancies between the two codes are $\sim 4\%$ for the $1\sigma$ errors and $\sim 2\%$ for the un-marginalised constraints. We note that the outcome of the two codes has been compared for several intermediate steps, different settings, and different probe combinations, always leading to an excellent agreement.

\begin{figure}
  \includegraphics[width=\linewidth]{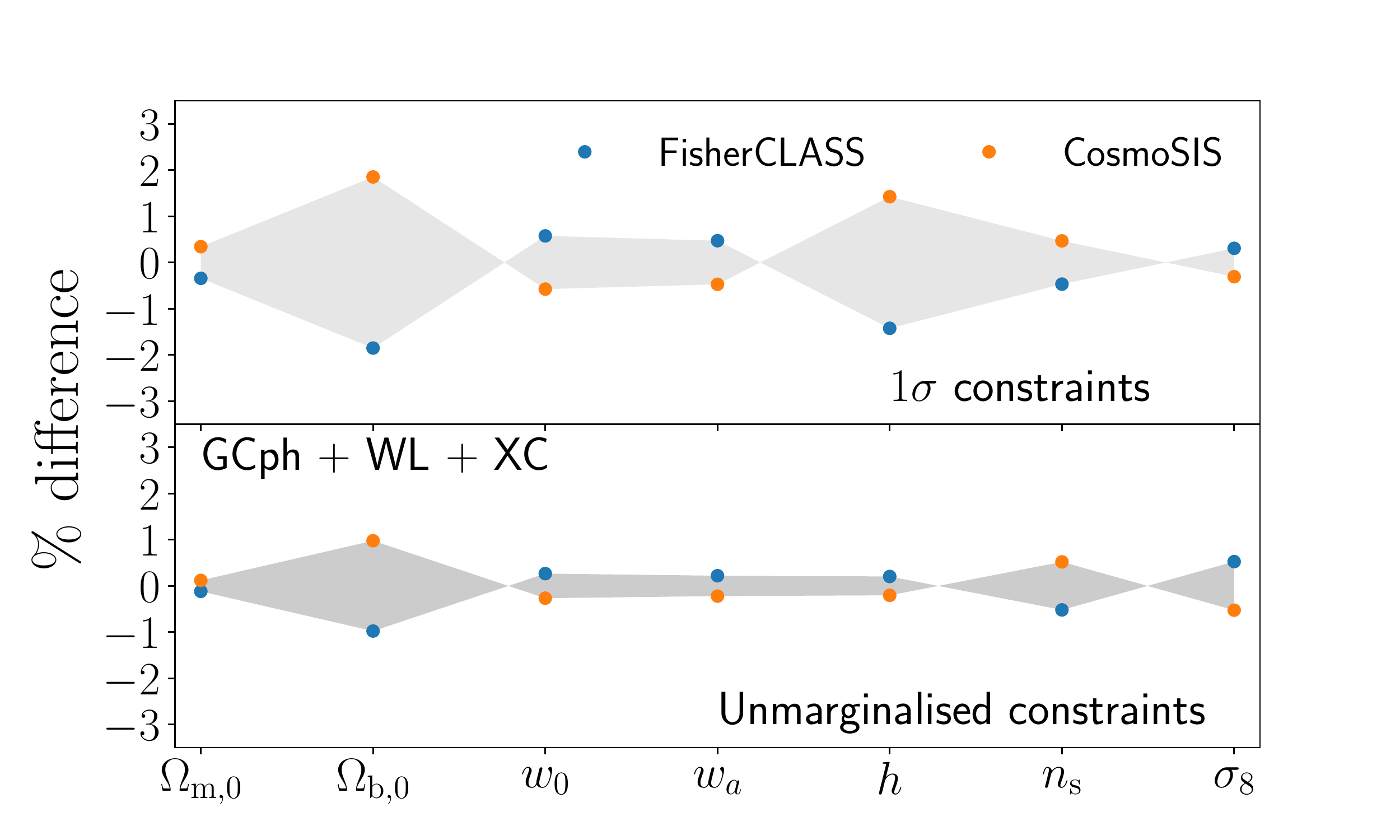}
  \caption{Percentage difference in the $1\sigma$ uncertainties (top panel) and un-marginalised constraints (bottom panel) for the probe combination $\text{GCph} + \text{WL} + \text{GGL}$. This analysis includes  ten and three  nuisance parameters for the galaxy bias and intrinsic alignment contributions, respectively, which are marginalised over in the $1\sigma$ constraints.
  }
  \label{fig:code-valid-1}
\end{figure}

\section{Fitting functions for $b(z)$ and $s(z)$}
\label{ap:fits}
We also fitted the galaxy bias and the local count slope found in the Flagship simulation with simple third-order polynomials. 
We found the following coefficients for the best fit:
\bea
s(z) &=& s_0 + s_1z+s_2z^2+s_3z^3\,,\\
b(z) &=& b_0 + b_1z+b_2z^2+b_3z^3\,,
\eea
with
\be
\begin{array}{cccc}
s_0=0.0842\,, & s_1= 0.0532\,, & s_2 = 0.298\,, & s_3 = - 0.0113\,, \\
b_0=0.5125\,, & b_1= 1.377\,,  & b_2 = 0.222\,, & b_3 = -0.249\,.
\end{array}
\ee

In Fig.~\ref{f:fit} we compare our best fit with the Flagship simulation measurements.
In our calculations we did not use these fits, but we present them here for convenience. The Flagship specifics have been estimated for the survey binning described in \Cref{sec:flag}, and therefore the fitting functions are adapted to this specific configuration. 

\begin{figure}
  \includegraphics[width=\linewidth]{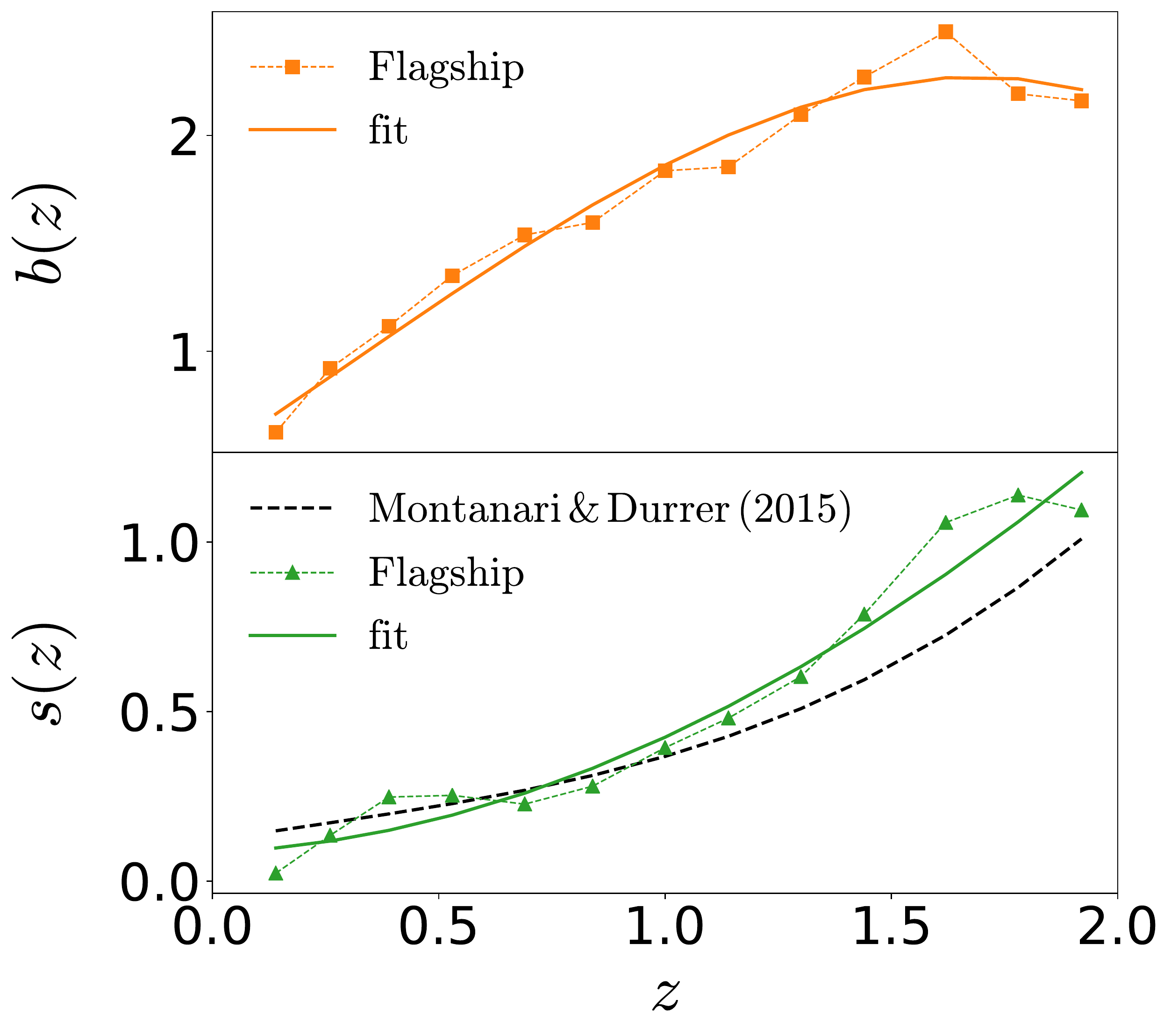}
\caption[]{\label{f:fit} Fit (continuous lines) to the galaxy bias (top panel) and the local count slope (lower panel) together with the simulation results. For the local count slope, we also plot the theoretical function for $s(z)$ derived in~\cite{Montanari:2015rga} for comparison (dashed black line). }
\end{figure}

\end{appendix}
\end{document}